%% file: main.tex
\documentclass[sigconf,edbt]{acmart-edbt2019}

\usepackage{booktabs} 

\setcopyright{none}

\usepackage{color}
\usepackage{multirow}
\usepackage{subfigure}

\usepackage{algorithm}
\usepackage{algpseudocode}

\newcommand{\fix}[1]{[\textcolor{blue}{#1}]}

\newcommand{\hide}[1]{}

\newcommand{\SP}{SP}

\begin{document}
%

\title{Flow Motifs in Interaction Networks}
\author{Chrysanthi Kosyfaki}

\orcid{1234-5678-9012}
\affiliation{%
  \institution{Dept. of Computer Science and Engeeniring}
  \city{University of Ioannina} 
  \state{Greece} 
}
\email{xkosifaki@cs.uoi.gr}

\author{Nikos Mamoulis}

\orcid{1234-5678-9012}
\affiliation{%
  \institution{Dept. of Computer Science and Engeeniring}
  \city{University of Ioannina} 
  \state{Greece} 
}
\email{nikos@cs.uoi.gr}

\author{Evaggelia Pitoura}

\orcid{1234-5678-9012}
\affiliation{%
  \institution{Dept. of Computer Science and Engeeniring}
  \city{University of Ioannina} 
  \state{Greece} 
}
\email{pitoura@cs.uoi.gr}

\author{Panayiotis Tsaparas}

\orcid{1234-5678-9012}
\affiliation{%
  \institution{Dept. of Computer Science and Engeeniring}
  \city{University of Ioannina} 
  \state{Greece} 
}
\email{tsap@cs.uoi.gr}

\input{abstract}

\maketitle

\input{introduction}

\input{relatedwork}

\input{definitions}

\input{algorithm}

\input{experiments}

\input{conclusion}

\bibliographystyle{ACM-Reference-Format}
\bibliography{references} 
\end{document}

%% file: abstract.tex
\begin{abstract}
Many real-world
phenomena
are best represented as interaction networks with dynamic structures (e.g., transaction networks, social networks, traffic networks). 
Interaction networks capture flow of data which is transferred between their vertices along a timeline.
Analyzing such networks is crucial toward comprehending
processes in them.
A typical analysis task is the finding of motifs, which are small subgraph patterns that repeat themselves in the network.
In this paper, we introduce \textit{network flow motifs}, a novel type
of motifs that model significant flow transfer among a set of vertices within a constrained time window.
We design an algorithm for identifying flow motif instances in a large graph. 
Our algorithm can be easily adapted to find the top-$k$ instances of maximal flow. In addition, we design a dynamic programming module that
finds the instance with the maximum flow.
We evaluate the performance of the algorithm on three real datasets and identify flow motifs which are significant for these graphs.
Our results show that our algorithm is scalable and that the real networks indeed include interesting motifs, which appear much more frequently than in randomly generated networks having similar characteristics.  
\end{abstract}


%% file: introduction.tex
\section{Introduction}
\begin{figure}[tbh]
  \centering
  \small
  \begin{tabular}{@{}c@{}c@{}c@{}c@{}}
  \includegraphics[width=0.34\columnwidth]{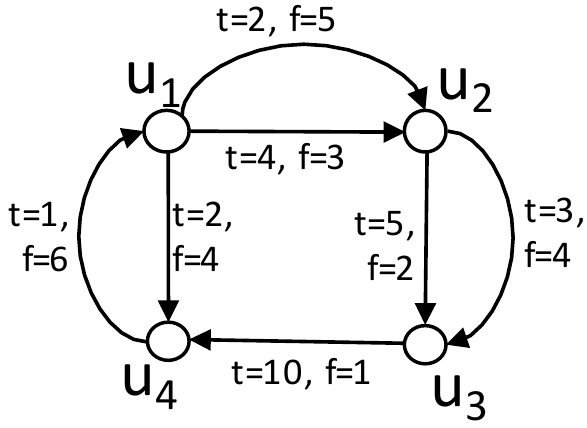}&
  \includegraphics[width=0.08\columnwidth]{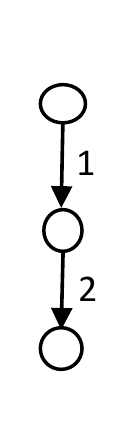}&
  \includegraphics[width=0.28\columnwidth]{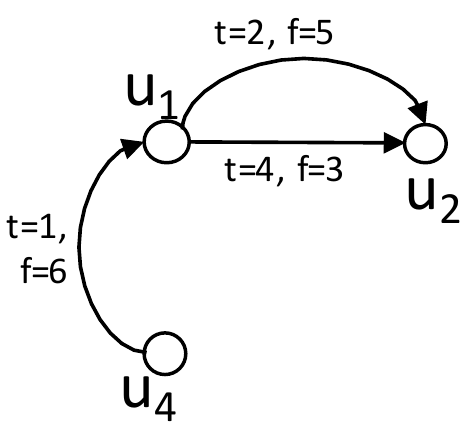}&
  \includegraphics[width=0.25\columnwidth]{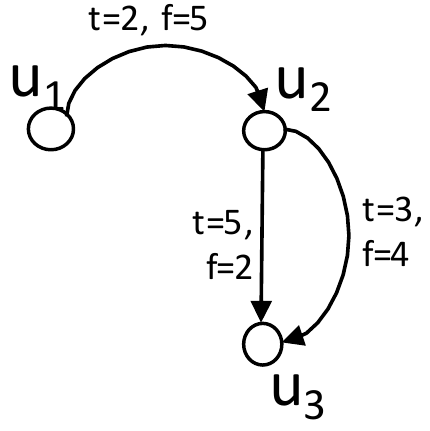}\\
    (a) multigraph&
         (b) motif&
              (c) instance 1&
                   (d) instance 2\\
  \end{tabular}
  \caption{Example of graph, motif, and instances}
  \label{fig:toyexample}
\end{figure}

{\em Interaction networks} 
include 
a large number of highly connected components that dynamically exchange
information.
Examples of such graphs are neural networks,
food webs, signal transfer pathways, the bitcoin network, social
networks, and traffic networks.
An interaction network captures {\em flow of data} (e.g., money,
messages, passengers,  etc.) which is transferred between
its vertices along a timeline. 
In such
a network, there could be multiple edges connecting the same pair of
vertices, modeling data exchange between them at different times. 
Figure \ref{fig:toyexample}(a) shows a small example of an interaction
network, where the vertices represent users who exchange money.
The edges are annotated by timestamped interactions; e.g., edge
$u_1u_2$ with label $t=2, f=5$ denotes that user $u_1$
sent 5 units of flow (money) to user $u_2$ at time $2$.

Interaction networks are a powerful and versatile model, and as such
they have been studied extensively in the literature \cite{DBLP:conf/cikm/ZhaoTHOJL10,li2018temporal,DBLP:conf/edbt/ZufleREF18}. In this paper, we consider the
problem of finding small
characteristic patterns in the networks, such as chains, triangles or cycles.
These patterns are called
{\em network motifs}. A motif is a subgraph that appears significantly
more often in a real network than in a
randomized network with similar characteristics \cite{Milo02networkmotifs:}.
Finding motifs is
a method of identifying functional properties of a network.
Previous work mainly focused on static motif patterns \cite{Milo02networkmotifs:, Yaverolu2014RevealingTH}.
Recently, there has been increasing interest in analyzing 
temporal networks  \cite{DBLP:journals/jcss/KempeKK02,DBLP:journals/corr/abs-1107-5646,DBLP:conf/cikm/ZhaoTHOJL10,DBLP:conf/wsdm/ParanjapeBL17,DBLP:conf/edbt/ZufleREF18}, where edges carry timestamps that signify the
time of interaction 
between vertices.
However, to the best of our knowledge, there is no previous work on
motif search that
considers the 
flow of data between connected nodes.
Motivated by this, we define the concept of {\em flow motifs} in
temporal interaction networks and study their identification. 

Our definition of flow motifs extends a well-accepted
definition of
temporal motifs
\cite{DBLP:conf/wsdm/ParanjapeBL17}. We define flow motifs as small
graphs whose edges are ordered; the order defines how the data
flows between the vertices. 
An instance of the motif is a subgraph of the interaction network, whose edges obey the total
order specified by the edges of the motif. 
Moreover, the time difference
between the temporally last and first edges should not
exceed a pre-defined threshold $\delta$ which is a parameter of the
motif.
These requirements are the same as in the temporal
motif definition of \cite{DBLP:conf/wsdm/ParanjapeBL17}, which however
disregards the data flow in interactions.
The distinctive feature of our flow motifs is that, in a flow motif instance, 
multiple edges of the graph can instantiate a single edge of the
motif, if they satisfy the order constraint with the edges that
instantiate the motif's previous and next edges.
The flow values in the edge-set that instantiates a motif edge are
{\em aggregated} to a single value, which captures the {\em total flow} passing
through the motif edge. The {\em minimum aggregated flow} at any motif
edge defines the flow of the instance. In order
for the instance to be valid, we require that its
flow exceeds a threshold $\phi$.

Consider again the interaction network of Figure
\ref{fig:toyexample}(a).
Assuming that the
motif of interest is a chain of three nodes (Figure
\ref{fig:toyexample}(b)), where the labels in edges specify the flow order and that $\delta=5$ and
$\phi=5$, the two subgraphs of Figures
\ref{fig:toyexample}(c) and \ref{fig:toyexample}(d) are instances of the motif
because the sets of edges mapped to each motif edge satisfy (i) the
time order constraint of the motif and (ii) thresholds $\delta$ and
$\phi$.  
For example, in Figure \ref{fig:toyexample}(d), both edges that connect $u_2$ to
$u_3$ are temporally after the edge that connects $u_1$ to $u_2$ and
their aggregated flow is 6 ($\ge \phi$); in addition, the time
difference between the temporally first and last edges in the instance
is $5-2=3$ ($\le \delta$).


Overall, a valid flow motif instance
should 
satisfy three requirements: (a) a \emph{structural constraint},
defined by the graph structure of the motif; (b) a \emph{temporal
  constraint} defined by the temporal window size $\delta$; (c) a
\emph{flow constraint} defined by the minimum flow value $\phi$.

Flow motifs correspond to frequently occurring sub-structures with
high activity that appear in short time windows. Finding instances of flow
motifs is of great importance in understanding interaction networks.
For instance, in networks that model money transfers, 
flow motifs
correspond to transaction patterns  involving significant flow of
money that appear more frequently than expected. 
Flow motif search is of particular interest to Financial intelligent units (FIUs); these are organizations which identify suspicious flow patterns that may suggest criminal behavior (e.g., money laundering). 
Belize FIU (fiubelize.org) and Hong Kong's JFIU (www.jfiu.gov.hk)
indicate as suspicious patterns which include `smurfing' (i.e., numerous small-volume transfers which aggregate to large amounts), cyclic transactions between parties, and chains of significant money transfers within limited time (e.g., payments out which are paid in on the same or previous day). 
In addition, bitcoin theft has been associated to flow patterns in \cite{DBLP:conf/imc/MeiklejohnPJLMVS13}.
In communication and
social networks, flow motifs may reveal common patterns of influence
\cite{DBLP:conf/sdm/LeskovecMFGH07,Gomez-RodriguezLK12}.
For example, the strength of the relationships between two social network
users is correlated with the frequency of online interactions between them
\cite{DBLP:conf/www/XiangNR10}. This implies that groups of users with
frequent communication between them within a short period have high
chance to influence each other.


Given a large interaction network,
we propose an algorithm that takes
as input a flow motif and efficiently finds its instances in the
network. Our algorithm operates in two phases. First, the structural
matches of the motif (disregarding temporal and flow information) are
identified. Then, for each structural match, we find the motif
instances which satisfy the temporal and flow constraints.
This is achieved by sliding a time window of the same length as the 
duration constraint of the motif and systematically finding the combinations of
edges that constitute motif instances.
Compared to motif search algorithms from previous work, our algorithm
is novel in that it considers the aggregated flow on multiple edges
that connect the same pair of nodes in the network during the
construction of the motif instances. Due to the large number of
possible edge combinations, the problem is harder compared to finding
instances of motifs, by disregarding flows and multiple edges. 
Our algorithm effectively uses the duration and flow constraints to prune the space.   
We also suggest a variant of the algorithm that identifies the
top-$k$ instances of an input flow motif with the highest 
flow. Finally, we propose a dynamic
programming module for the algorithm, 
for the problem of finding the motif instance with the
maximum flow. 

We evaluate the performance of the algorithm on three real datasets
of different nature (bitcoin user network, facebook network, and Passenger flow network).
We compare the performance of our algorithm to a baseline method which builds up motif instances by joining their components and demonstrate the superiority of our approach against this alternative method.
We also show that our tested flow motifs indeed appear more frequently in real networks than in
randomized networks having the same characteristics as the real ones.

In summary, this paper makes the following contributions:
\begin{itemize}
\item
We propose the novel concept of flow motif. To our knowledge, this is
the first work that defines and studies the search of flow motifs in
interaction networks. 
\item
We propose an efficient algorithm for finding flow motif instances in
large interaction networks and variants of it that identify the
instances of a motif with the maximum flow.
\item
We evaluate our approach using three real datasets, and demonstrate that
it scales well for large data.
\item
We investigate the significance of the tested motifs in the real
networks.
\end{itemize}
 

The rest of the paper is organized  as follows. Section
\ref{sec:related} describes work related to network flow motifs, which
are then formally defined in 
Section \ref{sec:def}.
Our motif search algorithm is presented in Section
\ref{sec:algorithm}.
Section \ref{sec:topk} shows how to extend our algorithm to find the $k$ instances of a given motif with the maximum flow.
In Section \ref{sec:exp}, we experimentally evaluate our algorithm and the significance of the motifs by using a randomization approach. Finally, in Section \ref{sec:conclusion}, we conclude our paper and give directions for future work.

%% file: relatedwork.tex
\section{Related Work}\label{sec:related}
There has been a lot of research interest in motif search and mining in
interaction networks \cite{DBLP:journals/bioinformatics/WernickeR06,DBLP:conf/icde/SemertzidisP16,DBLP:journals/corr/Holme15a}. In this section we summarize the most representative works in static and temporal networks.

\textbf{Static Networks.} Milo et al. \cite{Milo02networkmotifs:} 
introduced the concept of motifs and studied their identification in
large graphs. 
They defined a network motif as a \textit{``pattern of interconnections
  occurring in complex networks at numbers that are significantly
  higher than those in randomized networks''}. They
investigated motif discovery in directed networks, which do not carry temporal
information (i.e., the motifs do not consider the time when the
interactions took place).

FANMOD \cite{DBLP:journals/bioinformatics/WernickeR06} is an efficient tool for finding network motifs in static networks, up to a size of eight vertices. Given a subgraph size, the tool either enumerates all subgraphs of that size or samples them uniformly. The identified subgraphs are grouped into classes based on their isomorphism. The significance of each class is finally measured by counting their frequencies in a number of random graphs (generated by switching edges between vertices in the original network).

\textbf{Temporal Networks.} In {\em temporal networks}, the interactions between vertices are
labeled by the time when they happen. 
Fundamental definitions, concepts, and problems on temporal networks
are given in \cite{DBLP:journals/jcss/KempeKK02}. 
For instance, the concept of {\em time-respecting path} and
its relation to network flows are defined and studied here. 
 
Paranjape et al. \cite{DBLP:conf/wsdm/ParanjapeBL17} define motifs in
temporal networks
as small connected graphs, whose edges are
temporally ordered. Instances of a motif are subgraphs that 
structurally match the motif and their edges
obey the order. In addition, the time-difference between the
temporally last and
the first edges should not exceed a motif {\em duration} constraint
$\delta$. 
They propose a general algorithmic framework for
computing the number of motif instances in a graph and
fast algorithms that 
count
certain classes of temporal motifs.
Our network flow motifs are similar to the temporal motifs of
\cite{DBLP:conf/wsdm/ParanjapeBL17}, however, in our case (i) a motif
edge can be instantiated by multiple edges of the graph and (ii) we
introduce and consider a minimum flow requirement.  

Another work that defines and studies the enumeration of temporal
motifs is \cite{DBLP:journals/corr/abs-1107-5646}. In the context of this work, the
interactions between vertices are not instantaneous but they carry a
duration interval. 
Motifs are again subgraphs whose edges are temporally ordered. As
opposed to \cite{DBLP:conf/wsdm/ParanjapeBL17}, there is no
 $\delta$ threshold between the last and the first edge
 in a motif instance. Instead,
a maximum time-difference $\Delta t$
between consecutive edges in a motif instance
is allowed.

{\em Communication motifs} are suggested as a model for capturing
the structure of human interaction in networks over time.
Zhao et al.~\cite{DBLP:conf/cikm/ZhaoTHOJL10} studied the evolution of 
such behavioral patterns in social networks.
For any two adjacent interactions, the term {\em maximum flow} is used 
to characterize those interactions that are the most probable to belong 
to the same information
propagation path among any such adjacent interactions.
On the other hand,
in our context, flow refers to the data (e.g., money, messages, etc.)
being transferred from one node along network paths.
Another
work that studies behavioral patterns in social networks by defin-
ing and mining communication motifs between people in social
networks is \cite{DBLP:conf/sigmod/GurukarRR15}. A scalable mining technique (called COMMIT)
for communication motifs in interaction networks is proposed.

A recent
work that studies the structure of social networks and the temporal
relations between entities in them is \cite{DBLP:conf/edbt/ZufleREF18}.
Temporal pattern search is proposed as a tool in this direction.
In order to facilitate the efficient retrieval of pattern instances,
occurrences of small patterns are precomputed and indexed.
 
Motif discovery on Heterogeneous Information Networks (HiNs)
which carry temporal information was also recently studied~\cite{li2018temporal}.
In such graphs,
some nodes are associated to events (which
happened at a specific time).
A motif is then defined by a graph and a maximum temporal difference
between the events that instantiate its event nodes. 
As in the rest of previous work, any data flow on the edges of the
network is disregarded in the definition and search of motifs.

%% file: definitions.tex
\section{Definitions}\label{sec:def}
In this section, we formally define flow motifs and the graph wherein
they are identified.
Table \ref{table:notations} shows the notations used frequently in the paper.

\begin{table}[ht]
\caption{Table of notations}\label{table:notations}
\vspace{-1ex}
\centering
\small
\begin{tabular}{|c | c|}
\hline
Notations &Description\\ %
\hline  
$G_M(V_M,E_M)$ & graph structure of motif $M$\\ 
$\delta$&  duration constraint of a motif \\ 
$\phi$& flow constraint of a motif \\
$l(e)$ & order of edge $e$ in a motif $M$\\
$SP_M$& spanning path of motif $M$\\
$e_i$ or $\SP_M[i]$ & $i$-th edge of motif $M$\\
$\SP_M[i:j]$ & subpath $e_i\dots e_j$ of $SP_M$\\
$G(V,E)$ & input graph \\ 
$E(u,v)$ & set of edges in $G$ from $u$ to $v$\\
$f(e)$& flow on edge $e$\\ 
$t(e)$& timestamp of edge $e$ \\ 
$f(G_I)$ & flow of motif instance $G_I$\\
$G_T(V,E_T)$ & time-series graph equivalent to $G(V,E)$ \\ 
$(t,f)$ & flow interaction element on an edge of $E_T$\\
$R(u,v)$ & time series on edge $(u,v)\in E_T$\\
$R(e_i)$ & time series on edge of $E_T$ mapped to $e_i$\\
$S$ & set of structural matches of a motif\\
$G_s$ & structural match of a motif\\
\hline
\end{tabular}
\end{table}

The input to our problem is a directed multigraph $G(V,E)$, where each pair of
nodes $u,v \in V$ can be connected by any number of edges in $E$.
We denote by $E(u,v)$ the edge-set from $u\in V$ to $v\in V$.
Each edge $e \in E$ is annotated by a unique {\em timestamp} $t(e)$ in a continuous time domain $\mathcal{T}$
and a positive real number $f(e)$, called {\em flow}.

\begin{figure}[htb]
\centering
  \includegraphics[width=0.45\columnwidth]{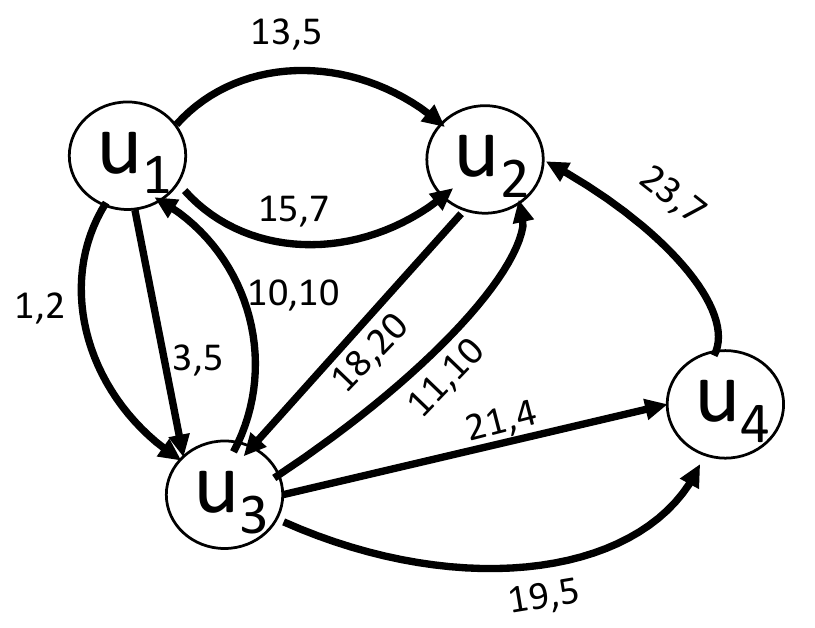}
  \caption{Example of an interaction graph (bitcoin user graph)}
  \label{fig:inputgraph}
\end{figure}

Figure \ref{fig:inputgraph} shows an example of an input graph $G$
from a real application,
where vertices correspond to users (addresses) of the bitcoin network and edges
correspond to transactions between them. Each edge is annotated by the
timestamp of the transaction followed by the transaction amount. 
For example, user $u_1$ at
timestamps $13$ and $15$ sent $5$ and $7$ bitcoins, respectively, to $u_2$.

\begin{definition}[Flow Motif]\label{def:motif}
A network flow motif 
$M$
is a triplet $(G_M,\delta,\phi)$ consisting of 
(i) a directed graph $G_M(V_M,E_M)$ with $m=|E_M|$
edges, where each edge $e$ is labeled by a unique number $\ell(e)$ in
$[1,m]$;
(ii) a value $\delta$, which  defines an upper-bound on the duration of the
motif; and (iii) a value $\phi$, which defines a lower bound on the flow of the motif.
\end{definition}

The labels of the edges in the motif graph $G_M$ define a total order
of the edges that models the direction of the flow in $G_M$. For
example, if $G_M$ consists of two edges $(u,v)$ and $(v,w)$ and we
have $\ell(u,v) = 1$ and $\ell(v,w) = 2$, this means that the flow in
the graph originates from node $u$, it is first transferred to $v$,
and then from $v$ to $w$.

\begin{figure}
\centering
\includegraphics[width=0.8\columnwidth]{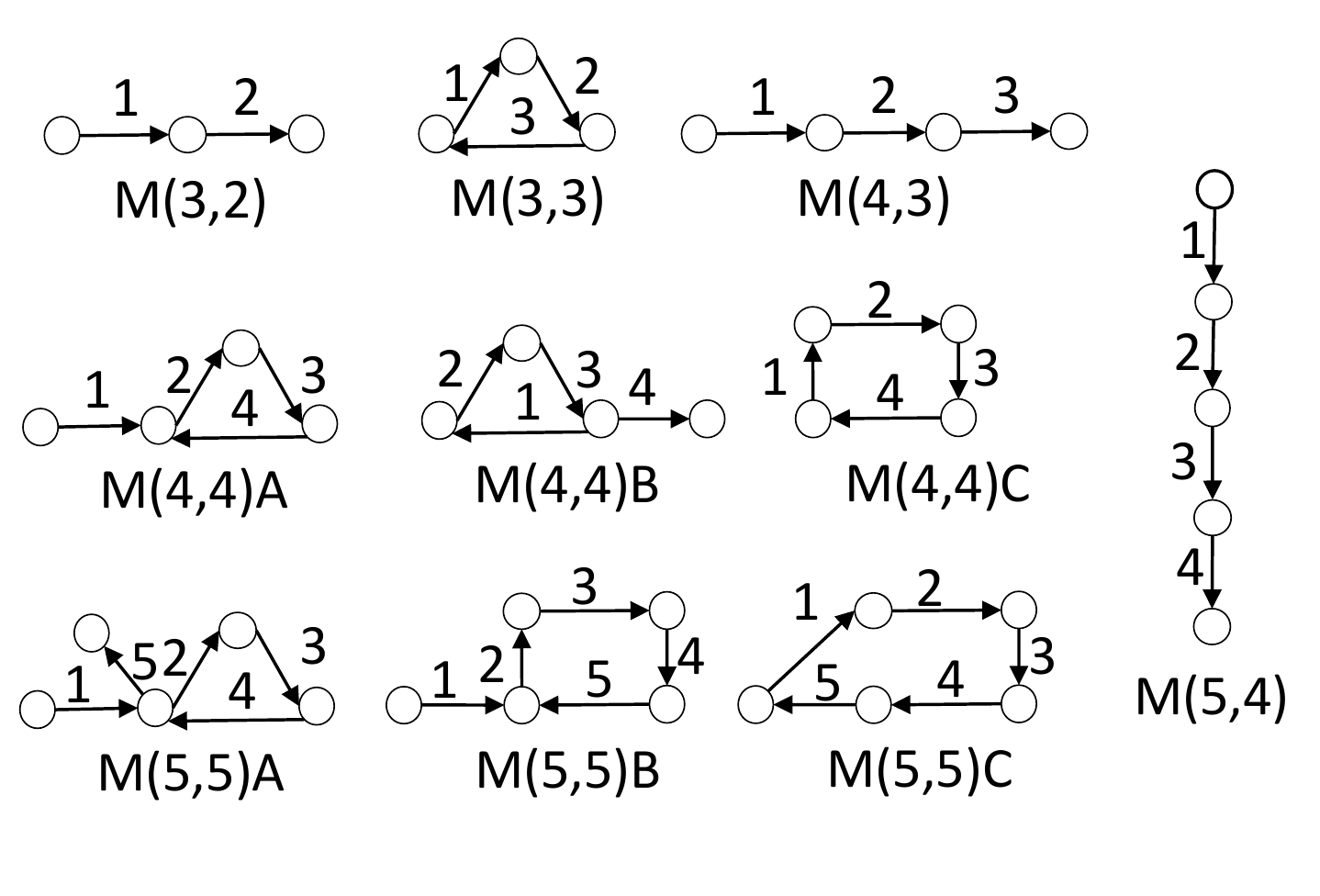}
\caption{Examples of motifs.}
 \label{fig:motifs} \vspace{-.5em}
\end{figure}

Figure \ref{fig:motifs}
shows some examples of
motifs (we only show the motif graphs $G_M$, but not the thresholds
$\delta$ and $\phi$).
The numbers in the parentheses denote the number of nodes and
edges in the motifs. For example, the motif labeled $M(3,3)$ models a cyclic
flow between three nodes.

We assume that the ordering of the edges according to their labels
defines a {\em path} in the graph $G_M$. We refer to this path as the
\emph{spanning path} of the motif, and we denote it as $\SP_M$. The
spanning path is not necessarily a simple path, i.e., there may be
repeated vertices in the path. We sometimes refer to a motif graph
$G_M$ by its spanning path $\SP_M=e_1e_2\dots e_m$, i.e., the
total order of its edges, where $e_i$ denotes the edge with label $i$.
For example, we may refer to motif $M(3,3)$ in Figure
\ref{fig:motifs}  by the sequence $\SP_{M(3,3)}=e_1e_2e_3$ of its
three edges.
In addition, we use $e_i$ or $\SP_M[i]$ to denote the $i$-th edge of the motif, and
$\SP_M[i:j]$ to denote the subsequence of edges $e_i\dots e_j$ along the
path.
We now define motif instances as follows.

\begin{definition}[Flow Motif Instance]\label{def:instance}
An instance of a motif $M = (G_M,\delta,\phi)$ in the graph $G(V,E)$ is a subgraph, $G_I(V_I,E_I)$, $V_I \subseteq V$, $E_I\subseteq E$ of $G$ with the following properties:
\begin{itemize}
    \item There is a bijection $\mu:V_M\rightarrow V_I$ from the vertex set of the motif graph $V_M$ to instance vertex set $V_I$.
    \item For every edge $(u,v) \in E_M$ there is a non-empty set of
      edges 
$E_I(\mu(u),\mu(v))$ 
in $G_I$, such that 
$E_I(\mu(u),\mu(v))
\subseteq E(\mu(u),\mu(v))$. In addition, $E_I = \bigcup_{(u,v)\in E_M}E_I(\mu(u),\mu(v))$.
    \item The edge-sets in $G_I$ are \emph{time-respecting}:
    For every pair of edges $(u,v)$ and $(v,w)$ in $E_M$, 
if $l(u,v)<l(v,w)$, then for every pair of edges 
$e_i\in E_I(\mu(u),\mu(v))$, 
$e_j\in E_I(\mu(v),\mu(w))$, 
$t(e_i)< t(e_j)$.
    \item The maximum time difference between any two edges in $E_I$ is at most $\delta$.
    \item The sum of flows of any edge-set in $E_I$ is at least $\phi$.
\end{itemize}
\end{definition}

\begin{figure}[htb]
  \centering
  \subfigure[maximal instance]{  
    \label{fig:maxinstance}
    \includegraphics[width=0.45\columnwidth]{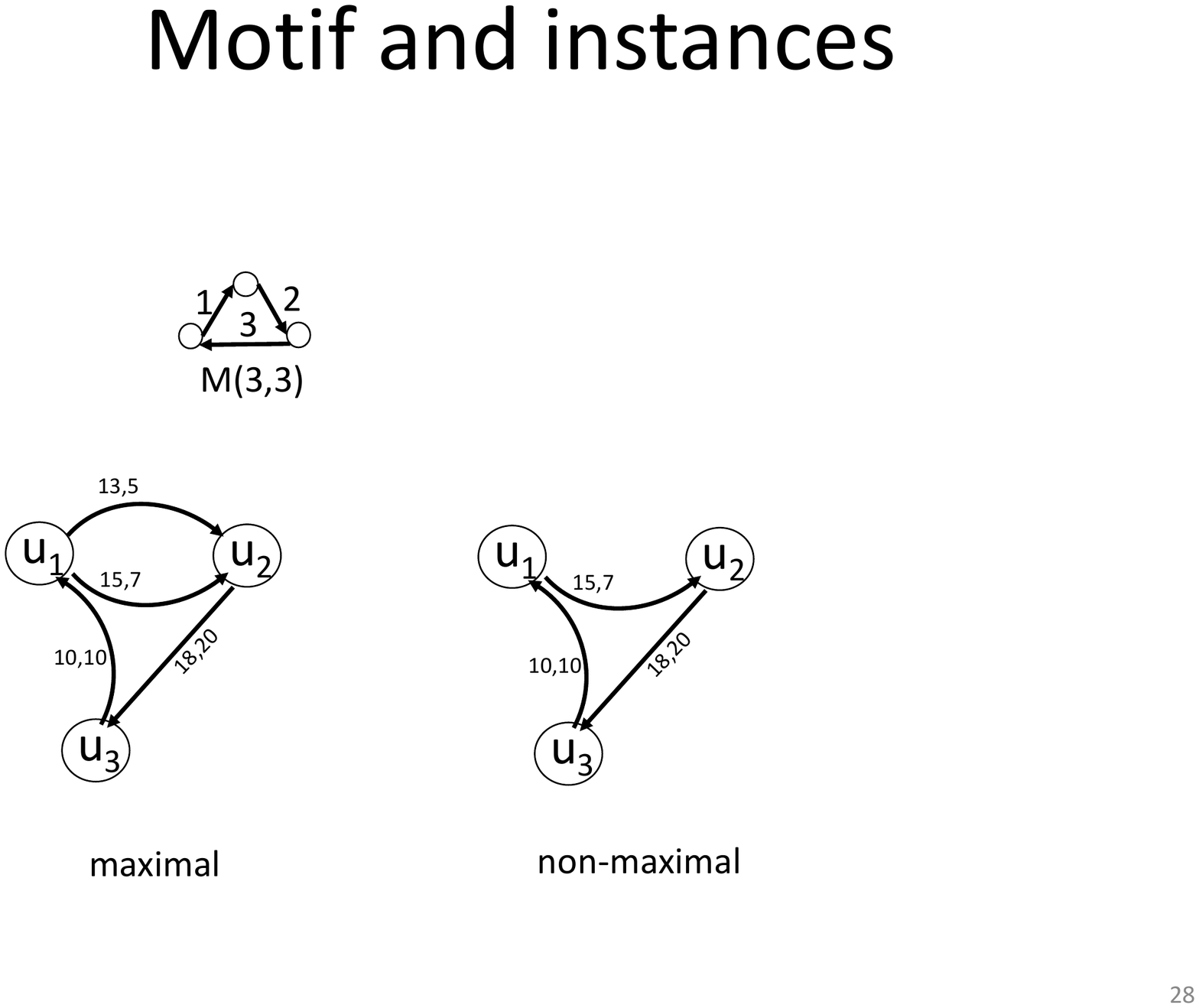}
    }
  \subfigure[non-maximal instance]{
   \label{fig:nonmaxinstance}
    \includegraphics[width=0.45\linewidth]{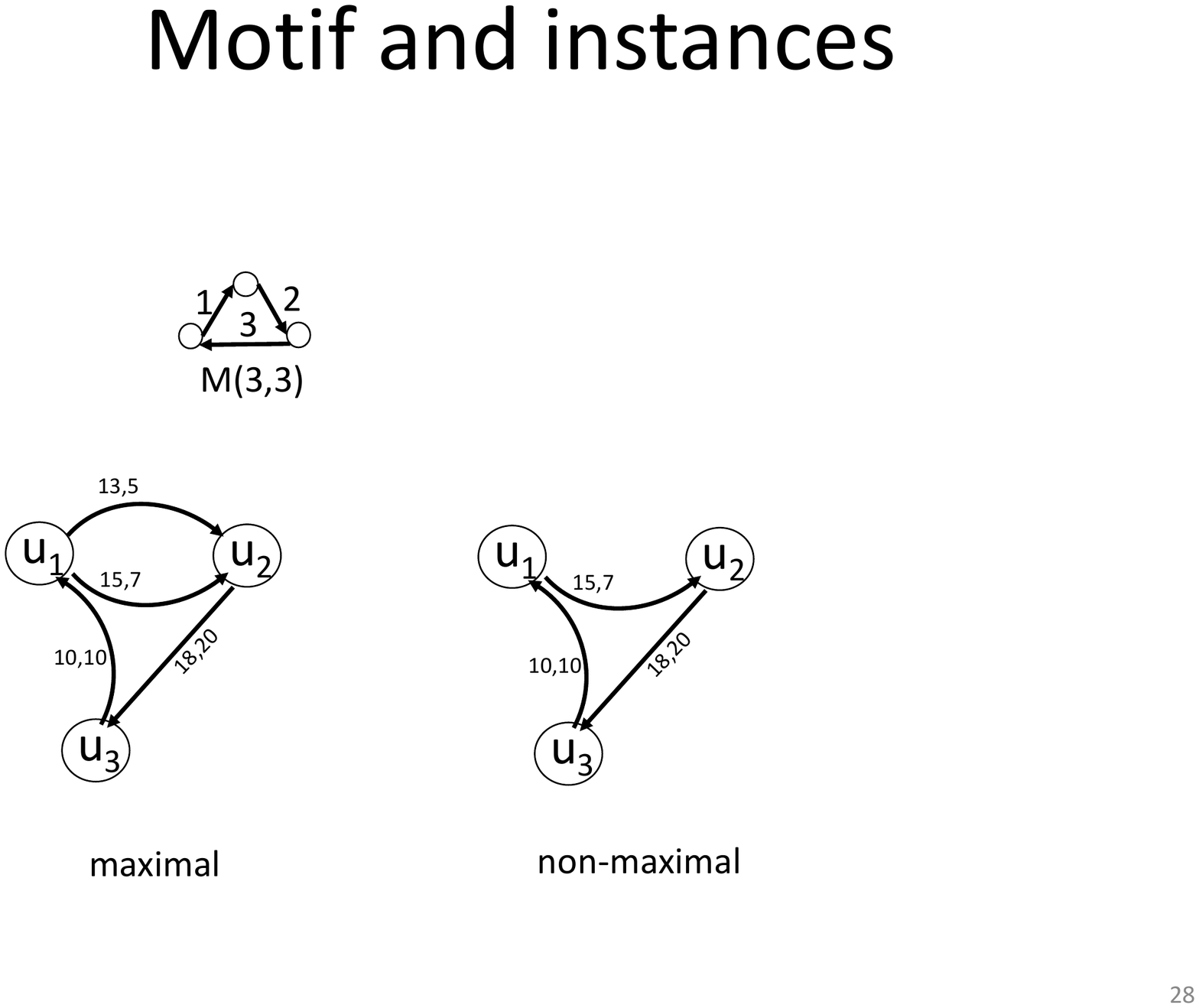}
    }
  \vspace{-0.1in}
  \caption{Examples of motif instances}
  \label{fig:instances}
\end{figure}

The first two conditions express 
a structural requirement on the matching subgraph,  the third and
fourth conditions temporal constraints, and the last condition a
minimum flow constraint.
Figure \ref{fig:maxinstance} shows an instance of $M(3,3)$ in the
graph of Figure \ref{fig:inputgraph}, assuming that $\delta=10$ and
$\phi=7$. $u_3,u_1$, and $u_2$ are mapped to the first, second, and
third node of $M(3,3)$ according to the order of its edges. $u_1$ and
$u_2$ in the instance are linked by two edges which are both
temporally after the edge(s) that link $u_3$ to $u_1$ and before the
edge(s) that link $u_2$ to $u_3$. The maximum time difference between any
two edges is 8 ($\le\delta$) and the aggregate flows on $E_I(u_3,u_1)$,
$E_I(u_1,u_2)$, and $E_I(u_2,u_3)$ are 10, 12, and 20,
respectively (i.e., each of them is at least $\phi$).
If we denote $M(3,3)$ by its spanning path $\SP_{M(3,3)}=e_1e_2e_3$,
we can express the instance of
Figure \ref{fig:maxinstance} by 
$[
e_1\leftarrow \{(10,10)\},
e_2\leftarrow \{(13,5),(15,7)\},
e_1\leftarrow \{(18,20)\}
]$.

For the ease of exposition, we define the flow $f(G_I)$ of an instance 
$G_I$ of motif $M$ as the minimum total flow among all edge-sets 
$E_I(\mu(u),\mu(v))$ which instantiate the edges $(u,v)$ of $M$.
Formally:
\begin{equation}\label{eq:instanceflow}
f(G_I)=\min_{(u,v)\in E_M}\sum_{e\in E_I(\mu(u),\mu(v))}f(e)
\end{equation} 

We now define the concept of motif instance maximality.

\begin{definition}[Instance Maximality]\label{def:instmax}
An instance $G_I(V_I,E_I)$ of a motif $M=(G_M,\delta,\phi)$ is maximal iff, 
the addition of one more edge to any edge-set 
$E_I(\mu(u),\mu(v))$ 
of $G_I$ from the corresponding edge-set $E(\mu(u),\mu(v))$ of $G$ violates the duration or flow constraints of the motif. 
\end{definition}

For example, assuming that $\delta=10$ and $\phi=7$, Figure
\ref{fig:nonmaxinstance} shows an instance of $M(3,3)$ in the
graph of Figure \ref{fig:inputgraph}, which is not maximal.
This is because
the addition of edge (13,5) to  $E_I(u_1,u_2)$ results in the valid
instance of  Figure \ref{fig:maxinstance}.
In this paper, we focus on finding maximal instances of motifs only,
because non-maximal ones are redundant and considering them can
mislead towards the importance of a motif. For example, if $\phi=0$,
all combinations of subsets of the edge-sets that form a valid motif
instance are also valid (but not maximal) instances. Considering
them would exponentially increase the total number of motif instances,
potentially over-estimating its importance.

%% file: algorithm.tex
\section{Finding Flow Motif Instances}\label{sec:algorithm}
\begin{figure}[htb]
  \centering
  \subfigure[multigraph]{
    \label{fig:multigraph}
    \includegraphics[width=0.45\columnwidth]{multigraphb.pdf}}
  \subfigure[time series graph]{
    \label{fig:flatgraph}
    \includegraphics[width=0.48\linewidth]{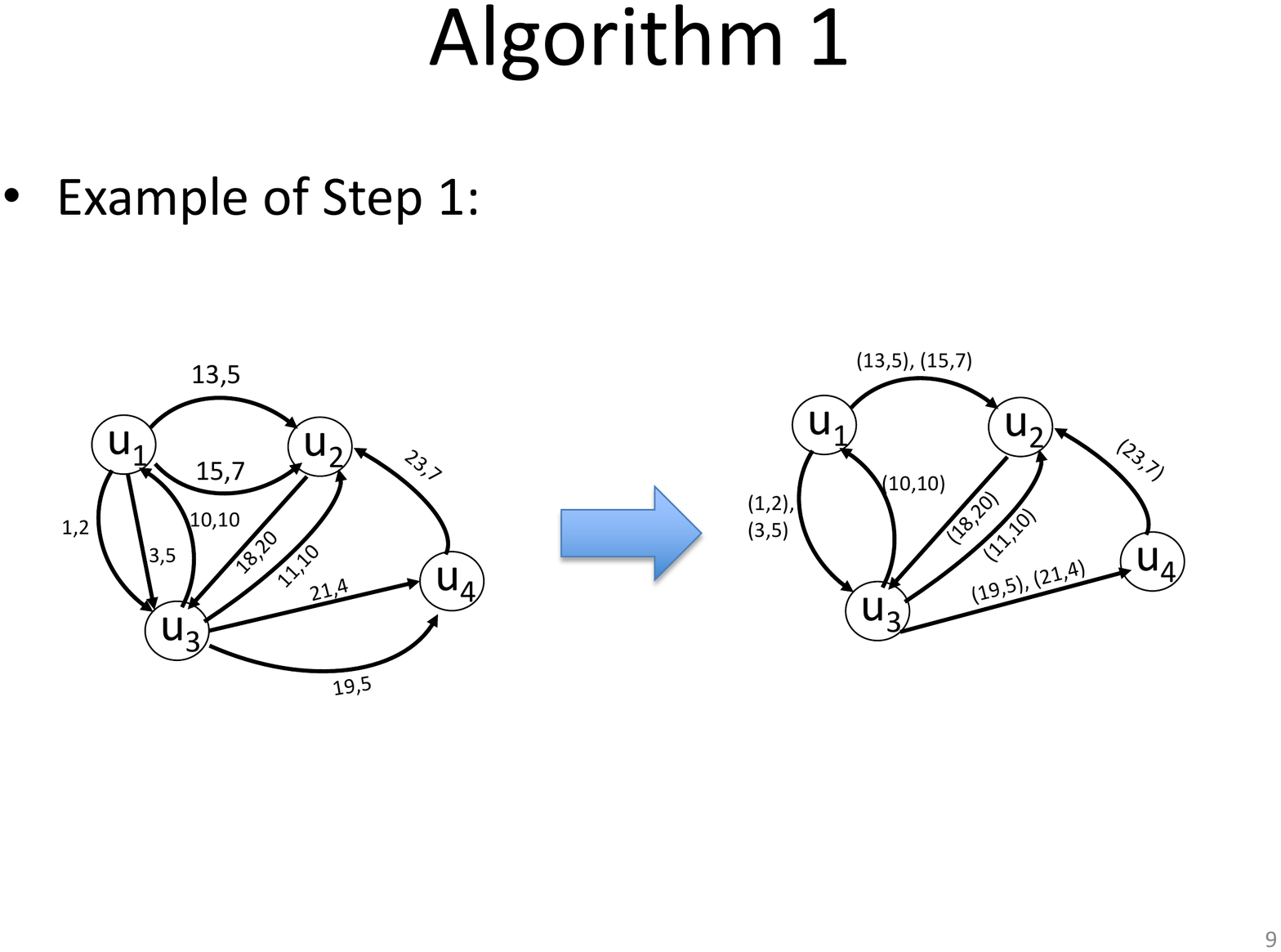}}

  \vspace{-0.1in}
  \caption{From a multigraph to a time series graph}
  \label{fig:step1}
\end{figure}

We now present an efficient algorithm for enumerating the maximal instances of a
given motif $M(V_M,E_M)$ in an input graph $G(V,E)$. For the ease of presentation,
we consider the input graph $G$ not as a temporal multi-graph, but as
a graph where all original edges from a vertex $u\in V$ to a vertex
$v\in V$ are {\em merged} to a single edge.
The single edge $(u,v)$ is associated with an 
\emph{interaction time-series} $R(u,v) = \{(t_1,f_1),(t_2,f_2),\dots,(t_m,f_m)\}$.
Each pair $(t_i,f_i)$ represents a \emph{flow interaction} occurring at time $t_i$ with flow transfer $f_i$ from $u$ to $v$.
The interaction time series is ordered in time.
Figure \ref{fig:step1} shows an
example of how the edges of a multigraph $G$ are merged to
time series. For example, the two edges from $u_1$ to $u_2$ are
considered as a single edge; the two edges with timestamps 13 and 15
are now considered as a time series on a single edge $(u_1,u_2)$. The
conversion of the multigraph to a graph does not have to be explicitly
performed; for each connected pair of vertices, it suffices to
consider their multiple edges ordered by timestamp.
We will use $G_T(V,E_T)$ to denote this graph and we will refer to it
as the \emph{time series graph}.

Our algorithm takes as input the multigraph $G(V,E)$
and a motif $M=(G_M,\delta,\phi)$, and finds all instances of $M$ in
$G$. The algorithm operates on the time series graph $G_T$ and works in two phases P1 and P2:
\begin{enumerate}
\item[P1] Find the set $S$ of all 
{\em structural matches}
  of graph $G_M$ in graph $G_T$, disregarding the labels on the edges and
  constraints $\delta$ and $\phi$.
\item[P2] For each $G_s\in S$, using the time series of the edges in
  $G_s$, find all instances of $M$ in $G_s$ (which should satisfy the duration and flow constraints defined by $\delta$ and $\phi$).
\end{enumerate}
We now elaborate on the two phases.

\begin{figure}[htb]
  \centering
  \includegraphics[width=0.95\columnwidth]{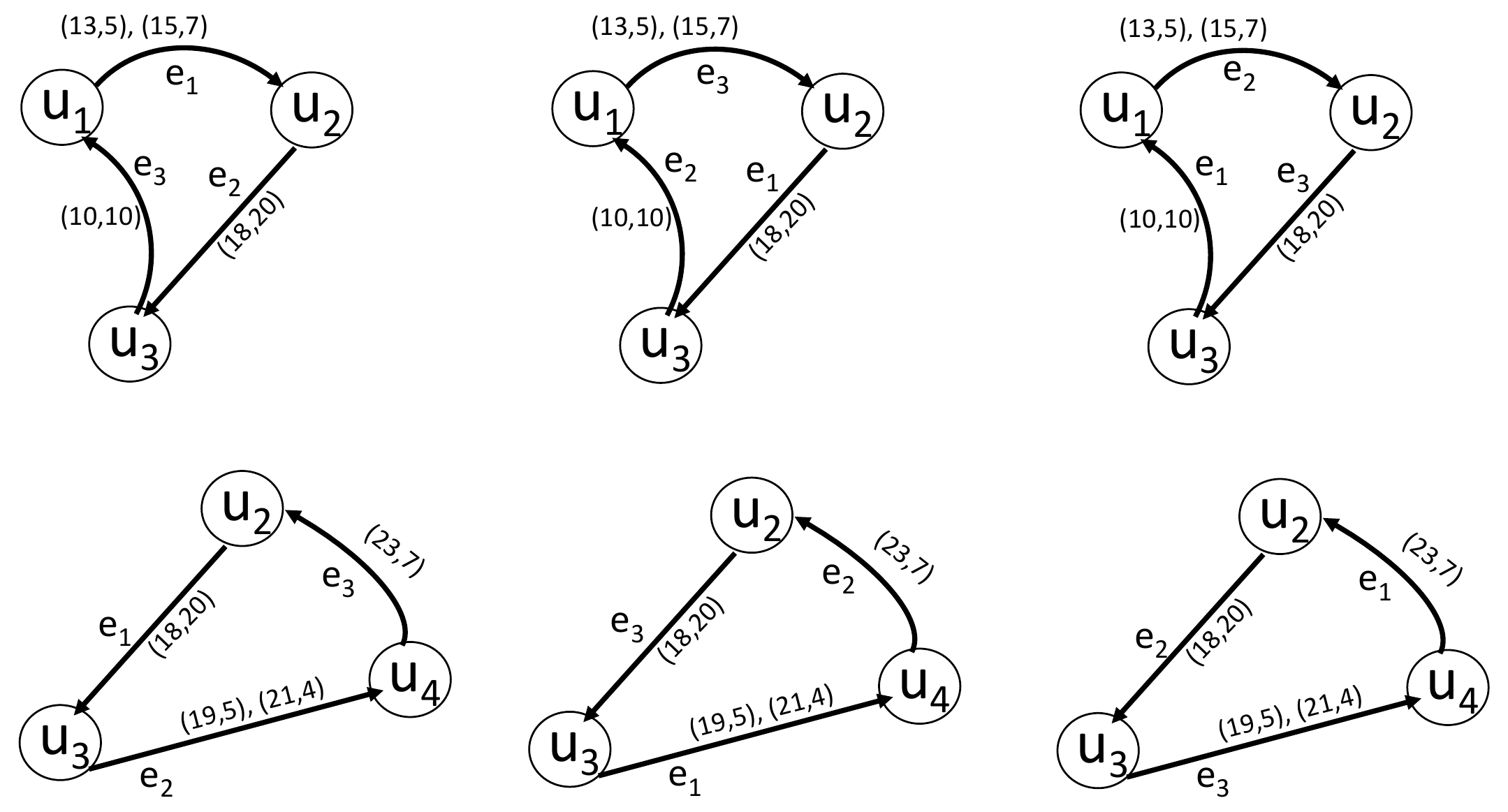}

  \vspace{-0.1in}
  \caption{Structural matches of $M(3,3)$ (phase P1)}
  \label{fig:flowagnostic}
\end{figure}

{\noindent\bf Phase P1:} To illustrate phase P1, as an example, consider the graph $G_T$ of Figure \ref{fig:flatgraph} and the
motif $M(3,3)$ shown in Figure \ref{fig:motifs}. Figure
\ref{fig:flowagnostic} shows all six structural  matches of
$M(3,3)$ in $G_T$ found in phase P1.
The labels $\{e_1,e_2,e_3\}$  on the edges of the matches indicate
the edges of the motif on which they are mapped.
For example, edge $(u_1,u_2)$ of the first match is mapped
to the first edge $e_1$ of the motif.


Algorithmically, for phase P1, any graph pattern matching algorithm for static graphs can be
used (e.g., \cite{DBLP:journals/bioinformatics/WernickeR06}).
In our implementation, we exploit the fact that the ordering of the edges defines
a path. Using a modified depth-first search algorithm on $G_T$, we can
extract all paths of length $|E_M|$ that are structural
matches of $G_M$ in $G_T$. Specifically, in a loop, we map every
node in $G_T$ to the first node in $G_M$ (i.e., the origin node of the
first edge in $G_M$) and recursively find all paths that originate
from that node and map to the spanning path $SP_M$ of $G_M$. For example,
for motif $M(3,3)$, the depth-first search algorithm should make sure that
the last vertex of the traversed path is the same as the first vertex
of the path. Hence, the algorithm on the graph $G$ of our running
example would identify path $u_1u_2u_3u_1$
as a match of $M(3,3)$.

{\noindent\bf Phase P2:} In phase P2, given the set of structural matches $S$, for
each $G_s\in S$, we process the time series on the edges of $G_s$ in
order to find valid flow motif instances.
In a nutshell, we slide a time window of length $\delta$ along the set of all $(t_i,f_i)$
interactions on the edges of $G_s$; for all sets of interactions
within $\delta$ time difference, we find all combinations thereof
which constitute valid motif instances.
Note that each structural match $G_s$ from phase P1 may produce
an arbitrary number of flow motif instances, as each time window
position can generate different instances depending on the combinations of edge flows we use.

\begin{figure}[tb]
  \centering
    \includegraphics[width=0.8\columnwidth]{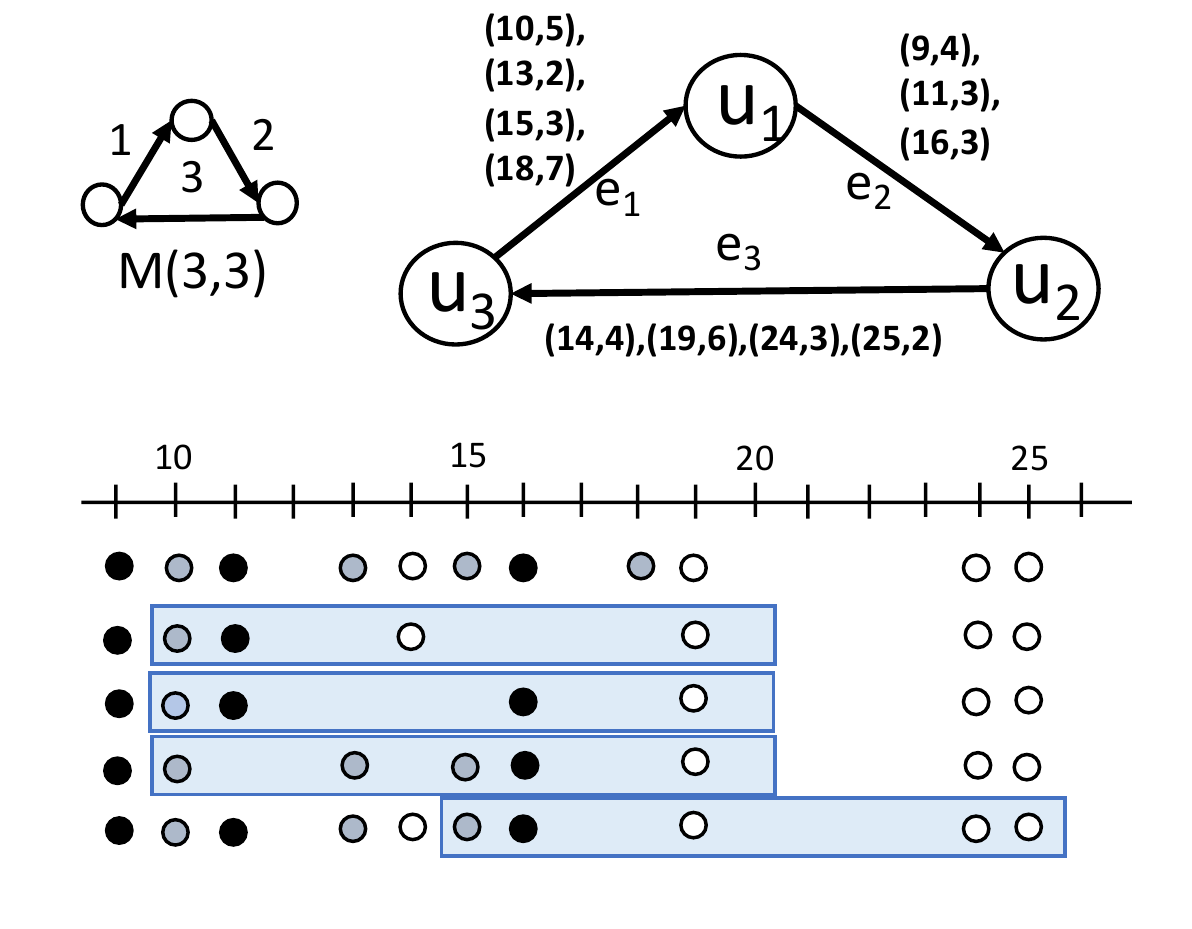}
  \caption{Example for Algorithm \ref{algo:instance_finding}}
  \label{fig:veriexam}
\end{figure}

To illustrate, consider again $M(3,3)$ (for $\delta=10$) and a
possible structural match, shown in Figure~\ref{fig:veriexam}. We will get
different flow motif instances depending on whether we consider window
$[10,20]$ or $[15,25]$. Furthermore, even for the specific time-window
$[10,20]$, we can get different flow motif instances depending on how
we combine the edges in this window. For example, one possible flow
motif instance is $[
e_1\leftarrow \{(10,5)\},
e_2\leftarrow\{(11,3),(16,3)\},
e_3\leftarrow\{(19,6)\}
]$,
while another flow motif instance is $[
e_1\leftarrow\{(10,5)\},
e_2\leftarrow\{(11,3)\},
e_3\leftarrow\{(14,4),(19,6)\}
]$. Note that the flow in the former case is 5, while in the latter is 3, meaning that the latter instance would be rejected for $\phi = 5$.

Algorithm \ref{algo:instance_finding} is applied in phase P2 to find
all instances of the motif $M$ in a match $G_s$ (found in phase P1). 
The algorithm slides a window $T$ of length $\delta$ over the time domain,
to find subsets of edges in $G_s$ that satisfy the duration constraint
$\delta$ and can generate maximal motif instances. 
Given a specific window $T$ we run procedure \textsc{FindInstances} in order to
generate all possible maximal flow-motif instances that satisfy the
flow constraint $\phi$. 
The procedure is recursive on the length $m$
of the
spanning path $\SP_M = e_1e_2\dots e_m$ of the motif.

\textsc{FindInstances} takes as input the graph instance $G_s$,
a spanning path $\SP$, a time-window $T$ and the threshold $\phi$.
Let $R(e_i)$ be the interaction time series on the edge of $G_s$ which
is mapped to edge $e_i$ of the motif. 
If the spanning path consists of a single edge $e_1$, then the procedure
finds the set $R_T(e_1)\subseteq R(e_1)$ of all elements in $R(e_1)$,
which are within the time-window $T$,
and aggregates their flow.
If the total flow $f(R_T(e_i))$ of these elements satisfies the
flow constraint $\phi$, the
edge-set of $G$ corresponding to $R_T(e_i)$ becomes an instance of $\SP$ and it is returned.
For longer spanning paths, the procedure considers again the first
edge $e_1 = \SP[1]$. For every prefix $T_p$ of the window $T$ that
contains instances of the edge $e_1$, it computes the set
$R_{T_p}(e_1)\subseteq R(e_1)$ of all $(t,f)$ interaction elements in $R(e_1)$
for which $t\in T_p$.
If $R_{T_p}(e_1)$ is non-empty and satisfies the flow constraint,
then \textsc{FindInstances} is recursively called on the rest of the
spanning path $\SP_{next} = \SP[2:m]$, with time window $T_{next} =
T-T_p$. This recursive call will return 
the set of valid instances within time-window $T_{next}$ for
the sub-motif defined by $\SP_{next}$. Each of these instances is
{\em concatenated} to $R_{T_p}(e_1)$ to create a new valid instance
for $\SP$.

\hide{
 of the all the time-series in $G_s$. We now focus on the subproblem of finding instances in a given
$s$. We present a solution for this subproblem (verification
module), described by  Algorithm \ref{algo:instance_finding}.
In a nutshell, we slide a window of length $\delta$ along the time
domain to capture the subset of edges in $s$ for which the timestamp
falls in the window. All positions of the window should be {\em
  maximal} in the sense that the $(t_i,f_i)$ pairs that belong to the last
edge of $s$ should change compared to the previous window position
(otherwise non-maximal motif instances could be produced) \fix{this to
  be elaborated further}.
For each position of this time window $T$, we apply a recursive
procedure (lines \ref{lin:pro:start}--\ref{lin:pro:end}),
which finds all motif instances in $s$ that are generated from
$(t_i,f_i)$ pairs that belong to the window.
The procedure considers all prefixes $T_p$ of $T$ for which the set $E$ of
$(t_i,f_i)$ pairs in the first edge $s[0]$ of $s$ changes compared to
the previously considered prefix. For each such prefix, the flows
$f_i$ in the $(t_i,f_i)$ pairs of $E$ are aggregated and if their sum is at
least $\phi$ (i.e., they form a valid motif edge instance), the
procedure is recursively called for the suffix $s[1:]$ of $s$ which
excludes its first edge and the suffix $T-T_p$ of the time window to
generate valid instances for the corresponding motif instance. These
instances are concatenated with  $E$ to produce valid instances for
$s$.
}

The condition at line  \ref{lin:pro:phi2}
of the algorithm helps us to
find invalid prefixes of the motif instances early. In other words, if
a sub-series $R_{T_p}(e_1)$
which is candidate for instantiating a motif edge does not
qualify $\phi$, we do not consider the possible instances that include
the elements of $R_{T_p}(e_1)$ as an instance of $e_1$. Hence, the search space is effectively
pruned.

\begin{algorithm}
\begin{algorithmic}[1]
\small
\Require $\delta$, $\phi$, time window $T$, structural match $G_s$
\State $\mathcal{I} \leftarrow \emptyset$\; \Comment{set of instances
  of $G_s$ in $T$}
\For{each maximal time window $T$ that satisfies $\delta$}
  \State $\mathcal{I} \leftarrow \mathcal{I}~\cup~$\textsc{FindInstances}$(G_s,\SP_M,T,\phi)$
\EndFor
\State
\Return $\mathcal{I}$
\Statex
\Procedure{FindInstances}{$G_s,\SP, T,\phi$} \label{lin:pro:start}
\State $\mathcal{I} \leftarrow \emptyset$\; \Comment{set of instances of $G_s$ in $T$}
\If{$length(SP)=1$}
  \State $R_T(e_1) \leftarrow$ all $(t,f)$ elements of $R(e_1)$ in $T$\;
  \If{$f(R_T(e_1)) \ge \phi$}\label{lin:pro:phi1} \Comment{$\phi$ condition check}
     \State  add $R_T(e_1)$ to $\mathcal{I}$
\EndIf
\Else
\For{each prefix $T_p$ of time window $T$}\;
  \State $R_{T_p}(e_1) \leftarrow$ all $(t,f)$ elements of $R(e_1)$ in $T_p$\;
  \If{$f(R_{T_p}(e_1)) \ge \phi$}\label{lin:pro:phi2} \Comment{$\phi$ condition check}
      \State $\SP_{next} \leftarrow \SP[2:m]$\;\Comment{suffix of $\SP$}
      \State $T_{next} \leftarrow T-T_p$\;\Comment{suffix of $T$}
      \State $\mathcal{I}_{next} \leftarrow \textrm{\textsc{FindInstances}}(G_s,\SP_{next},T_{next},\phi)$\;
      \For{each $I\in \mathcal{I}_{next}$}
        \State add $R_{T_p}(e_1) \circ I$ to $\mathcal{I}$\;
      \EndFor
  \EndIf
\EndFor
\EndIf
  \State \Return  $\mathcal{I}$
\EndProcedure \label{lin:pro:end}
\end{algorithmic}
\caption{Instance finding module}
\label{algo:instance_finding}
\end{algorithm}

Figure \ref{fig:veriexam} illustrates the functionality of Algorithm
\ref{algo:instance_finding}. On top, the figure shows motif $M(3,3)$ and a
structural match $G_s$ of it, where each edge is labeled by the
time series of flows between the corresponding nodes (e.g., at time 10,
$u_2$ sent to $u_1$ a flow of 5). 
The elements on the edges of $G_s$ are
illustrated (as sequences of dots ordered by time) 
at the bottom of the figure, colored by the edge
they belong to (e.g., black for $e_2$).
The first row of dots includes all $(t,f)$ elements, i.e., the first
black dot corresponds to element $(9,4)$ on edge $(u_1,u_2)$, which is mapped to
the second edge $e_2$ of $M(3,3)$.
To find the motif instances that
comprise of nodes and edges in $G_s$, we slide a window of length
$\delta$ along the timeline. Assuming that $\delta=10$, the first
position of the sliding window is $[10,20]$. The algorithm finds all
prefixes of elements in $R(e_1)$ that fall in this window and for each
such prefix, it generates recursively the combinations of elements
from other edges that form valid instances (according to
$\delta$). For example, for the prefix $T_p=[10,10]$, which includes
just the first element $(10,5)$ from $e_1$, the 2nd and the 3rd line of
dots in the figure show the valid instances formed. Specifically,
these instances are
$[
e_1\leftarrow \{(10,5)\},
e_2\leftarrow \{(11,3)\},
e_3\leftarrow \{(14,4),(19,6)\}
]$
and
$[
e_1\leftarrow \{(10,5)\},
e_2\leftarrow \{(11,3),(16,3)\},
e_3\leftarrow \{(19,6)\}
]$.
Note that the $\phi$ constraint is applied at every prefix in order to
prune the search space if it is violated (e.g., if $\phi=5$, any instance
$[
e_1\leftarrow \{(10,5)\},
e_2\leftarrow \{(11,3)\},
\dots]$ would be rejected.
Note also that there is no instance which contains just the first two
elements of $e_1$ but not the third one, because there is no element
from $e_2$ which is temporally between $(13,2)$ and $(15,3)$.
Finally, note that the next position of the sliding window is
$[15,25]$ because the position $[13,23]$ which starts from the 2nd
element of $e_1$ does not include any new elements from $e_3$ compared
to the previous window position $[10,20]$; hence, considering window position
$[13,23]$ would result in redundant (i.e., non-maximal) instances and
this position is skipped.

Algorithm \ref{algo:instance_finding} does not miss any maximal
instances because it systematically explores the combinations of
edge-sets which are time-respecting and maximal within a window. Moreover, the
windows have maximal lengths and in each of them the produced instances
essentially include the temporally first $(t_i,f_i)$ element that maps to
$e_1$ and the temporally last $(t_i,f_i)$ element that maps to $e_m$.
At least one of these pairs changes in the next window position;
therefore, instances produced at different windows do not violate the
maximality condition.


\noindent
{\bf Complexity Analysis.}
In the worst case, for each $G_s$ and each time window, we should
consider all combinations of edges in $G$ that instantiate the
edges of the motif. For example, when $\phi=0$, prefix-based pruning
cannot be applied. In the worst case, $G_s =G$ and the edges in $G$ 
ordered by timestamp are assigned to the sequence of motif edges
in a round-robin fashion. That is, the temporally first edge of $G$ is
mapped to $e_1$, the second to $e_2$, etc.
In this case, assuming the loosest
possible constraints $\delta=\infty, \phi=0$, the number of
combinations of pairs to be considered (which all form valid motif
instances) is $O(|E|/m)^{m}$, i.e., exponential to the number of
edges $m$ in the motif. In addition, the number of structural matches is
also exponential to $m$. In practice, $G_T$ is sparse (or $V$ is
small) and the constraints $\delta$ and $\phi$ help in pruning
combinations of edges that do not form instances early, which renders
the algorithm scalable, as we will show in the experimental evaluation.



\section{Top-k flow motif search}\label{sec:topk}
Setting an appropriate value for the parameters $\delta$ and $\phi$
could be hard for non-experts of the domain. Parameter $\delta$ is
intuitively easier to be set to a time constraint that makes sense to
the application (for example, the analyst could be interested in
patterns of bitcoin transactions which happen within an hour or
day). On the other hand, $\phi$ is less intuitive, as too large
values could result in too few or zero instances, whereas too small
values could result in thousands of instances which may overwhelm the
user. One solution to this problem is to replace the $\phi$ constraint
by a ranking of the motif instances $G_I$ with respect to their flow
(see Equation \ref{eq:instanceflow}).
In other words, we may opt to search for the $k$
instances $G_I$ of the motif (with $\phi=0$) that satisfy
$\delta$, which have the maximum flow $f(G_I)$.


To solve this top-$k$ flow motif search problem, we can 
use our algorithm
with a small
number of modifications. Phase P1 is identical; we should still find the set
$S$ of all structural matches. Then, for each $G_s\in S$, we
apply phase P2, by making the following changes to Algorithm
\ref{algo:instance_finding}.
First, we keep track in a priority queue (heap) the top-$k$
instances in terms of their minimum flow so far. Second, in place of
$\phi$,
we use the flow $f(G_I^k)$ of the $k$-th instance $G_I^k$
so far as a dynamic (floating) threshold. 

\subsection{Finding the top motif instance}\label{sec:top1}
For the special case, where $k$=1, the top-1 motif instance search
problem can potentially be solved faster with the help of a dynamic
programming (DP) algorithmic module. Recall that the objective of procedure
\textsc{FindInstances} in Algorithm
\ref{algo:instance_finding} is to find the motif instances in a
structural match $G_s$, within a
time window $T$, which qualify $\phi$. We can replace this module by a
dynamic programming algorithm that finds the instance of maximum flow
within $T$. This DP module can be described by Algorithm
\ref{algo:dp}.

\begin{algorithm}
\begin{algorithmic}[1]
\small
\Require $\delta$, time window $T$, structural match $G_s$
\State $maxflow \leftarrow 0$ \Comment keeps track of max flow found
at any instance
\For{each maximal time window $T$ that satisfies $\delta$}
  \ForAll{timestamps $t_i$ in $T$}
	\State compute $Flow([t_1,t_i],1)=flow([t_1,t_i],1)$
  \EndFor
  \For{$\kappa=2$ to $n$}
         \ForAll{timestamps $t_i$ in $T$}
		\State compute $Flow([t_1,t_i],\kappa)$ by Eq. \ref{eq:dp}
        \EndFor
  \EndFor
  \State $maxflow = \max\left (maxflow, Flow([t_1,t_{\tau},m)\right )$
\EndFor
\State \Return $maxflow$
\end{algorithmic}
\caption{DP module for top-1 instance search}
\label{algo:dp}
\end{algorithm}

Specifically, let $[t_1,t_2,\dots,t_{\tau}]$ be the sequence of
timestamps in $T$ for which there is a $(t,f)$ interaction element in $G_s$.
Let $M_\kappa$ be the prefix of $M$ which includes its fist $\kappa$ edges only
and $Flow([t_1,t_i],\kappa)$ be the flow of the top-$1$ motif instance of
$M_\kappa$ in the time window $[t_1,t_i]$.
Then,
$Flow([t_1,t_i],\kappa)$ can be recursively computed as follows:
\begin{equation}\label{eq:dp}
\small
Flow([t_1,\!t_i],\!\kappa)\!\!=\!\!\!\max_{1<j\le i}\!\!\{\min(Flow([t_1\!,\!t_{j\!-\!1}],\!\kappa\!-\!1),\!flow([t_j,\!t_i],\!\!\kappa))\!\},\!
\end{equation}
where $flow([t_j,t_i],\kappa)$ is the total flow of all $(t,f)$ elements of the
time series $R(e_{\kappa})$ on the $\kappa$-th edge of $G_s$, whose timestamps are in the time interval $[t_j,t_i]$.
The $Flow([t_1,t_i],1)$ array is initialized by scanning the elements of the first edge of $G_s$ in $T$.
Then, for each $\kappa>1$,  $Flow([t_1,t_i],\kappa)$ is computed using
array  $Flow([t_1,t_i],\kappa-1)$.
Finally, $Flow([t_1,t_{\tau}],m)$ corresponds to the top-$1$ flow of
any motif instance in $G_s$ within time window $T$. By applying this
algorithm for every window $T$, we can find the top instance in
$G_s$. Repeating this for each $G_s$ gives us the  top-$1$ instance of
$M$ in $G$.


Table \ref{table:dpexam} shows the steps of the DP module
in the course of finding the top-1 instance
in time window $[10,20]$ (assuming that $\delta$=10) for 
the structural match of $M(3,3)$ shown in Figure \ref{fig:veriexam}.
The first row shows the values of $Flow([t_1,t_i],1)$ for the first edge
of the motif and for all values of $t_i$ (i.e., columns of the table).
(Recall that the starting timestamp $t_1$ of the time window is $10$.)
The second row shows,
for the first two edges of the motif,
the value of $Flow([t_1,t_i],2)$ for all values of $t_i$, as well as the
value of $t_j$, which determines $Flow([t_1,t_i],2)$.
For all $t_i$, the value of $t_j$ that maximizes the flow is 11 and for $t_i\ge 16$ the flow becomes $min(5,3+3)=5$.
Finally, the last row shows the maximum flow for the best arrangement of $(t,f)$ pairs to all three edges of the motif, for all prefixes of the time window.
Note that the last value corresponds to the entire window and contains
the flow of the best instance of the entire motif in $[10,20]$, which
is 5.
The cells of the matrix in bold show how the top-1
instance, i.e.,  $[
e_1\leftarrow \{(10,5)\},
e_2\leftarrow\{(11,3),(16,3)\},
e_3\leftarrow\{(19,6)\}
]$, can be identified.

\begin{table}[ht]
\caption{Example of the DP module} \label{table:dpexam}
\centering
\scriptsize
\begin{tabular}{|l|r|r@|@{~}c@{~}|@{~}c@{~}|@{~}c@{~}|@{~}c@{~}|@{~}c@{~}|@{~}c@{~}|}
\hline
$t_i$&10&11&13&14&15&16&18&19\\
\hline
 $\kappa$=1&{\bf 5}&5&7&7&7&7&10&10\\
\hline
 $\kappa$=2&&3 ($t_j$=11)&3 ($t_j$=11)&3 ($t_j$=11)&3 ($t_j$=11)&5 ($t_j$=11)&{\bf 5 ($t_j $=11)}&5 ($t_j$=11)\\
\hline
 $\kappa$=3&&&0 ($t_j$=13)&4 ($t_j$=14)&4 ($t_j$=14)&4 ($t_j$=14)&4
($t_j$=14)&{\bf 5 ($t_j$=19)}\\
\hline
\end{tabular}
\end{table}

\noindent
{\bf Complexity Analysis.}
For each $G_s$ and each time window, we should
consider all binary splits of the window at each iteration (i.e., for
each edge in $M$).
Hence the time complexity is $O(\tau^2|E|)$, where $\tau$ is the number
of timestamps in $T$ for which there is an $(t_i,f_i)$ element in $G_s$.
The space complexity is $O(\tau\cdot |E|)$ because we only need all
$Flow([t_1,t_i],\kappa-1)$ for $\kappa-1$ when we process the $\kappa$-th edge.
The overall time complexity per structural match in $S$
is $O(|S|\delta \tau^2|E|)$, since the number of
windows to be considered is
$O(\delta)$.
The number of structural matches $|S|$ is exponential to $m$, as discussed
in our previous analysis.


\noindent
{\bf Extensibility.}
The algorithm can be applied to solve top-1 problems at a finer
granularity.
In particular, it can be used to find the top-1 instance for each structural match $G_s$.
This may be useful if we want to compare the sets of entities 
that constitute the structural instances (e.g., groups of bitcoin users)
based on their max-flow interactions.
In addition, we might be interested in finding the top-1 instance for each position of
the sliding time window $T$.
This can be used in analysis tasks that compare the volume of interactions (according to the motif structure) at different periods of time.

%% file: experiments.tex
\section{Experimental Evaluation}\label{sec:exp}
The goal of our experimental evaluation is twofold: 
test the performance and scalability of our algorithms 
and study the significance of flow motifs.
We implemented the algorithm presented in Section \ref{sec:algorithm}
and its two variants proposed in Section \ref{sec:topk}
(top-$k$ instance search, dynamic programming module for top-$1$ search).
As a baseline, we also implemented an alternative motif instance finding method based on finding and joining instances of motif components in a hierarchical manner.

We evaluate the performance of all these methods on three real networks, to be described in Section \ref{sec:datasets}.
We measure the efficiency and scalability of the tested methods as a function of the problem parameters $\delta$ and $\phi$ on the motif structures shown in Figure \ref{fig:motifs}. These graphs model representative flows of interaction that could be of interest to data analysts (e.g., $M(3,3)$ corresponds to cyclic transactions in a money-exchange network, $M(4,3)$ corresponds to chains of region-to-region movements in a passenger flow network).
We also assess the statistical significance of the tested motifs in three real graphs.
All algorithms were implemented in Python3 and we ran all the experiments on a machine with an Intel Xeon CPU E5-2620 prossesor running Ubuntu 18.04.1 LTS.

\subsection{Dataset Description}\label{sec:datasets}



We used three
datasets extracted from real interaction networks: the \textbf{Bitcoin}
  network, the \textbf{Facebook} network and a {\bf Passenger} flow
network. Table \ref{table:datasets} shows statistics of the datasets.
The third column is the distinct number of node pairs $(u,v)\in V$, for which 
there is at least one edge (i.e., interaction) from $u$ to $v$. This number equals to the number $|E_T|$ of edges in the corresponding time-series graph $G_T$.
We now provide more details about them.

\begin{table}[ht]
\caption{Statistics of Datasets}
\vspace{-0.3cm}
\centering
 \small
\begin{tabular}{@{}|@{~}c@{~}|@{~}c@{~}|@{~}c@{~}|@{~}c@{~}|@{~}c@{~}|@{}}
\hline
Dataset &\#nodes& \#connected node pairs  &\#edges & Avg. flow per edge \\

\hline
Bitcoin & 24.6M&88.9M&123M& 4.845\\

Facebook & 45800&264000 &856000&3.014 \\

Passenger &289&77896 &215175&1.933\\[0.2ex]

\hline
\end{tabular}
\label{table:datasets}
\end{table}


\begin{figure*}[t!]
\subfigure[Bitcoin Network]{
    \label{fig:exp:join_algo_bitcoin}
    \includegraphics[width=0.32\textwidth]{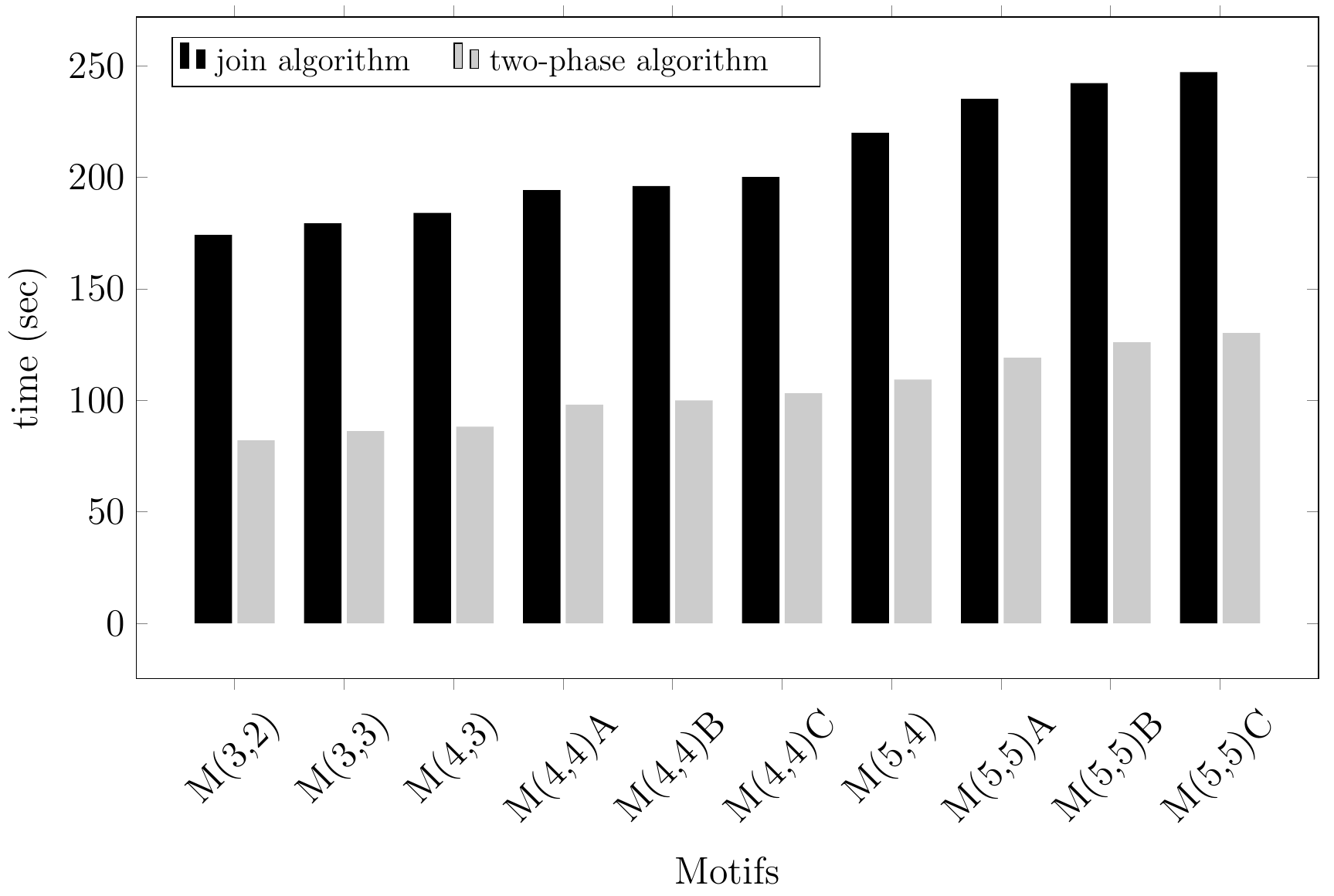}
    }\!\!\!\!
  \subfigure[Facebook Network]{
    \label{fig:exp:join_algo_fb}
    \includegraphics[width=0.32\textwidth]{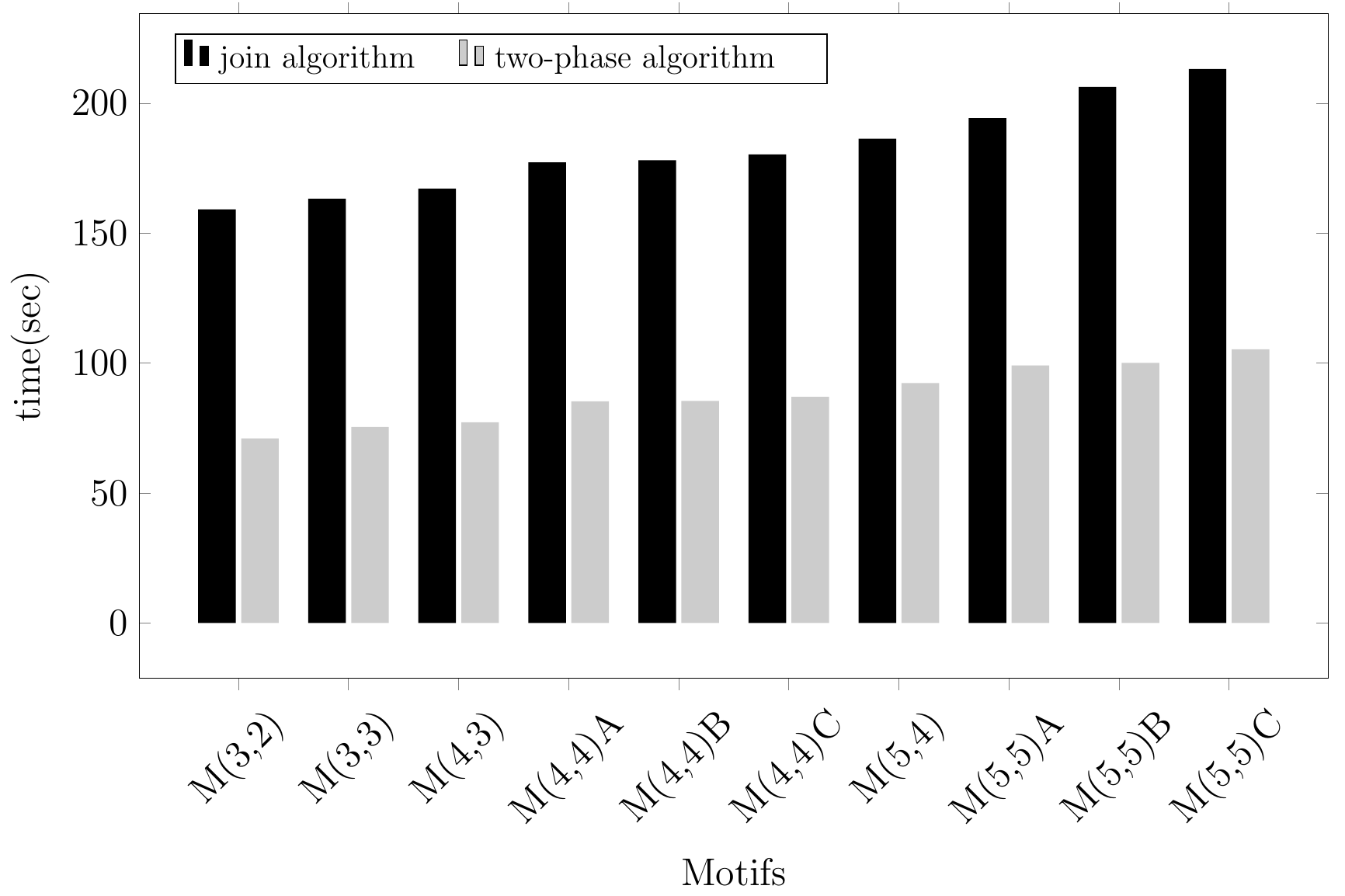}
    }\!\!\!\!
  \subfigure[Passenger Network]{
    \label{fig:exp:join_algo_traffic}
    \includegraphics[width=0.32\textwidth]{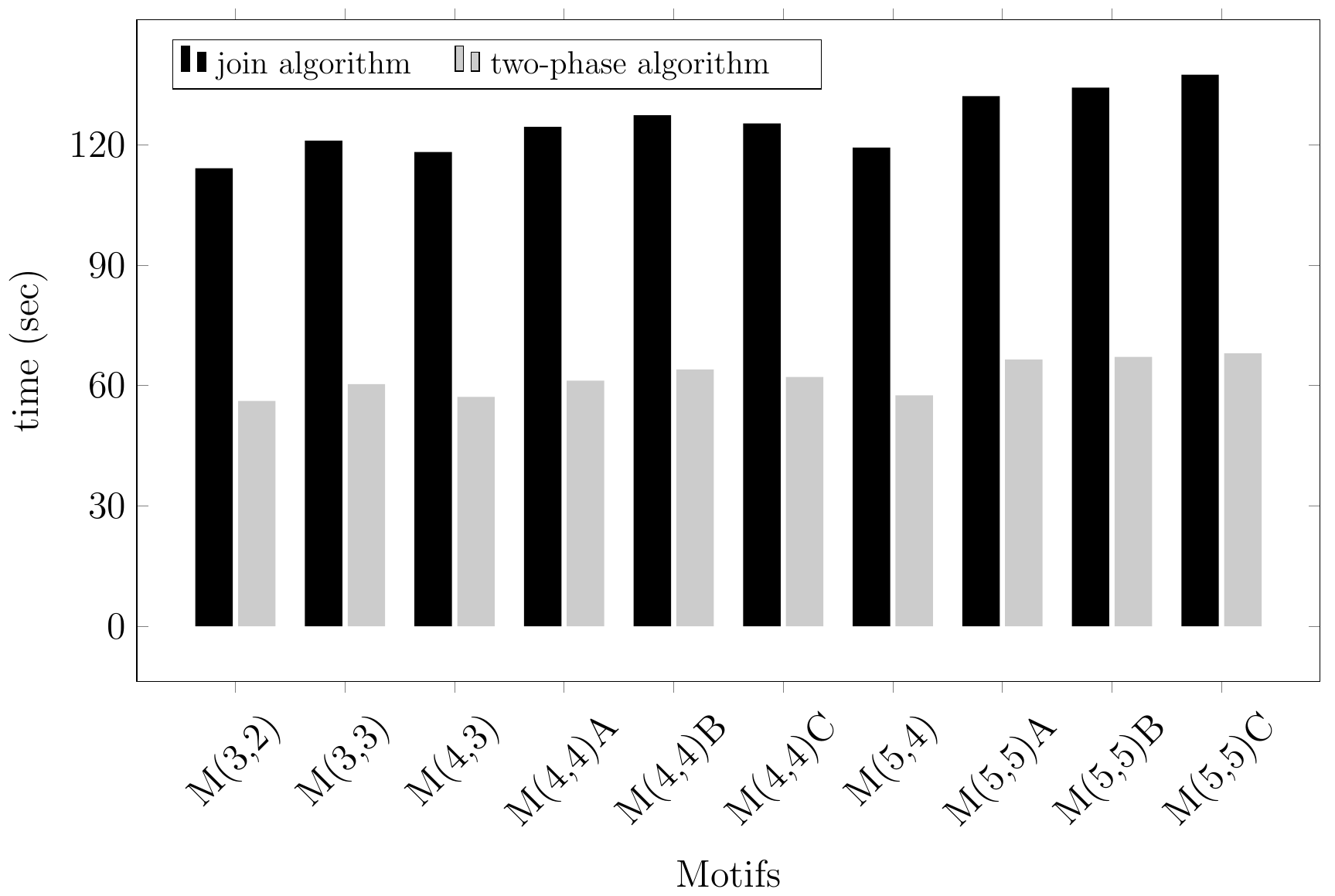}
    }
  \vspace{-0.1in}
  \caption{Our two-phase algorithm vs. the join algorithm}
  \label{fig:exp:join_algo}
\end{figure*}

\textbf{Bitcoin network.}
We downloaded all transactions in the bitcoin blockchain \cite{Nakamoto_bitcoin:a}
in the period February 1st 2014 to November 30 2014
and converted them to a bitcoin {\em user graph}.%
\footnote{data obtained from http://www.vo.elte.hu/bitcoin}
Nodes correspond to users and
for each transaction of $f$ bitcoins in the blockchain from user
$u$ to user $v$ at time $t$, we added an edge from $u$ to $v$ with label $(t,f)$.
Since the same bitcoin user may control and use multiple addresses, we applied a well-known heuristic \cite{DBLP:conf/complexnetworks/RemyRM17,DBLP:journals/corr/KondorPCV13} to {\em merge} addresses that are considered to belong to the same user to a single network node. Specifically, we merged addresses that appear together as input in the same transaction.
We did not take into account insignificant transactions with amounts under 0.0001 BTC.
Bitcoin is a relatively sparse graph and the cases of two nodes
being connected by multiple edges is rare.
Finding motif instances in the Bitcoin network can help towards understanding complex interactions between users and can possibly help toward identifying suspicious transactions like money laundering and bitcoin theft \cite{DBLP:conf/imc/MeiklejohnPJLMVS13}.

\textbf{Facebook network}:
We consider Facebook as an interaction network between users.
We divide the time into 30-second intervals $[t_s,t_e)$ and
for each pair of users $u$ and $v$ we aggregate
all interactions from $u$ to $v$
and add an edge from $u$ to $v$ with label $(t_s,f)$,
where $f$ is the total number of interactions from $u$ to $v$ in this interval.
We consider as interactions the posts of likes by $u$ targeting $v$ or
the messages sent from $u$ to $v$.
We created the Facebook user network using data from April 2015 to October 2015; the same dataset is used in \cite{DBLP:journals/ccr/RoyZBPS15}.
The Facebook network is relatively sparse and each pair of connected nodes
have about four edges on average.
Motif search on this graph can help in analyzing influence
\cite{DBLP:conf/sdm/LeskovecMFGH07,Gomez-RodriguezLK12}
and finding important interactions among users \cite{DBLP:conf/nips/McAuleyL12}.

\textbf{Passenger flow network}: We processed trips of yellow
taxis in NYC in January
2018.\footnote{obtained from
http://www.nyc.gov/html/tlc/html/about/trip\_record\_data.shtml}.
Each record includes the pick-up and drop-off taxi zones (regions)
the date/time of the pick-up and drop-off, and the number of
passengers inside the taxi. Using these records, we created an
interaction network where the nodes are the taxi zones; for each
record, we generate an edge that links the corresponding nodes and
carries the timestamp of the activity (i.e., the pickup time) and the corresponding flow (i.e., the number of passengers).
This Passenger flow network is dense; in addition, each pair of connected nodes
have about three edges on average.
Motif instances found in this passenger flow graph can help in understanding the flow of movement between different regions on a map.

\subsection{Efficiency and Scalability}
In this section, we evaluate the efficiency and scalability of our algorithm when
applied to find the instances of the motifs depicted in Figure \ref{fig:motifs}.
The default values for the duration constraint $\delta$ are $600$ sec., $600$ sec.,
and $900$ sec. on Bitcoin, Facebook, and Passenger, respectively. 
These value represent realistic time intervals for the corresponding applications.
The corresponding
default values for $\phi$ are $5$, $3$, and $2$, respectively.

\subsubsection{Comparison to a competitor}\label{sec:competitor}
In our first set of experiments, we compare our algorithm with an
alternative motif instance finding algorithm which is based on
progressively finding and joining instances of motif
subgraphs. 

Specifically, this {\em join algorithm} starts by accessing each
edge $(u,v)$ of the time series graph $G_T$ and finding all time-intervals
of length at most $\delta$ and their aggregated flows. For each such interval
$[t_s,t_e]$ a quintuple $(u,v,t_s,t_e,f)$ is generated. These tuples are kept in two tables; $C_1$ sorts them by starting vertex $u$ and $C_2$ sorts them by ending vertex $v$.
In the next step, $C_2$ and $C_1$ are merge-joined to find all pairs $(c_2,c_1)$
having $c_2.u=c_1.v$ and also satisfying $c_1.t_e-c_2.t_s\le \delta$.
The set $P$ of all these tuple pairs constitute results of all sub-motifs
of $M$ which include two consecutive edges.
In the next step, $P$ is self-joined again to produce instances of sub-motifs 
of $M$ with three consecutive edges.
This is done by 
finding pairs $\{(c_2,c_1), (c'_2,c'_1)\}$ of couples
in $P$ for which $c_1=c'_2$ and $c'_1.t_e-c_2.t_s\le \delta$.
The next steps are applied in a similar manner until the instances of the
entire motif $M$ are constructed. Note that for each motif or sub-motif that closes a cycle (e.g., $M(3,3)$), we check the additional condition that the starting vertex
of the first motif edge in the instance is the same as the target vertex of the last edge.
At each step, we apply a merge join for
the production of sub-motif instances, after having sorted the tuples produced in the previous step accordingly.

Figure \ref{fig:exp:join_algo} compares the runtime cost of the join algorithm with
that of our two-phase algorithm presented in Section \ref{sec:algorithm}.
For all motifs, we used the default values for $\delta$ and $\phi$.
Note that our two-phase algorithm is typically twice as fast as the
join algorithm. This is attributed to the fact that the join algorithm produces a
large number of intermediate results (i.e., sub-motif instances), which are
avoided by our method. Note that many of these sub-motif instances do not end up
as components of any instance of the complete motif, so their generation is
redundant. 
In the rest of this section, we do not include additional
comparisons with the join algorithm since it was always found to be slower than
our approach.

\begin{table*}[ht]
  \small
  \caption{Number of structural matches and runtime in phase P1 of
    motif search}\label{table:phase1}
  \vspace{-1.5mm}
  \begin{tabular}{|l|l|c|c|c|c|c|c|c|c|c|c|}
    \hline
    &Motif&M(3,2)&M(3,3)&M(4,3)&M(4,4)A&M(4,4)B&M(4,4)C&M(5,4)&M(5,5)A&M(5,5)B&M(5,5)C\\
    \hline
    \multirow{2}{*}{Bitcoin}&Instances&634K&485K&484K&210K&205K&213K&145K&122K&124K&121K\\
    &Time (sec)&47.02&49.23&50.15&57.05&60&61.16&64.35&69.11&73.02&75.15\\
    \hline
    \multirow{2}{*}{Facebook}&Instances&415K&276K&272K&113K&113K&114K&97K&90K&91K&90K\\
    &Time(sec)&40.02&43.43&44.21&48.45&49.32&49.01&52.33&50.12&52.07&54.31\\
    \hline
\multirow{2}{*}{Passenger}&Instances&27893&16455&25778&14877&14569&14903&22134&12345&12567&12009\\
&Time(sec)&19.14&21.33&22.15&26.22&29.03&29.11&25.04&30.45&31.14&32\\
    \hline
  \end{tabular}
\end{table*}

\begin{figure*}[t!]
  \includegraphics[width=0.88\textwidth]{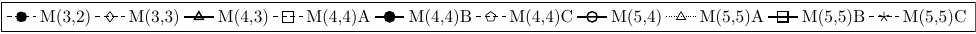}
\subfigure[Bitcoin Network]{
    \label{fig:exp:bitcoin_phi=5}
    \includegraphics[width=0.32\textwidth]{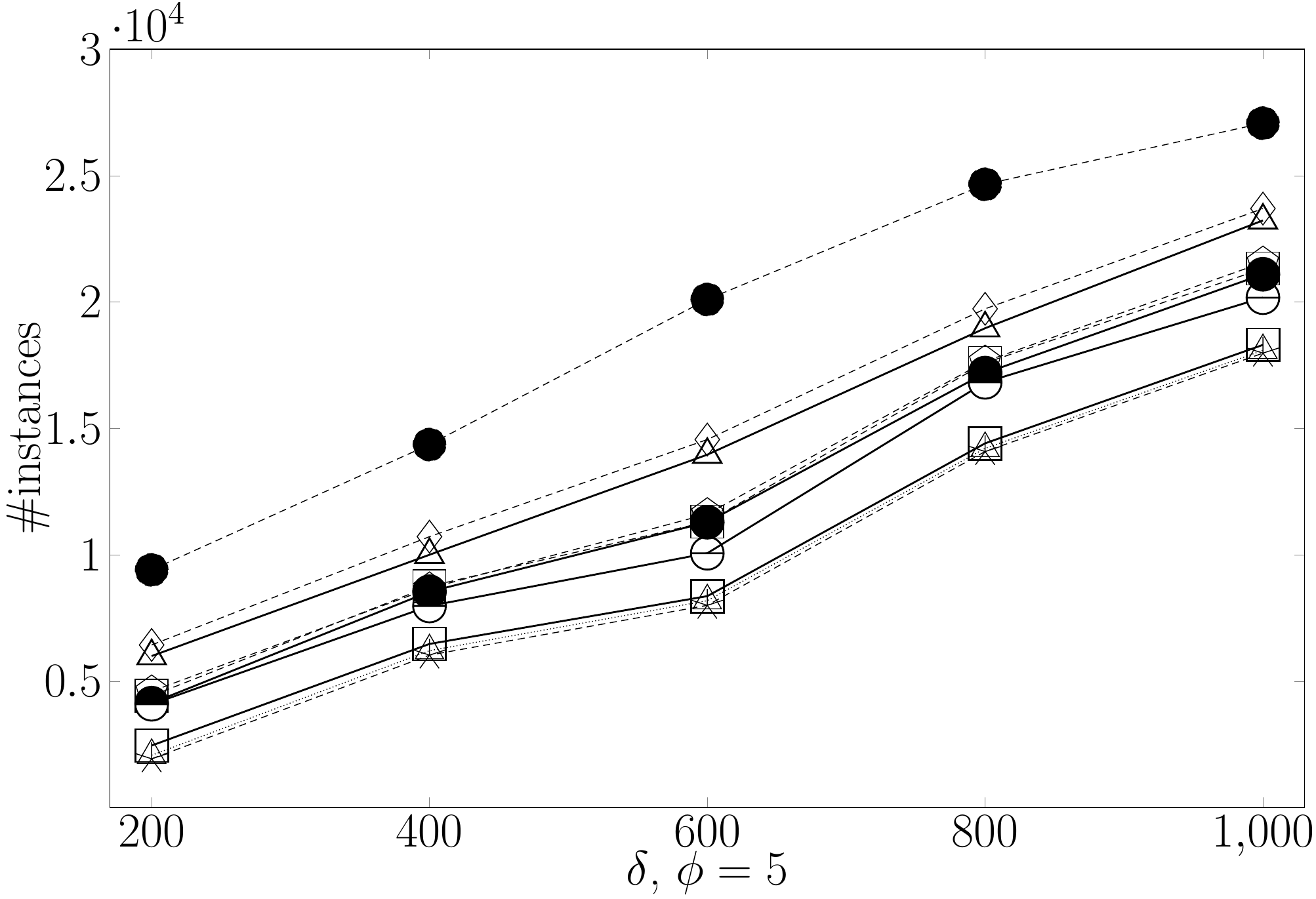}
    }\!\!\!\!
  \subfigure[Facebook Network]{
    \label{fig:exp:fb_phi=3}
    \includegraphics[width=0.32\textwidth]{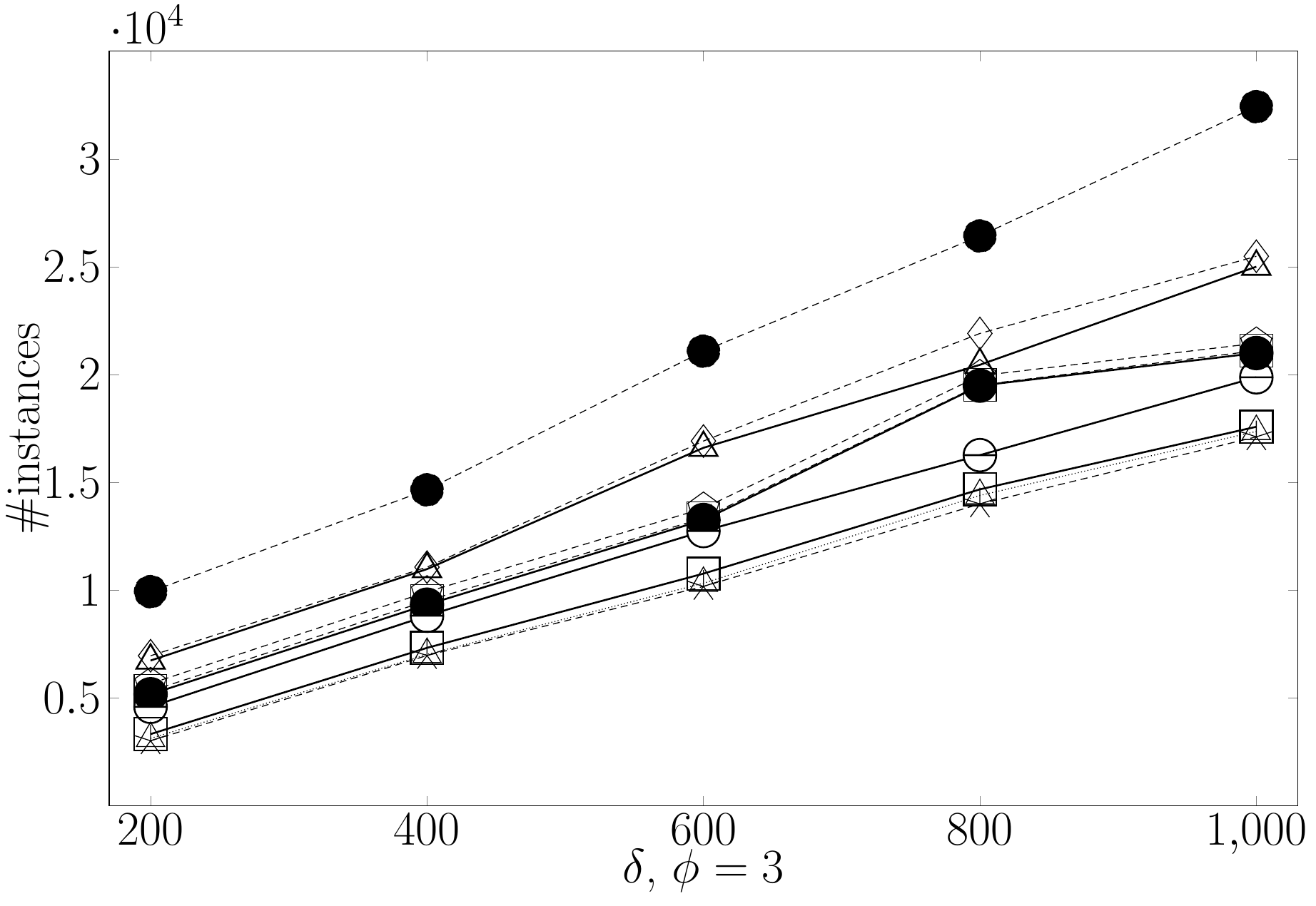}
    }\!\!\!\!
  \subfigure[Passenger Network]{
    \label{fig:exp:traffic_phi=2}
    \includegraphics[width=0.32\textwidth]{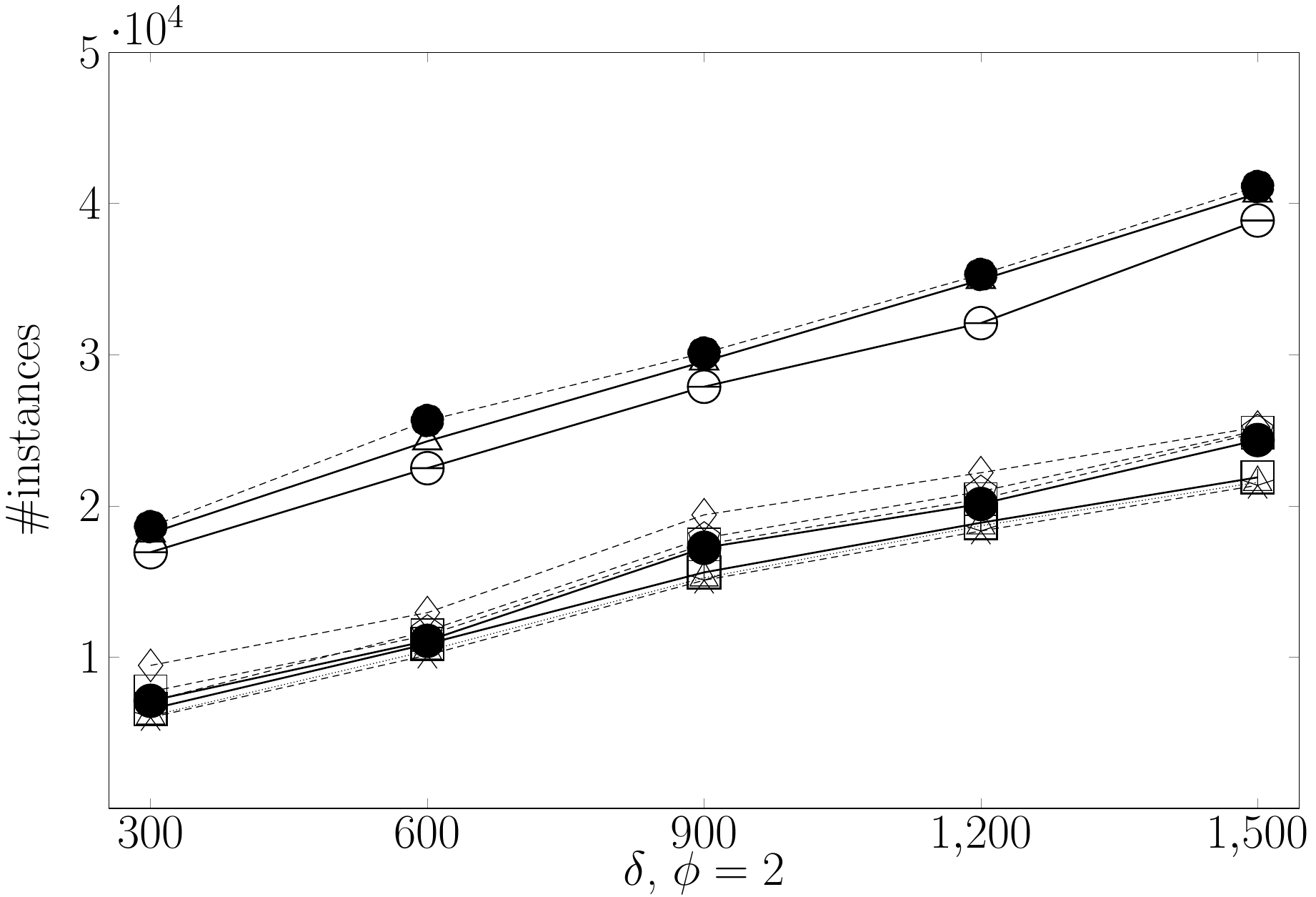}
  }
  \vspace{-1mm}
  \subfigure[Bitcoin Network]{
    \label{fig:exp:bitcoin_scala_delta}
    \includegraphics[width=0.32\textwidth]{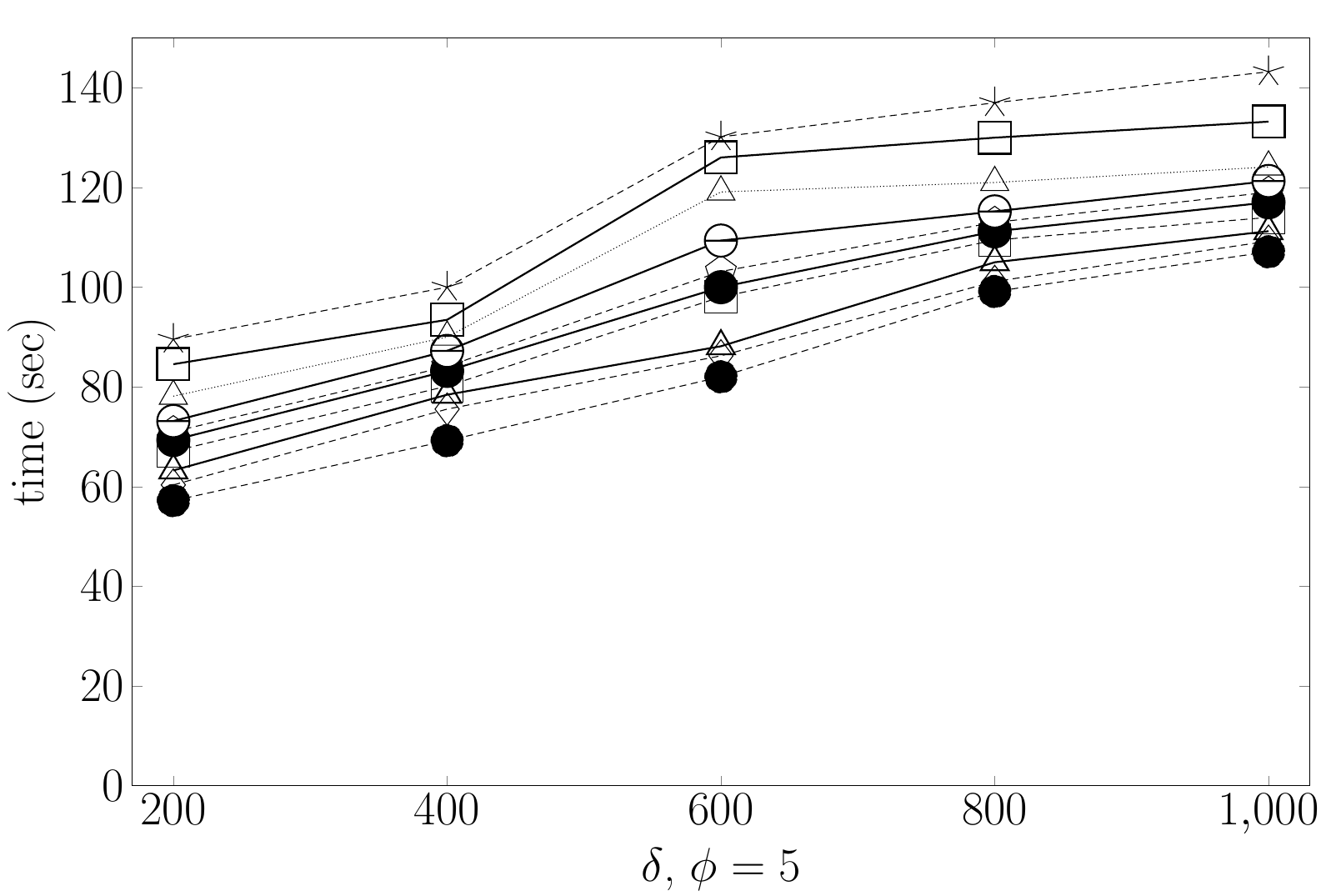}
    }\!\!\!\!
  \subfigure[Facebook Network]{
    \label{fig:exp:fb_scala_delta}
    \includegraphics[width=0.32\textwidth]{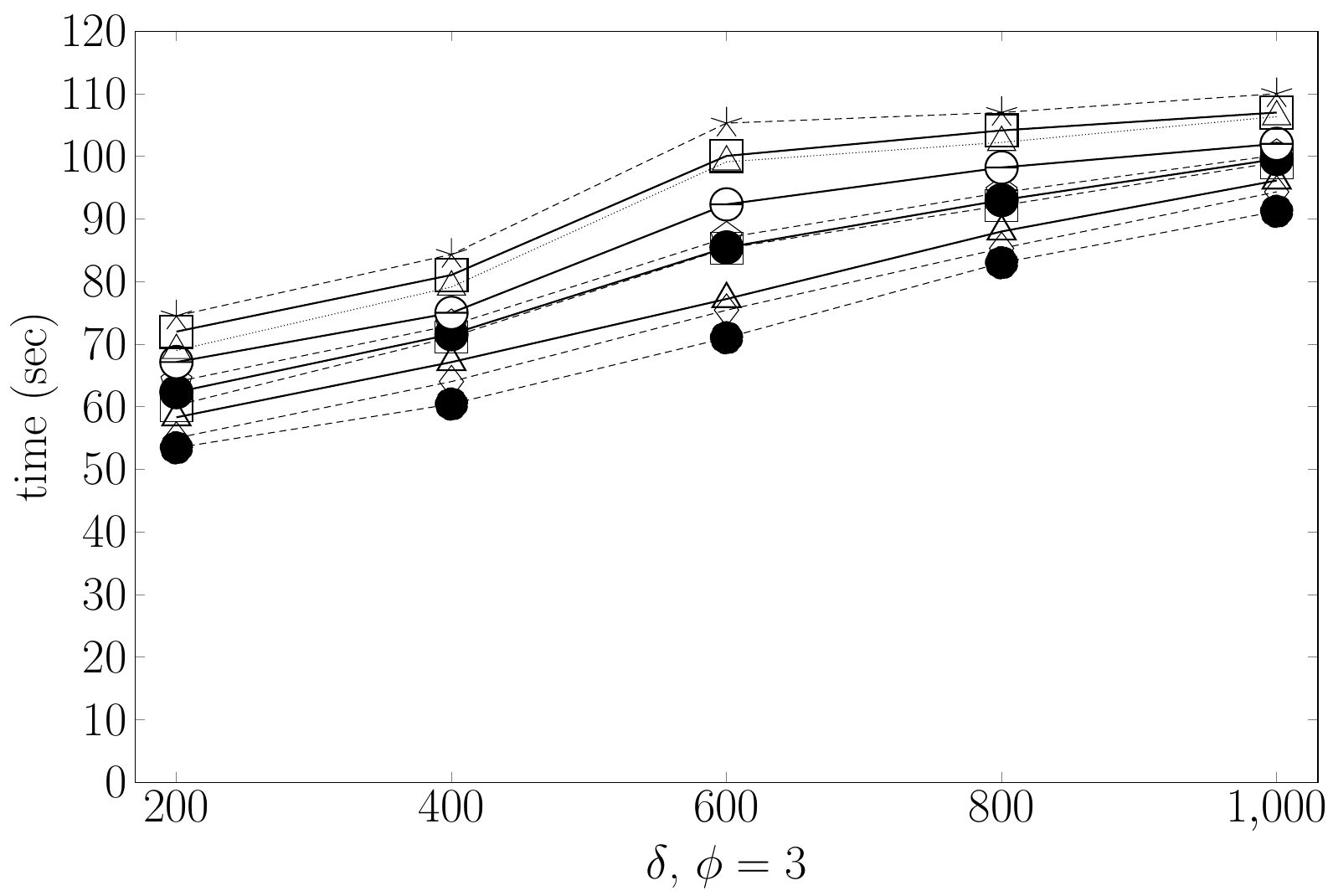}
    }\!\!\!\!
  \subfigure[Passenger Network]{
    \label{fig:exp:traffic_scala_delta}
    \includegraphics[width=0.32\textwidth]{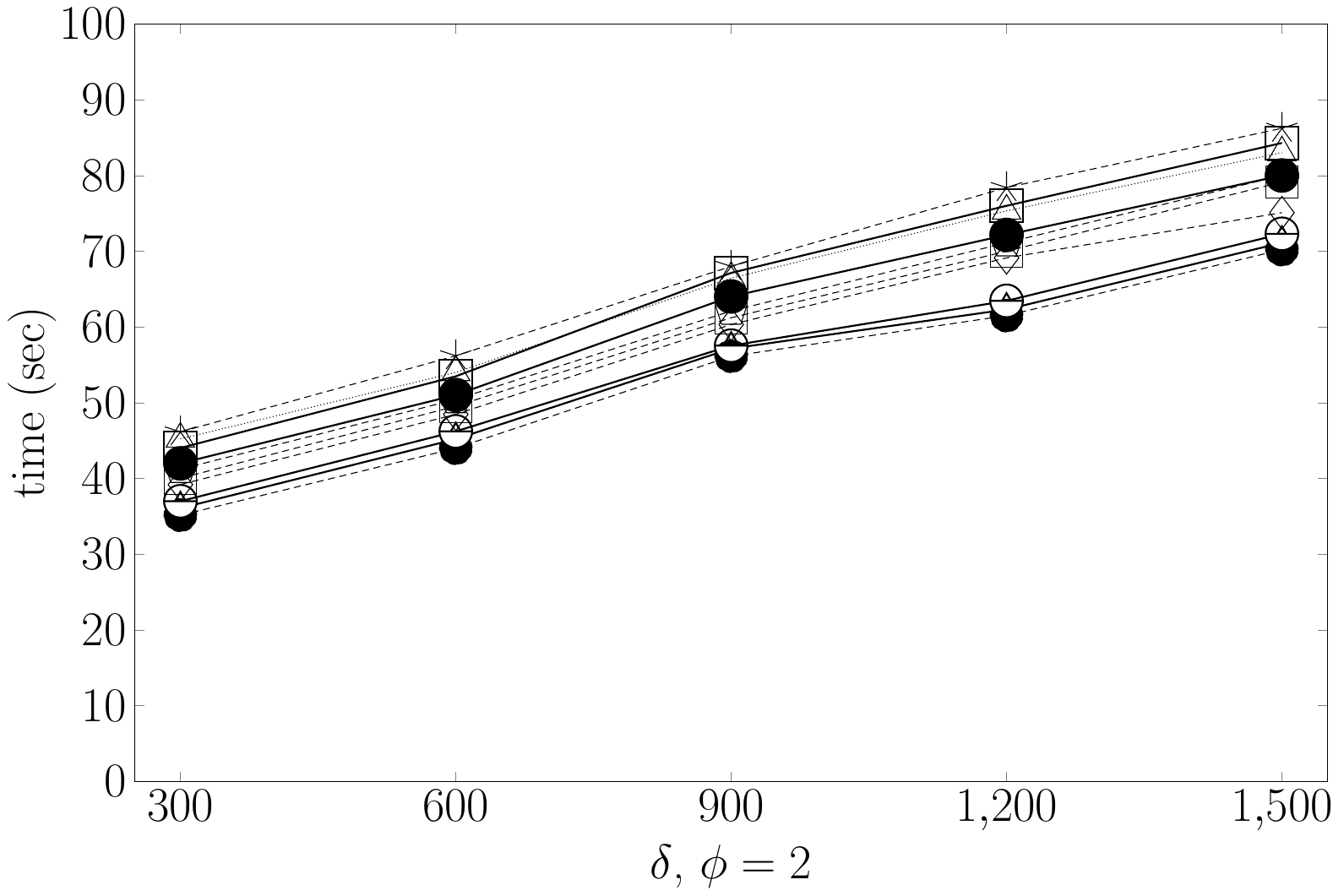}
    }
  \vspace{-0.1in}
  \caption{Number of instances and time for different values of $\delta$}
  \label{fig:exp:delta}
\end{figure*}

\begin{figure*}[t!]
\subfigure[Bitcoin Network]{
   \label{fig:exp:inst_bitcoin_delta=600}
    \includegraphics[width=0.32\textwidth]{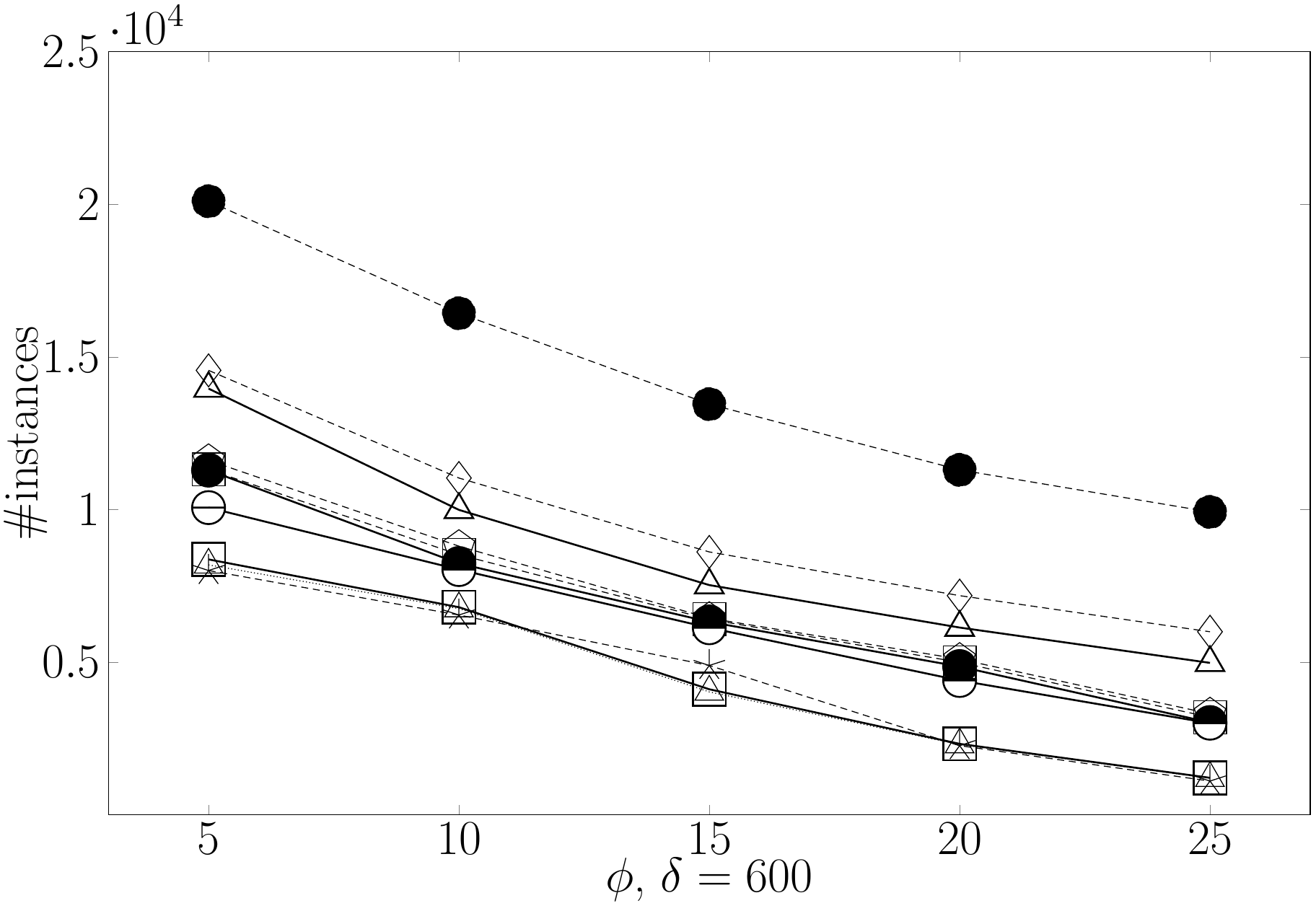}
    }\!\!\!\!
  \subfigure[Facebook Network]{
   \label{fig:exp:inst_fb_delta600}
    \includegraphics[width=0.32\textwidth]{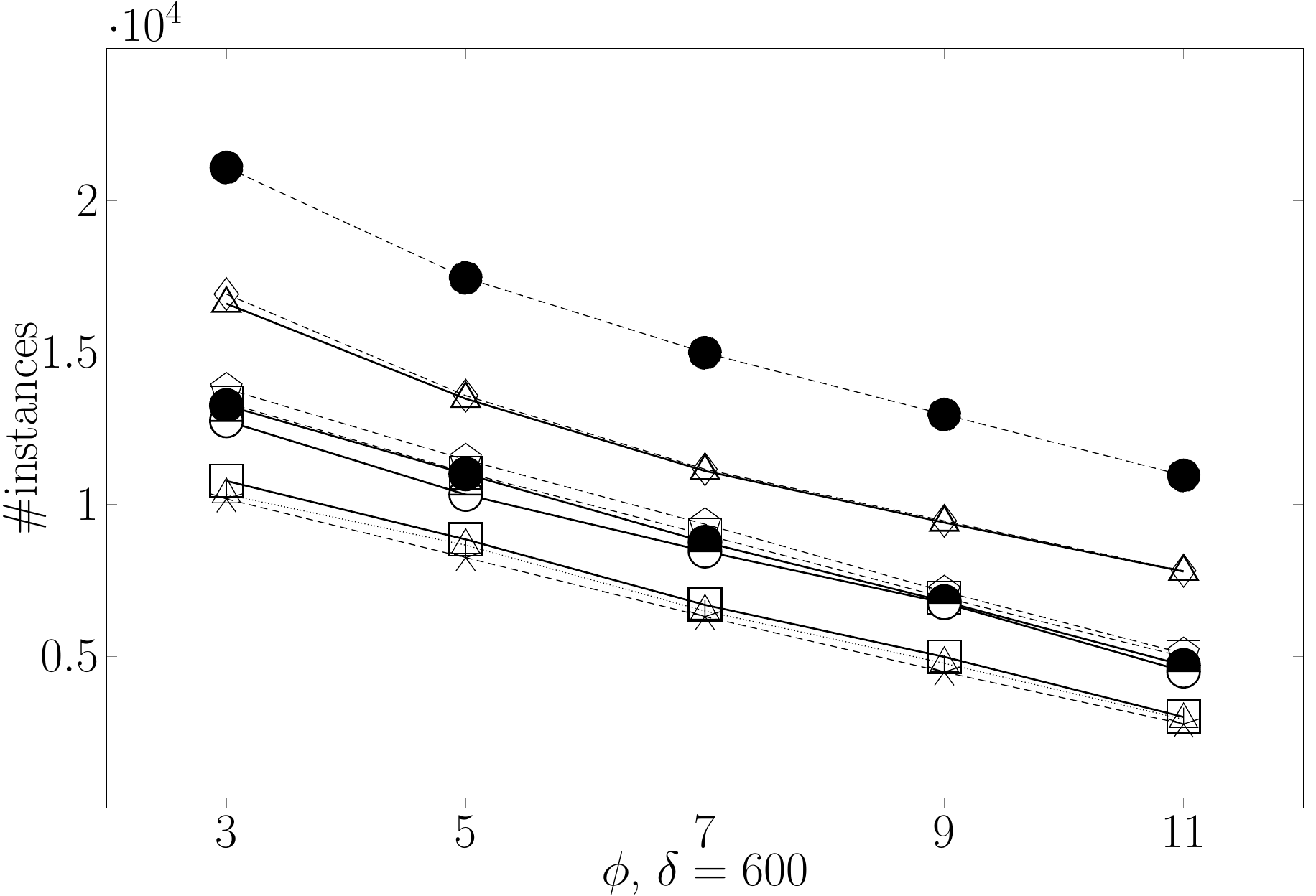}
    }\!\!\!\!
  \subfigure[Passenger Network]{
   \label{fig:exp:flow_topk_fb}
    \includegraphics[width=0.32\textwidth]{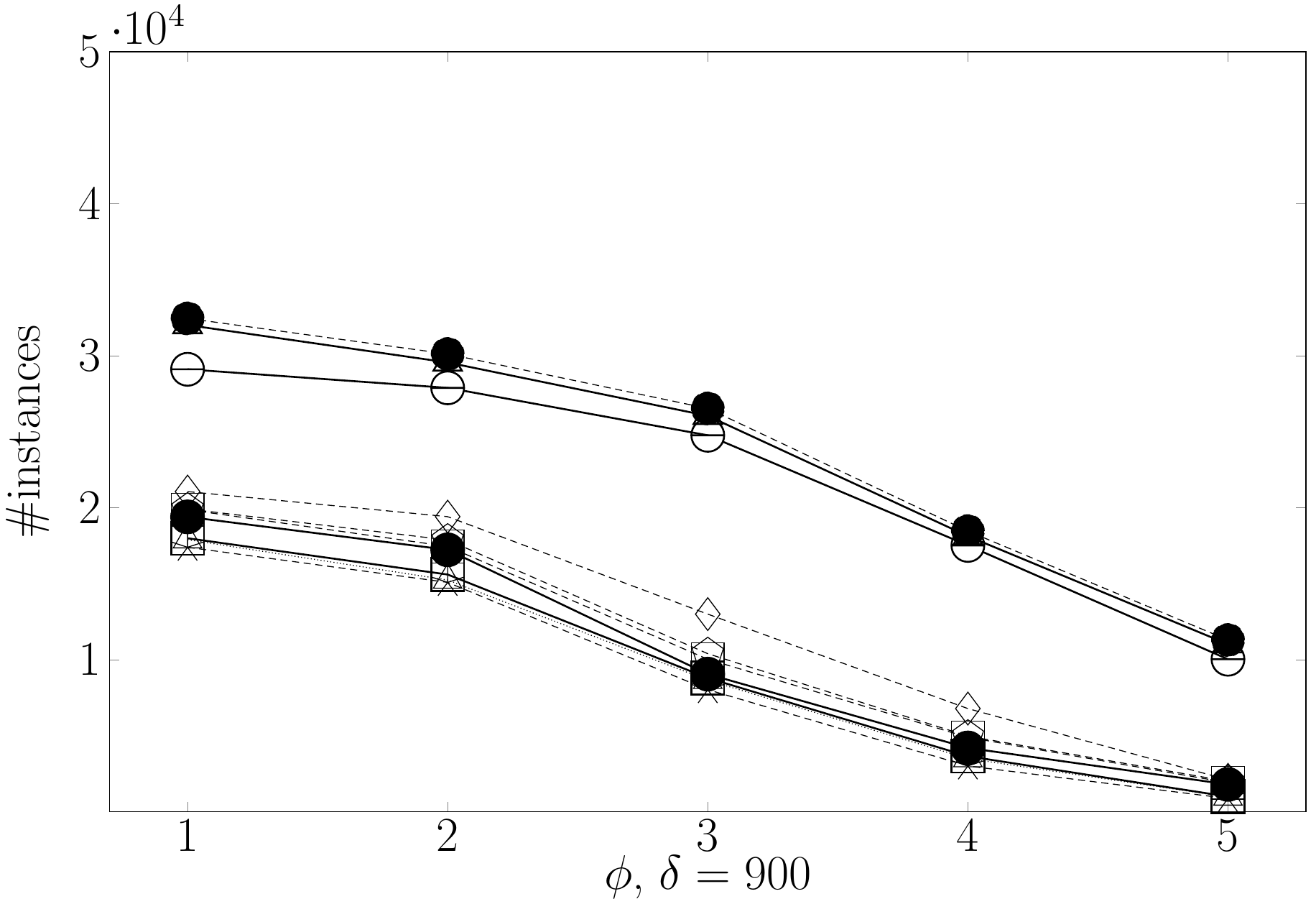}
  }
  \vspace{-1mm}
  \subfigure[Bitcoin Network]{
    \label{fig:exp:bitcoin_scala_phi}
    \includegraphics[width=0.32\textwidth]{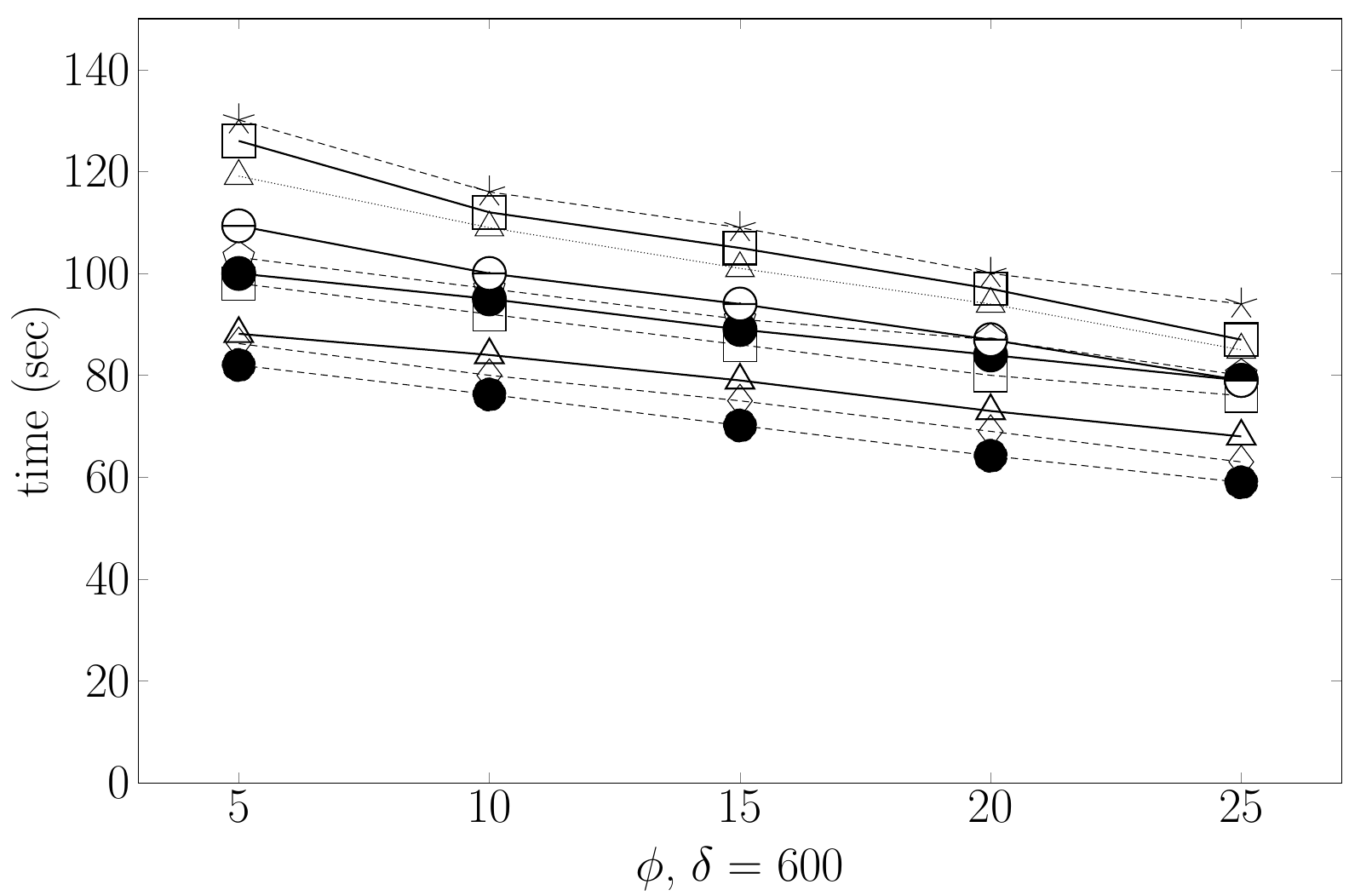}
    }\!\!\!\!
  \subfigure[Facebook Network]{
    \label{fig:exp:fb_scala_phi}
    \includegraphics[width=0.32\textwidth]{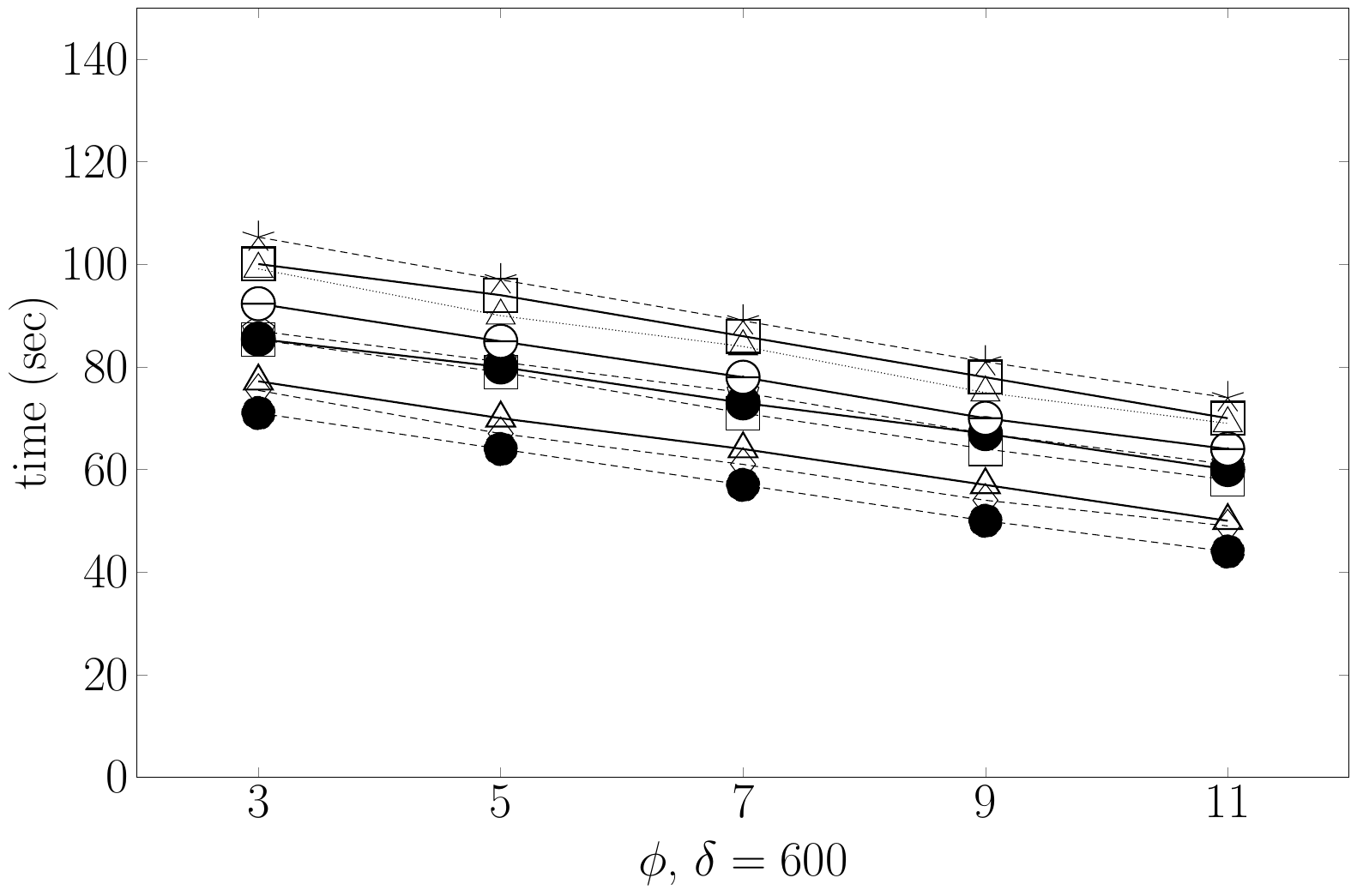}
    }\!\!\!\!
  \subfigure[Passenger Network]{
    \label{fig:exp:traffic_scala_phi}
    \includegraphics[width=0.32\textwidth]{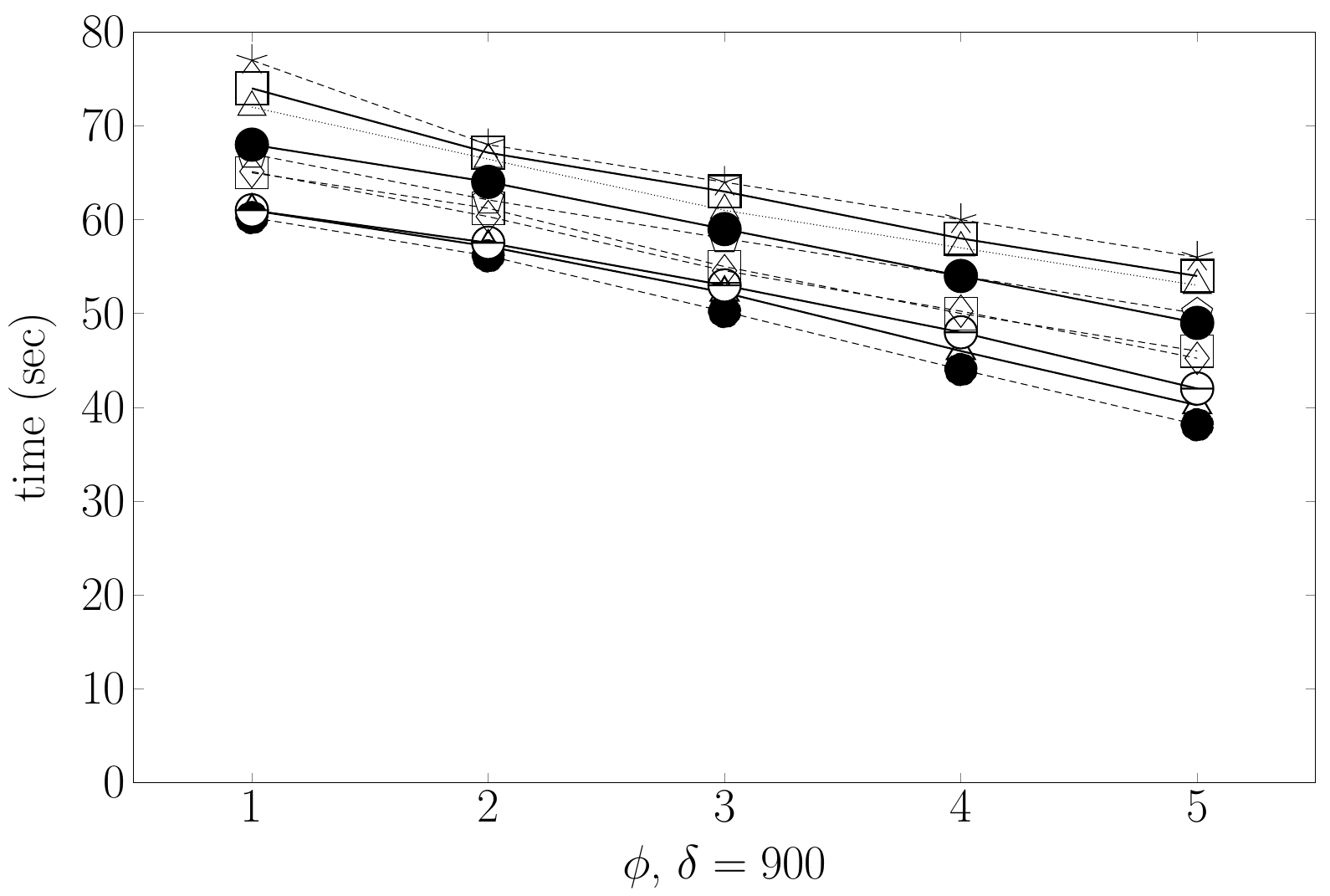}
    }
  \vspace{-0.1in}
  \caption{Number of instances and time for different values of $\phi$}
  \label{fig:exp:phi}
\end{figure*}

\hide{
\begin{figure*}[t!]
  \includegraphics[width=0.88\textwidth]{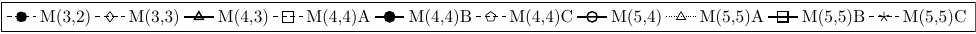}
\subfigure[Bitcoin Network]{
    \label{fig:exp:bitcoin_scala_phi}
    \includegraphics[width=0.32\textwidth]{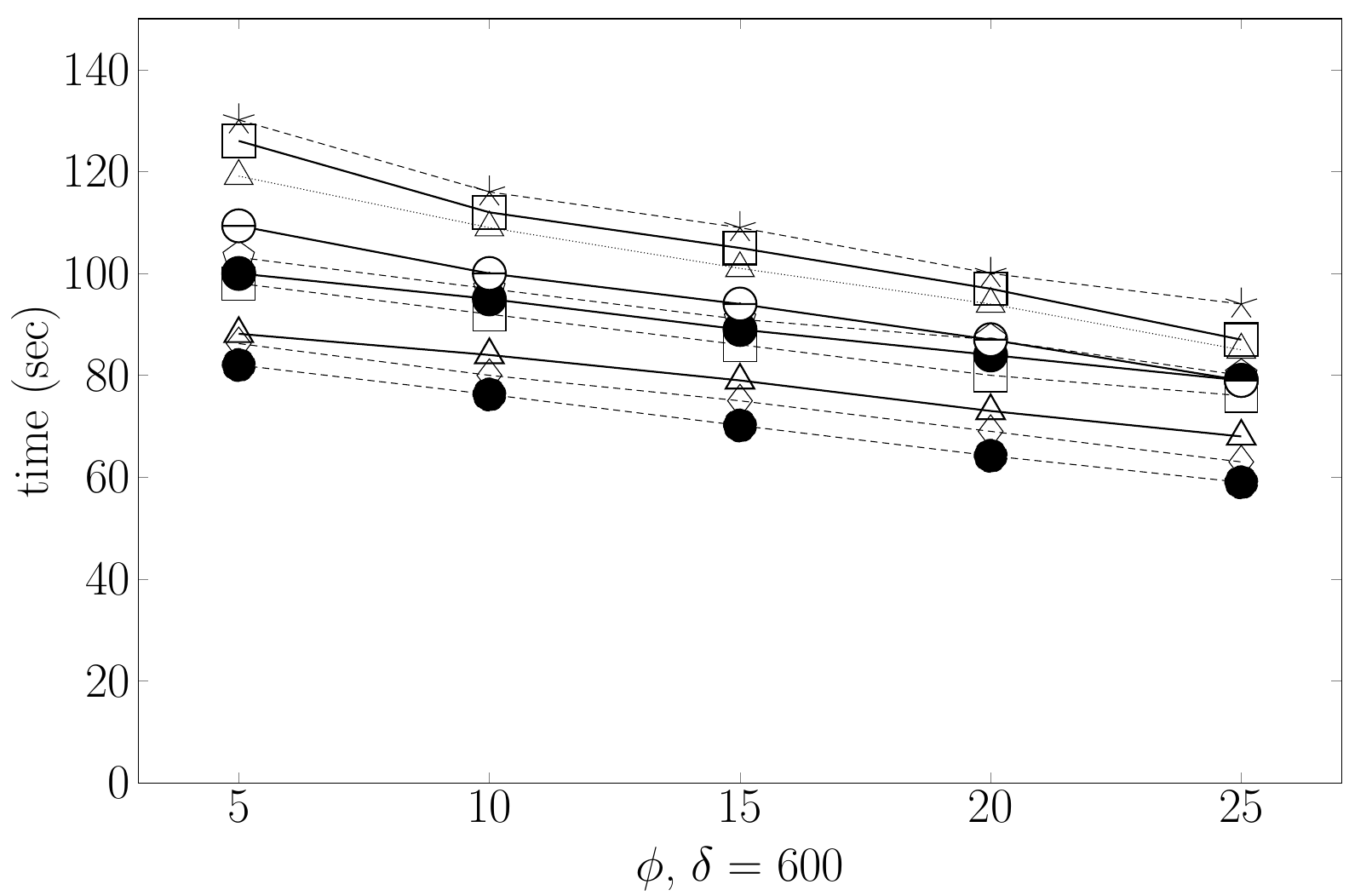}
    }\!\!\!\!
  \subfigure[Facebook Network]{
    \label{fig:exp:fb_scala_phi}
    \includegraphics[width=0.32\textwidth]{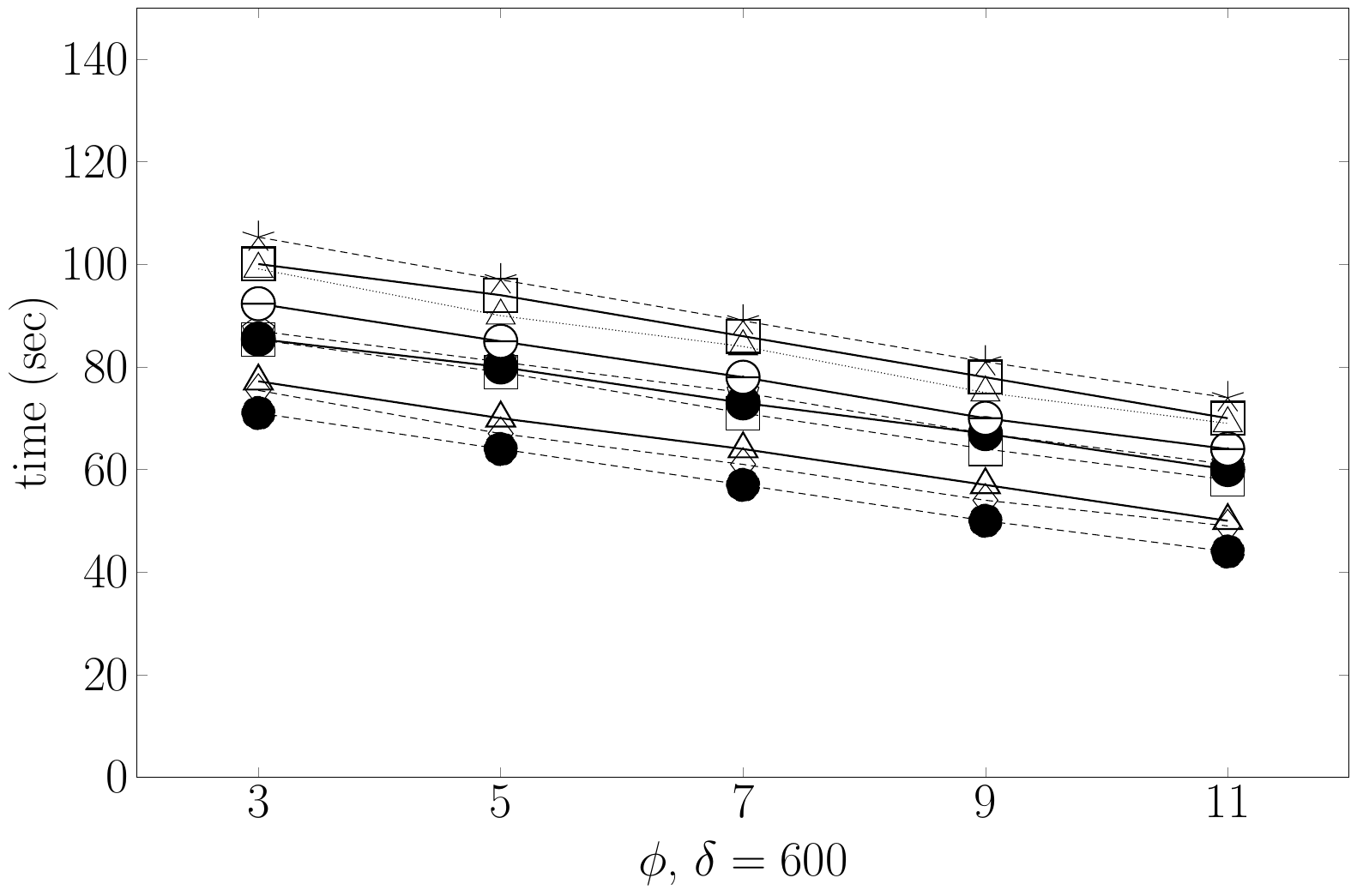}
    }\!\!\!\!
  \subfigure[Passenger Network]{
    \label{fig:exp:traffic_scala_phi}
    \includegraphics[width=0.32\textwidth]{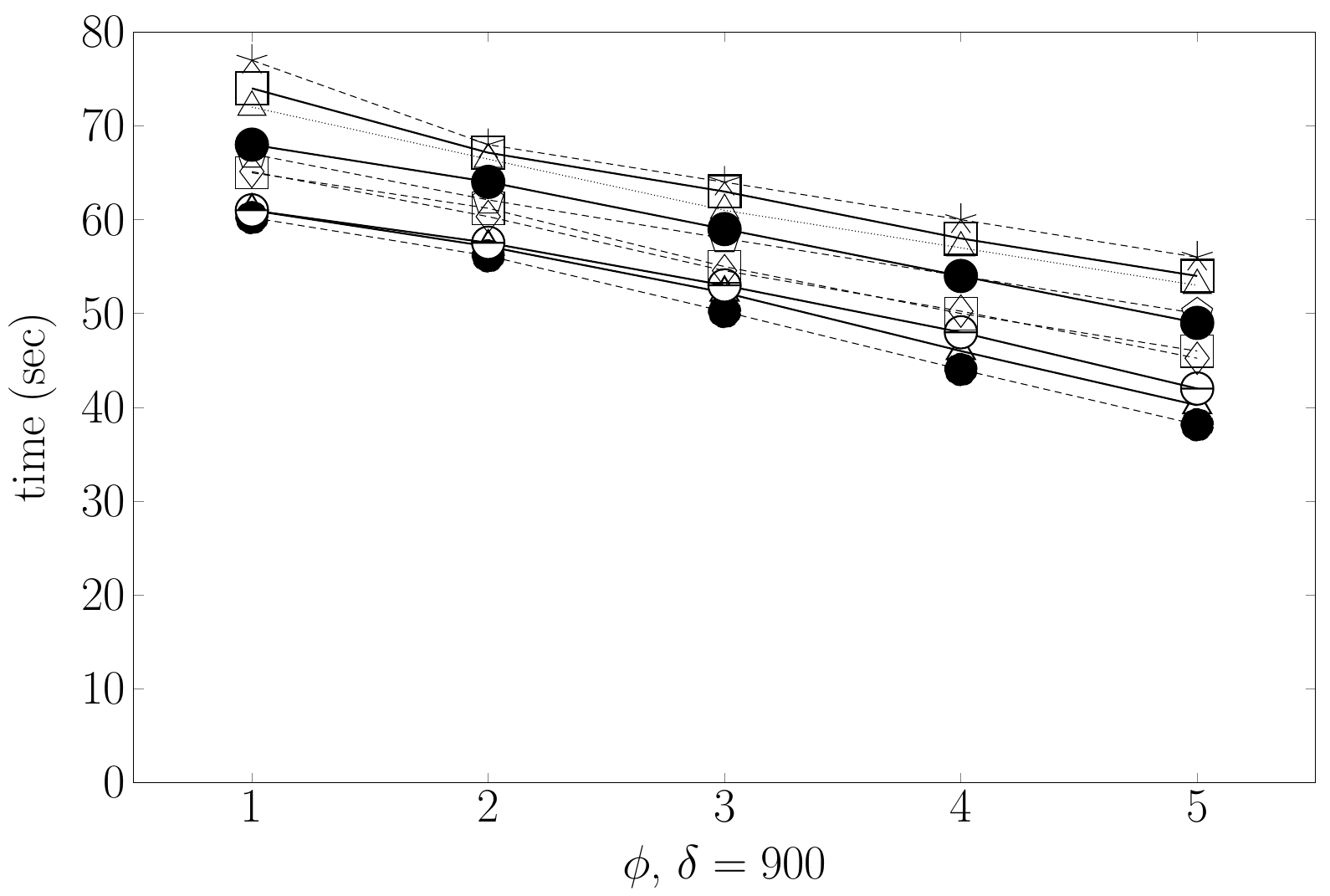}
    }
  \vspace{-0.1in}
  \caption{Scaling $\phi$}
  \label{fig:exp:scaling}
\end{figure*}
}


\subsubsection{Sensitivity to $\delta$ and $\phi$}\label{sec:deltaphi}
The next set of experiments evaluate the performance of our algorithm on
the different datasets and motifs, for various values of the constraints
$\delta$ and $\phi$. Table \ref{table:phase1} shows the number of structural
matches found and the time spent by the algorithm just for its first phase, which is independent of the $\delta$ and $\phi$ values
(since these constraints are not used when searching for the structural matches).
This cost constitutes a lower bound for our algorithm. Naturally,
more complex motifs require more time but they also have fewer structural matches.

Figures \ref{fig:exp:delta} and \ref{fig:exp:phi} show the number of instances and
total runtime of our algorithm for different values of $\delta$ (in seconds) and $\phi$. When we vary
$\delta$, we set $\phi$ to its default value and vice versa.
As expected, in all cases, when $\delta$ increases the number of instances and
the runtime increases. The algorithm scales well as its cost increases at a lower pace
compared to the results found. 

When comparing the different motifs, note that the
simpler ones (e.g., $M(3,2)$ and $M(3,3)$) naturally have more
instances and are cheaper to search compared to
the more complex ones (e.g., $M(5,5)A$).
The relative order between the motifs is similar in the Bitcoin and Facebook networks.
In both networks cyclic flow is quite common;
i.e., motifs containing cycles have a similar number of instances as motifs
without cycles having the same number of edges.
On the other hand, in the Passenger network, acyclic motifs dominate in terms of
number of instances. This is expected, as it is relatively rare that passengers
move between regions on a map forming cycles compared to moving along a
chain of different regions.

The behavior is also consistent to our expectation when $\phi$ varies; the number of
instances and the runtime drop when $\phi$ increases.
The algorithm becomes faster because partial motif instances that do not
qualify $\phi$ are pruned early.

\begin{figure*}[t!]
  \includegraphics[width=0.88\textwidth]{plots+tables/legend.pdf}
\subfigure[Bitcoin Network]{
   \label{fig:exp:top-k_bitcoin}
    \includegraphics[width=0.32\textwidth]{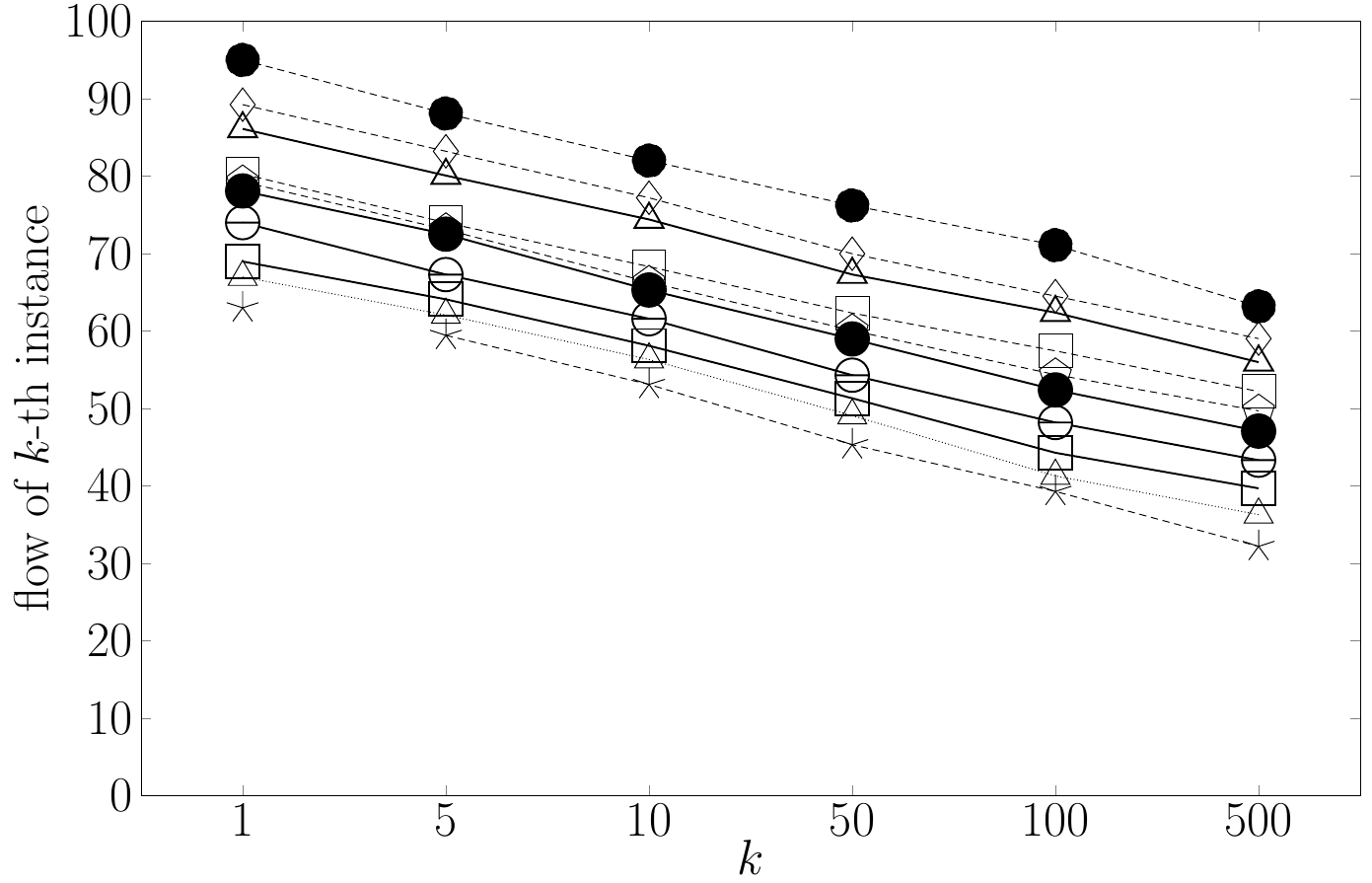}
    }\!\!\!\!
  \subfigure[Facebook Network]{
   \label{fig:exp:top_k_facebook}
    \includegraphics[width=0.32\textwidth]{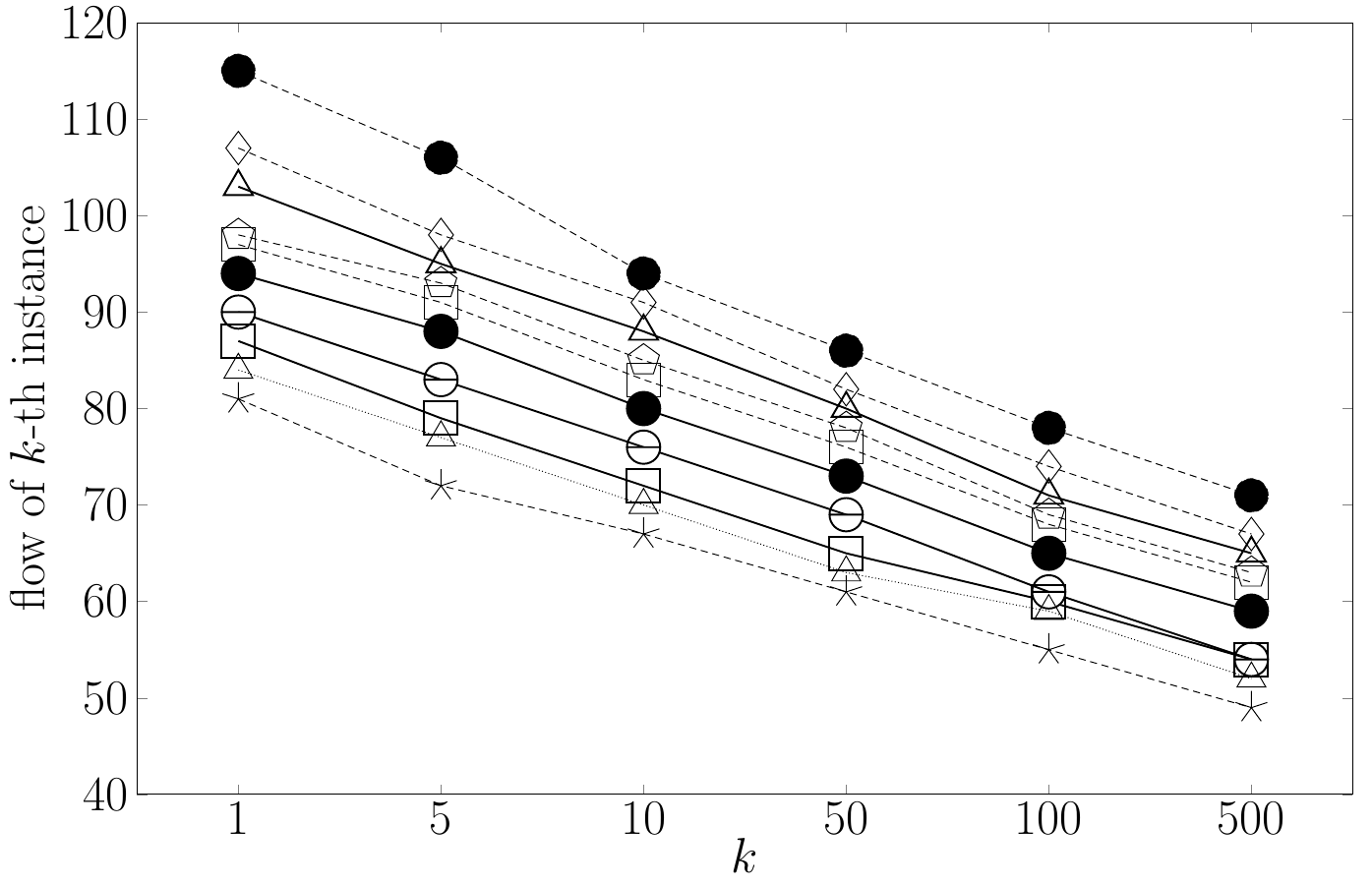}
    }\!\!\!\!
  \subfigure[Passenger Network]{
   \label{fig:exp:top_k_traffic}
    \includegraphics[width=0.32\textwidth]{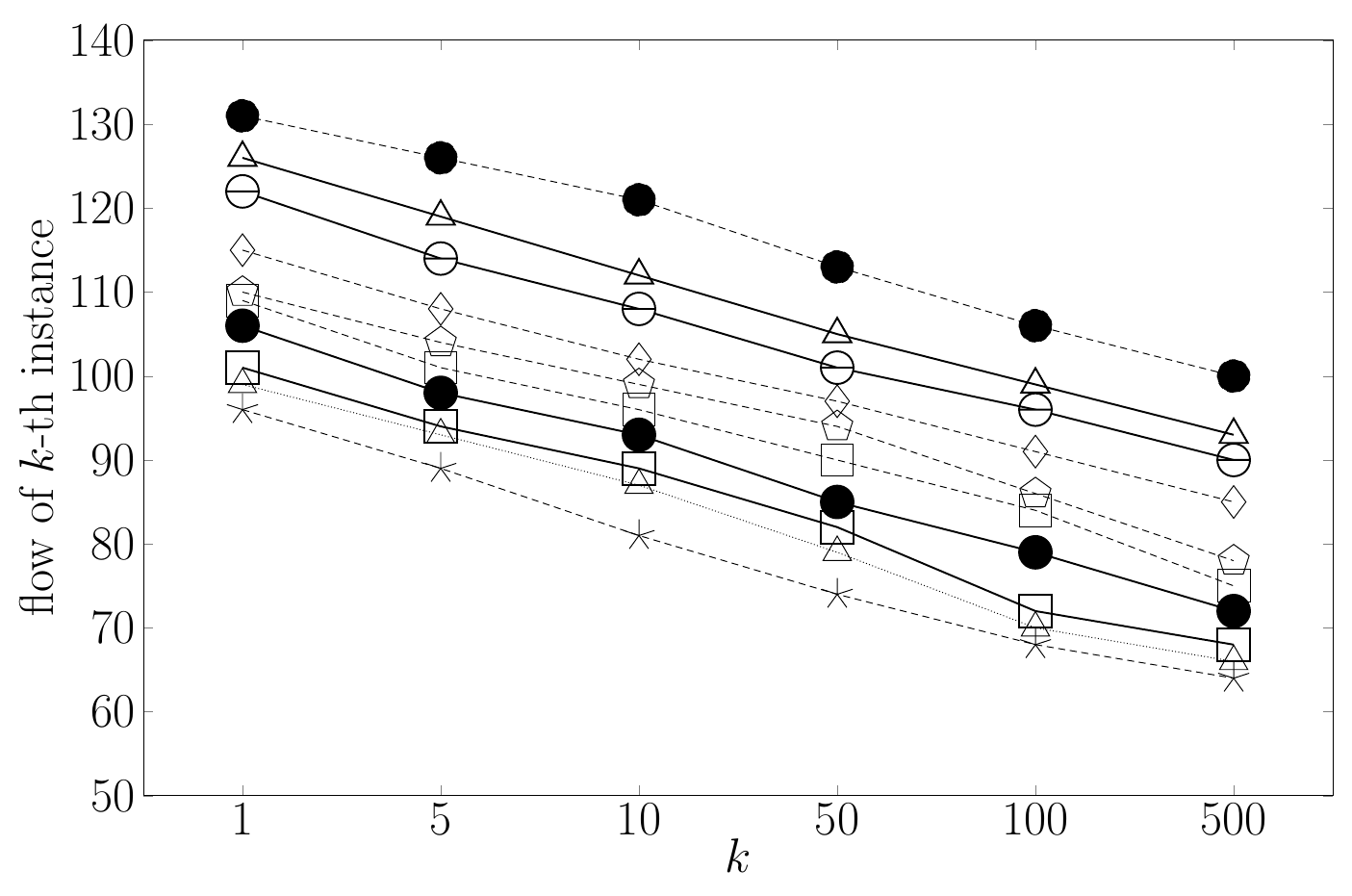}
    }
  \vspace{-0.1in}
  \caption{Flow of $k$-th instance}
  \label{fig:exp:topk}
\end{figure*}

\begin{figure*}[t!]
\subfigure[Bitcoin Network]{
   \label{fig:exp:bitcoin_dp}
    \includegraphics[width=0.32\textwidth]{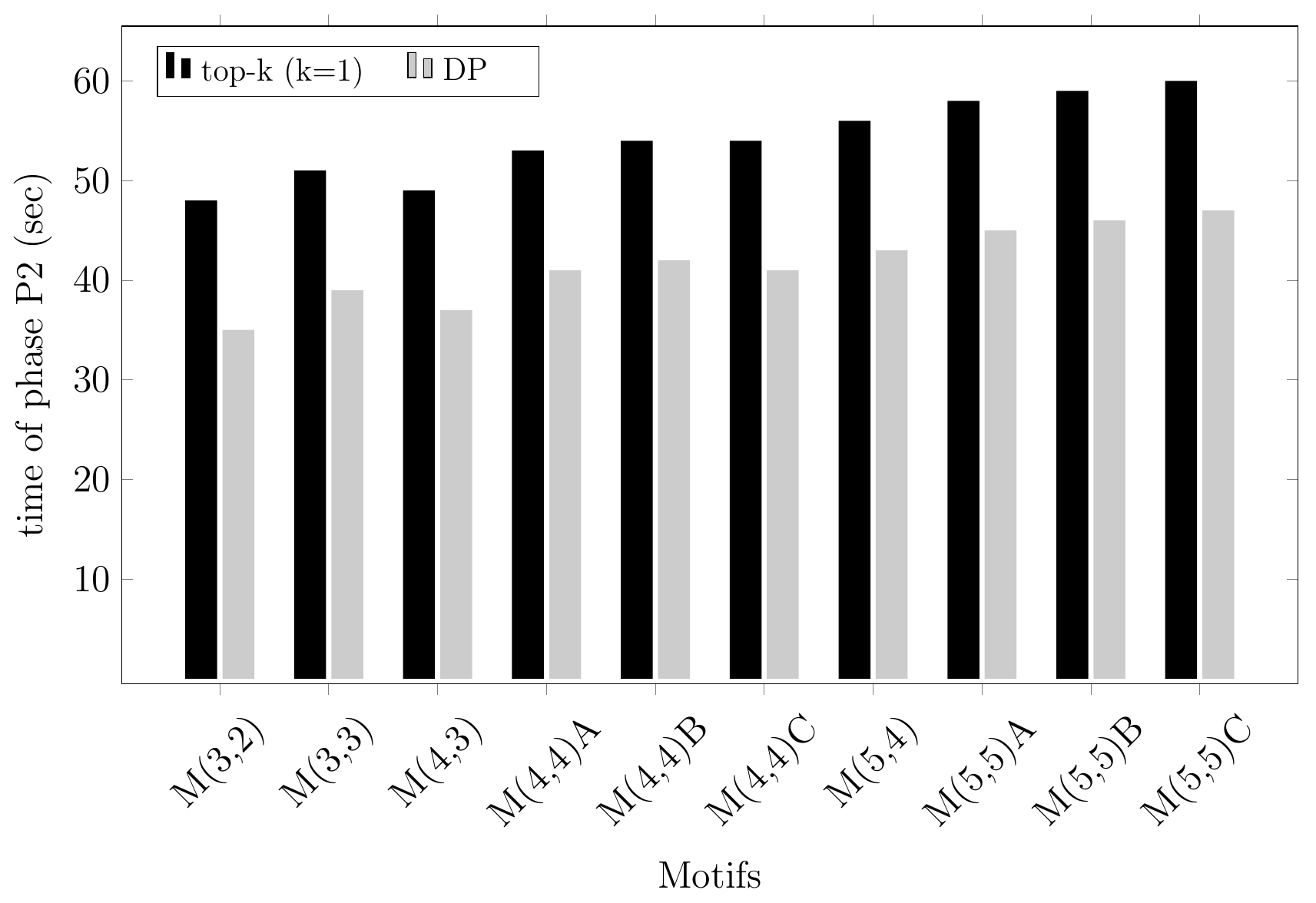}
    }\!\!\!\!
  \subfigure[Facebook Network]{
   \label{fig:exp:fb_dp}
    \includegraphics[width=0.32\textwidth]{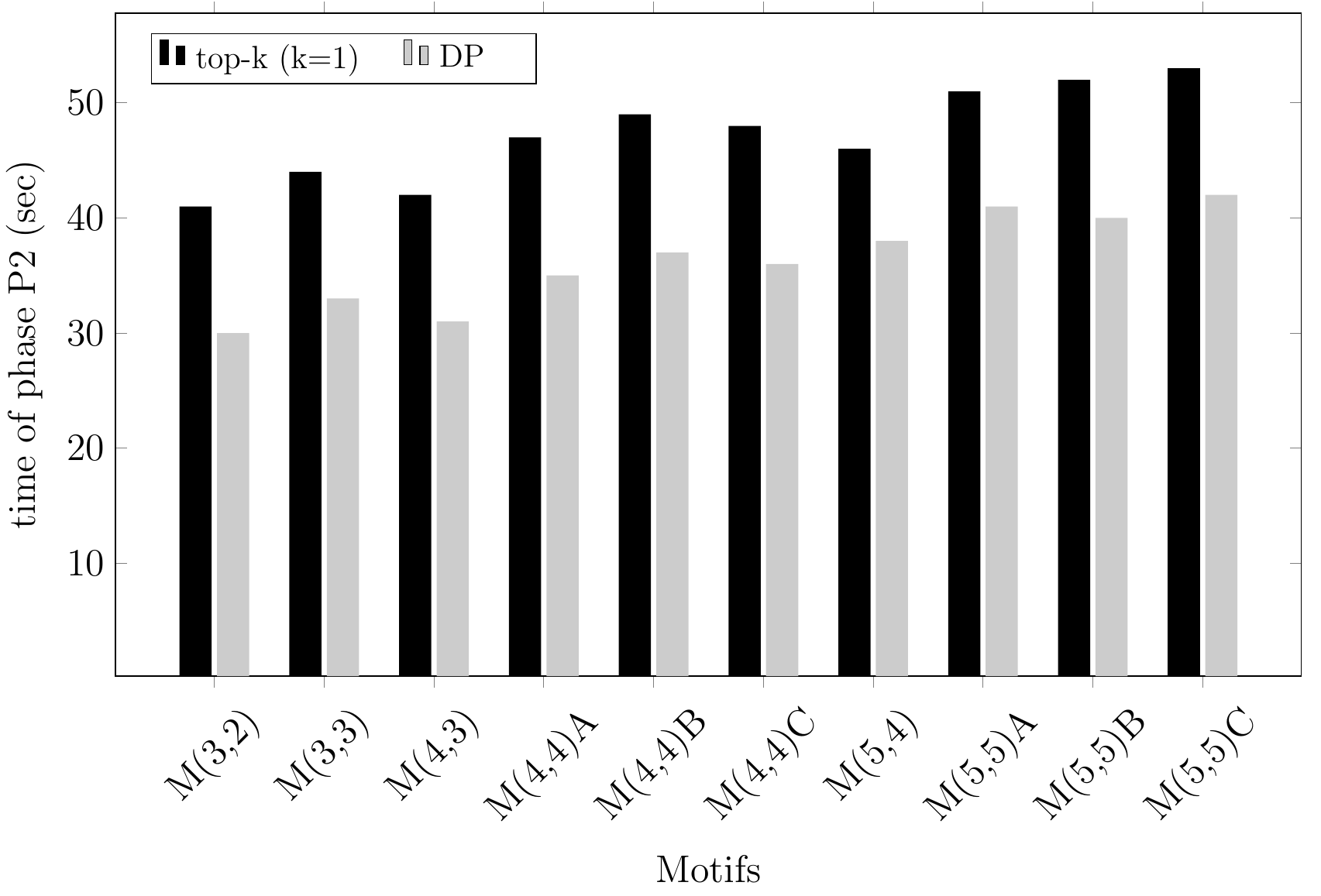}
    }\!\!\!\!
  \subfigure[Traffic Network]{
   \label{fig:exp:traffic_dp}
    \includegraphics[width=0.32\textwidth]{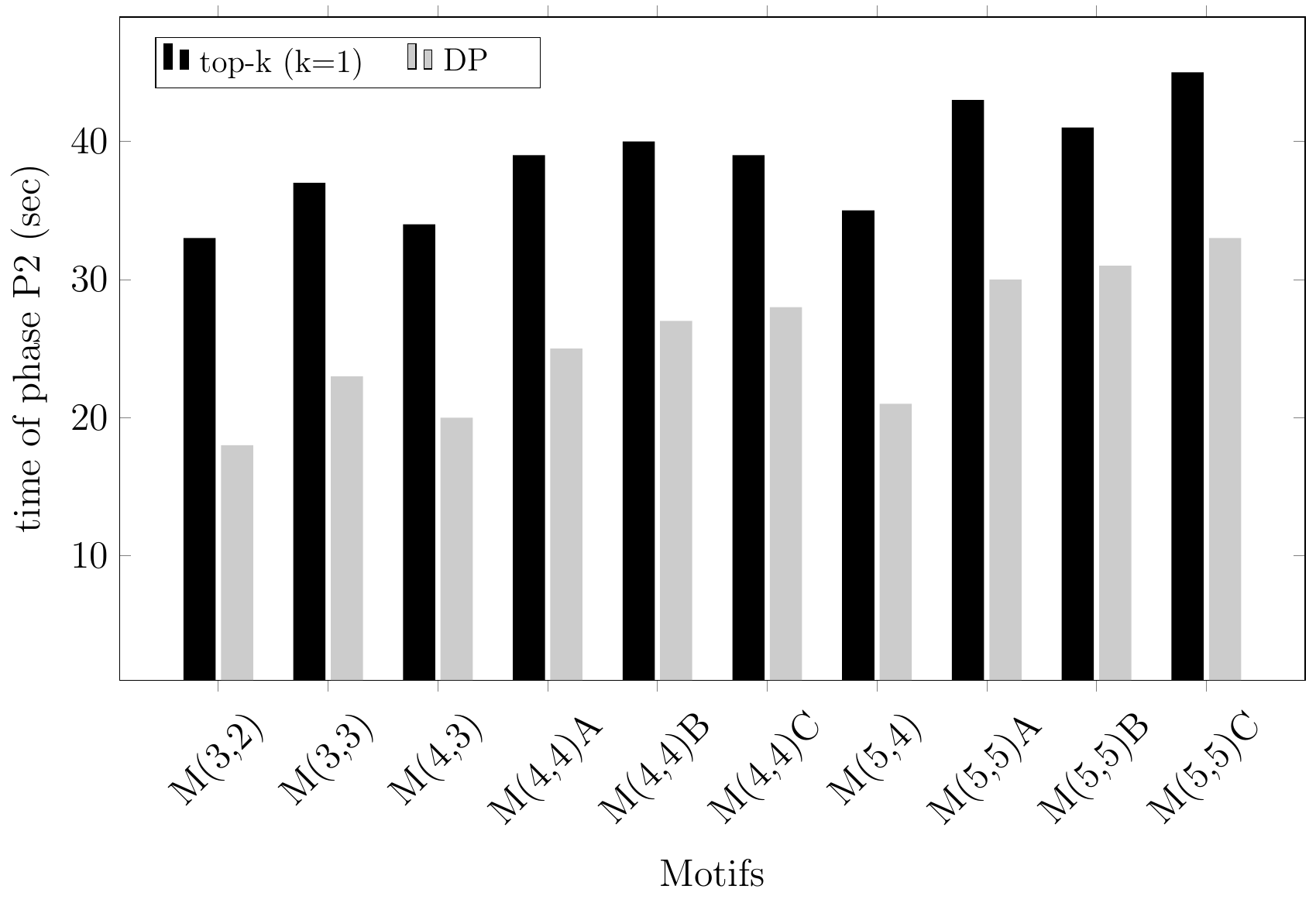}
    }
  \vspace{-0.1in}
  \caption{Efficiency of the dynamic programming module}
  \label{fig:exp:dp}
\end{figure*}

\subsubsection{Top-$k$ flow motif instance search}\label{sec:deltaphi}
We now evaluate the results and the performance of top-$k$ motif search
on the three datasets, when
using the default values of $\delta$.
In the first experiment, we run the version of our algorithm
which finds the top-$k$ motif instances that have the maximum flow.
For each run, we record the flow of the $k$-th instance in Figure \ref{fig:exp:topk}.
As expected, the flow of the $k$-th instance drops as $k$ increases;
the drop rate decreases when $k$ becomes large (note that the x-axis is not linear).
In the second experiment, we compare the runtime of the general top-$k$ algorithm
with its version that employs the dynamic programming module proposed
in Section \ref{sec:top1}. The barcharts show that the second phase of the algorithm
benefits from the use of dynamic programming (the runtime drops 20\% to 40\%). The improvement is better on the Passenger network.

\begin{figure*}[t!]
  \includegraphics[width=0.88\textwidth]{plots+tables/legend.pdf}
\subfigure[Bitcoin: instances per dataset]{
    \label{fig:exp:scala:bit:inst}
    \includegraphics[width=0.32\textwidth]{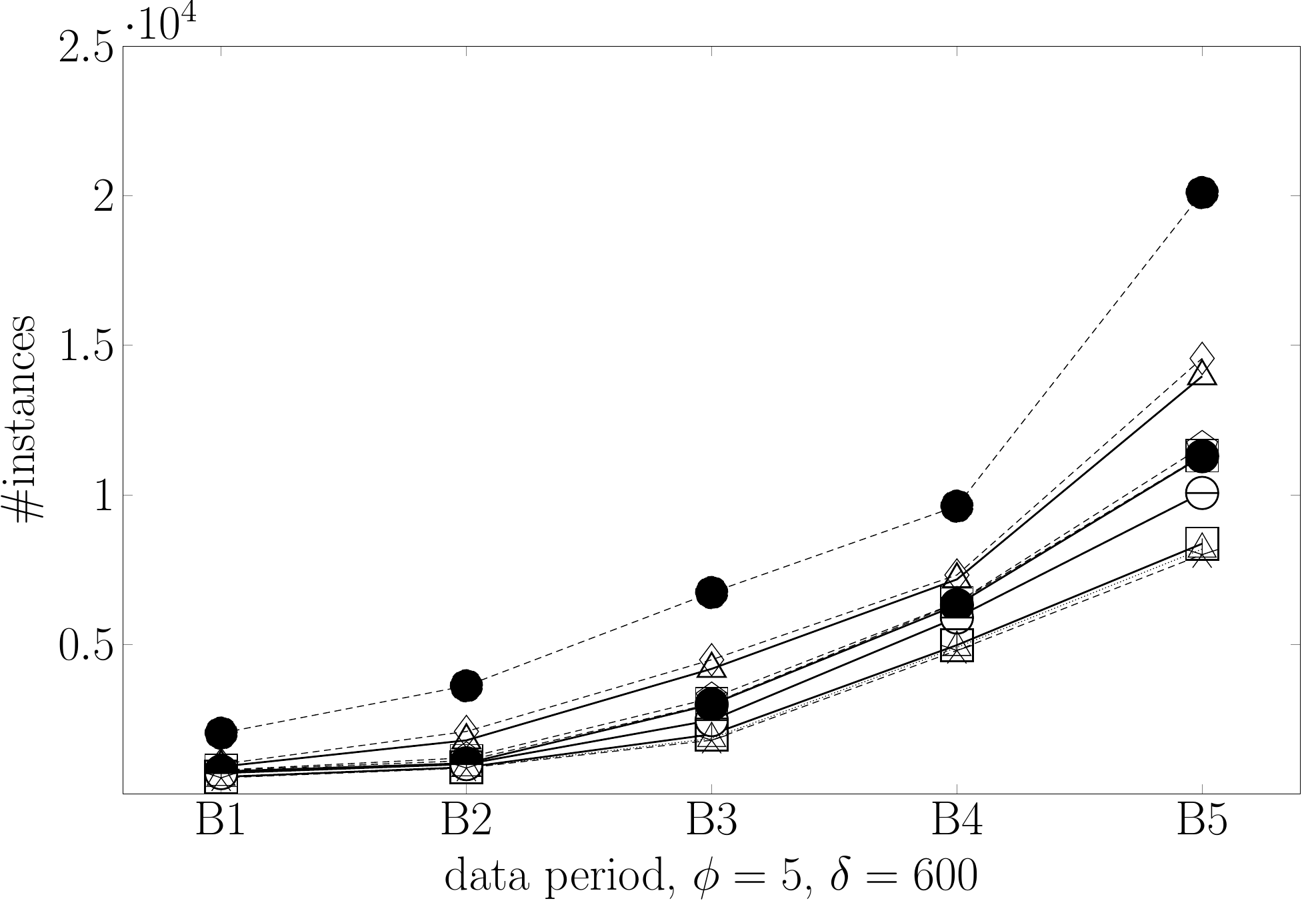}
    }\!\!\!\!
  \subfigure[Facebook: instances per dataset]{
    \label{fig:exp:scala:fb:inst}
    \includegraphics[width=0.32\textwidth]{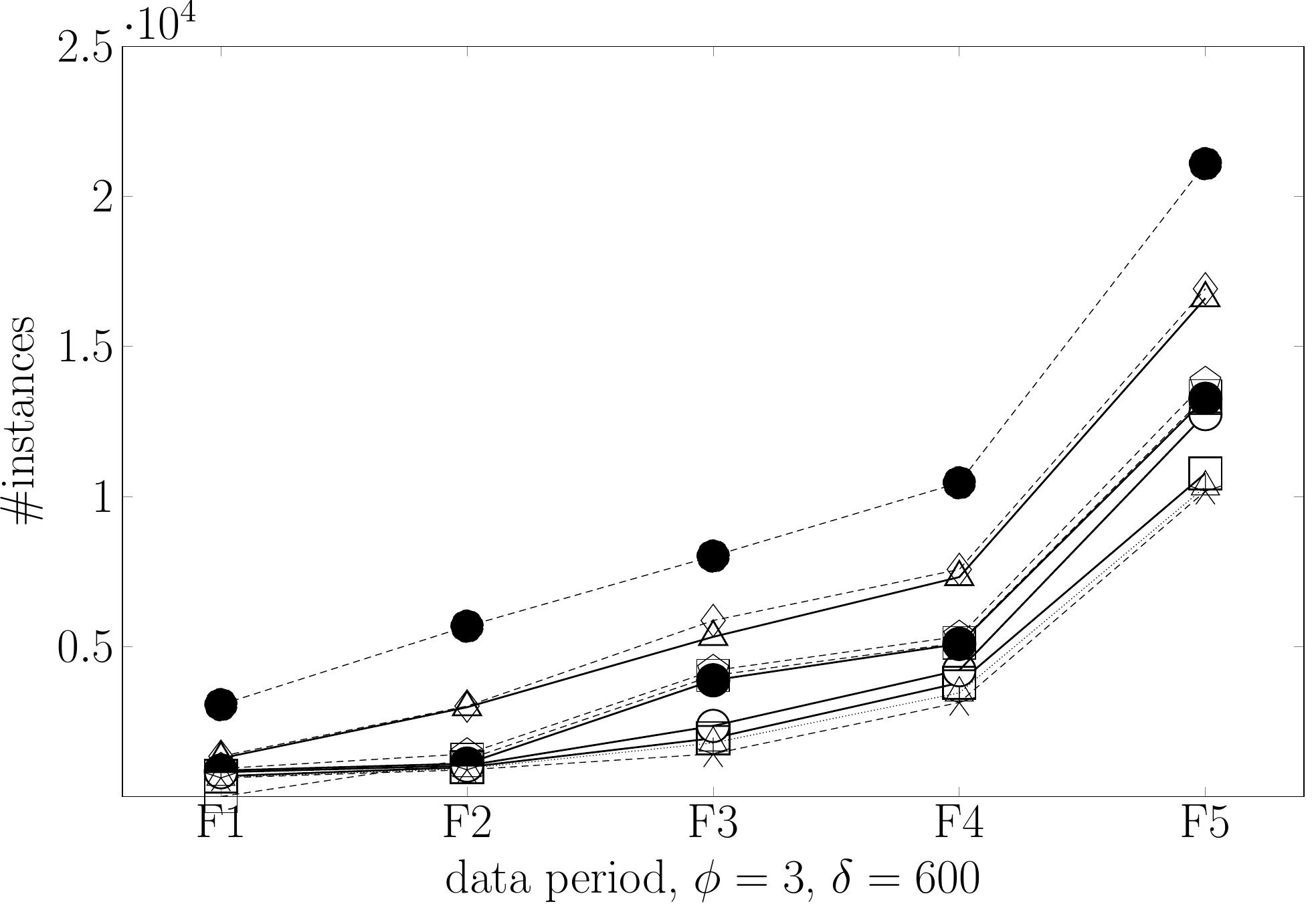}
    }\!\!\!\!
  \subfigure[Passenger: instances per dataset]{
    \label{fig:exp:scala:traffic:inst}
    \includegraphics[width=0.32\textwidth]{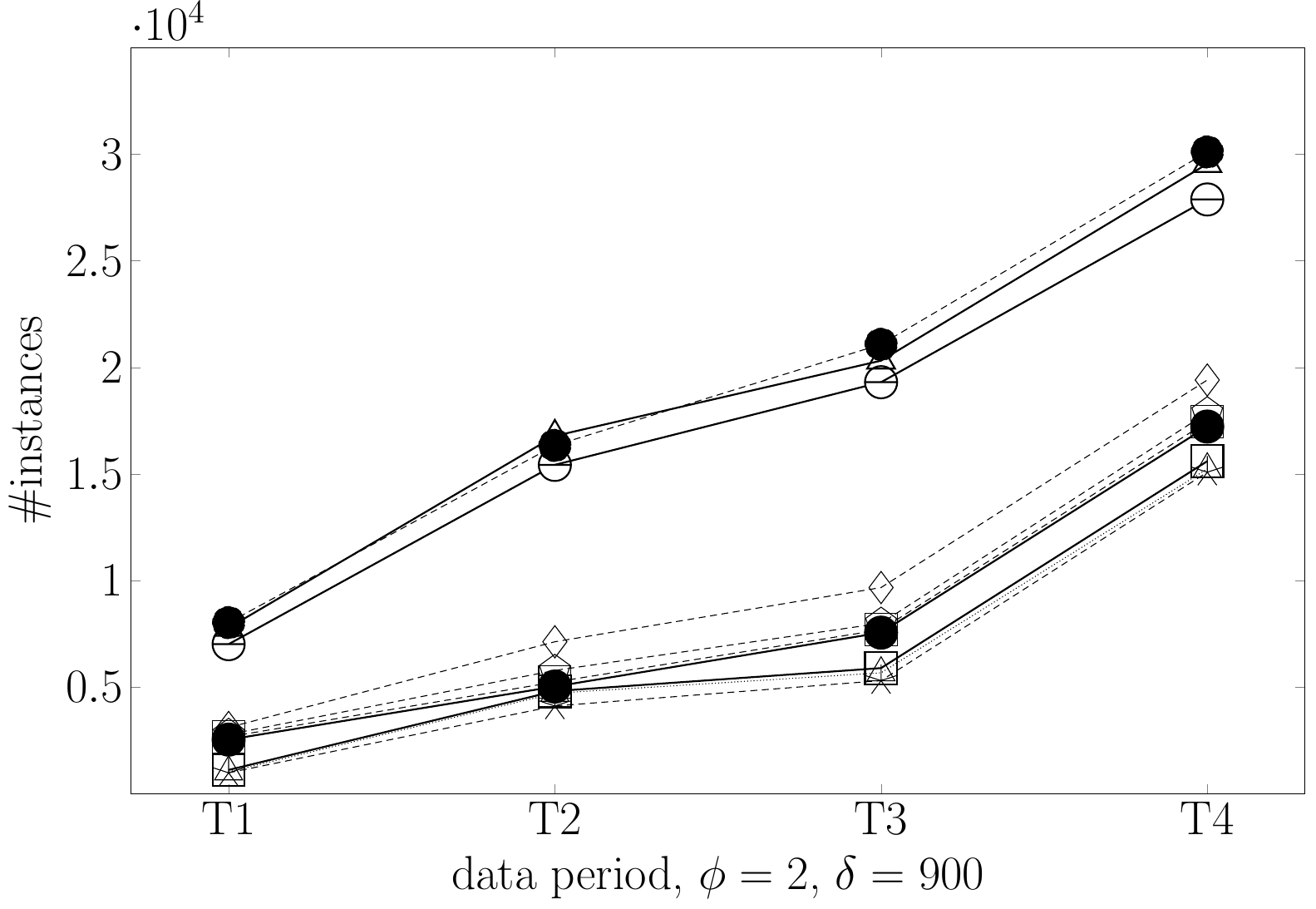}
    }
\subfigure[Bitcoin: time per dataset]{
    \label{fig:exp:scala:bit:time}
    \includegraphics[width=0.32\textwidth]{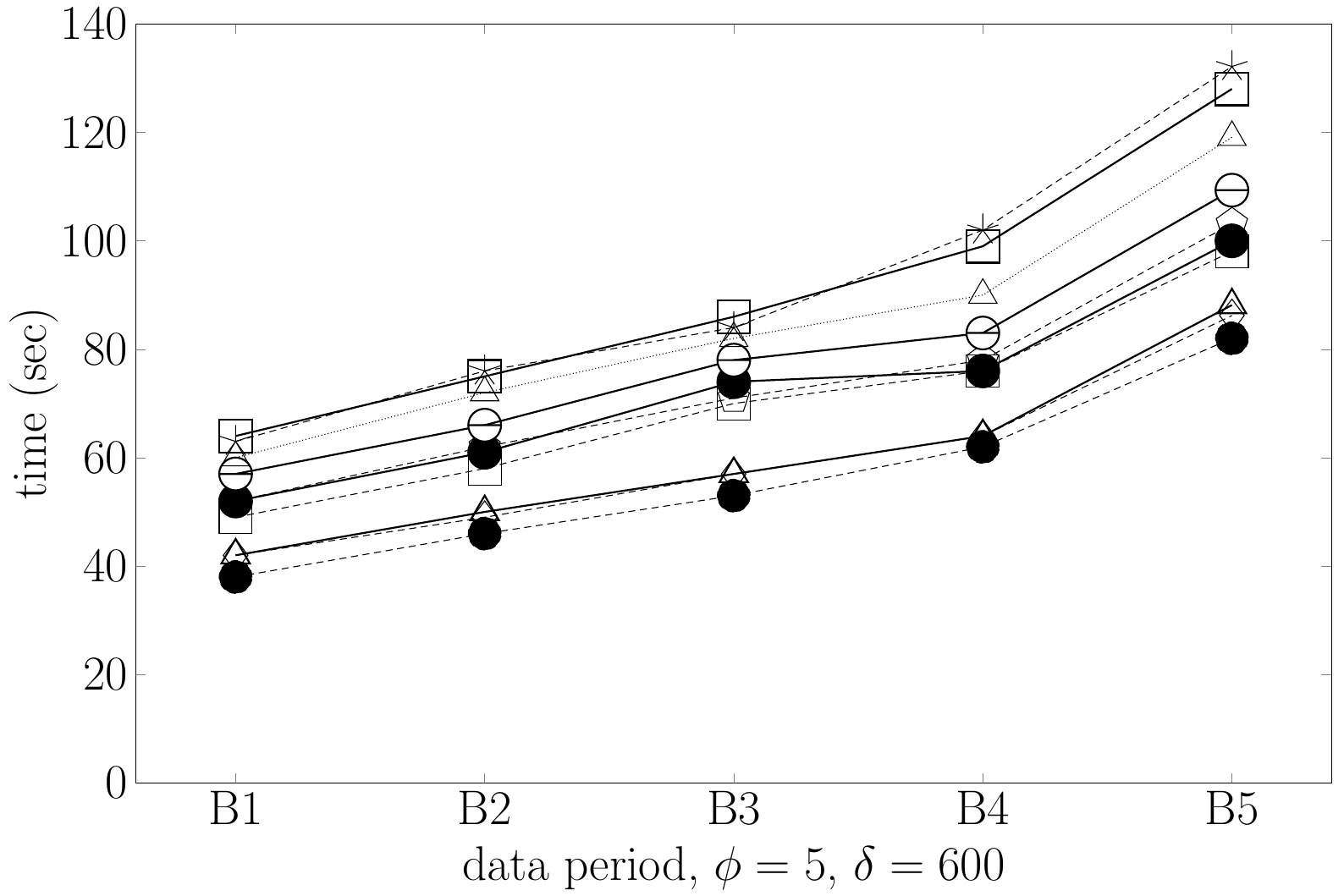}
    }\!\!\!\!
  \subfigure[Facebook: time per dataset]{
    \label{fig:exp:scala:fb:time}
    \includegraphics[width=0.32\textwidth]{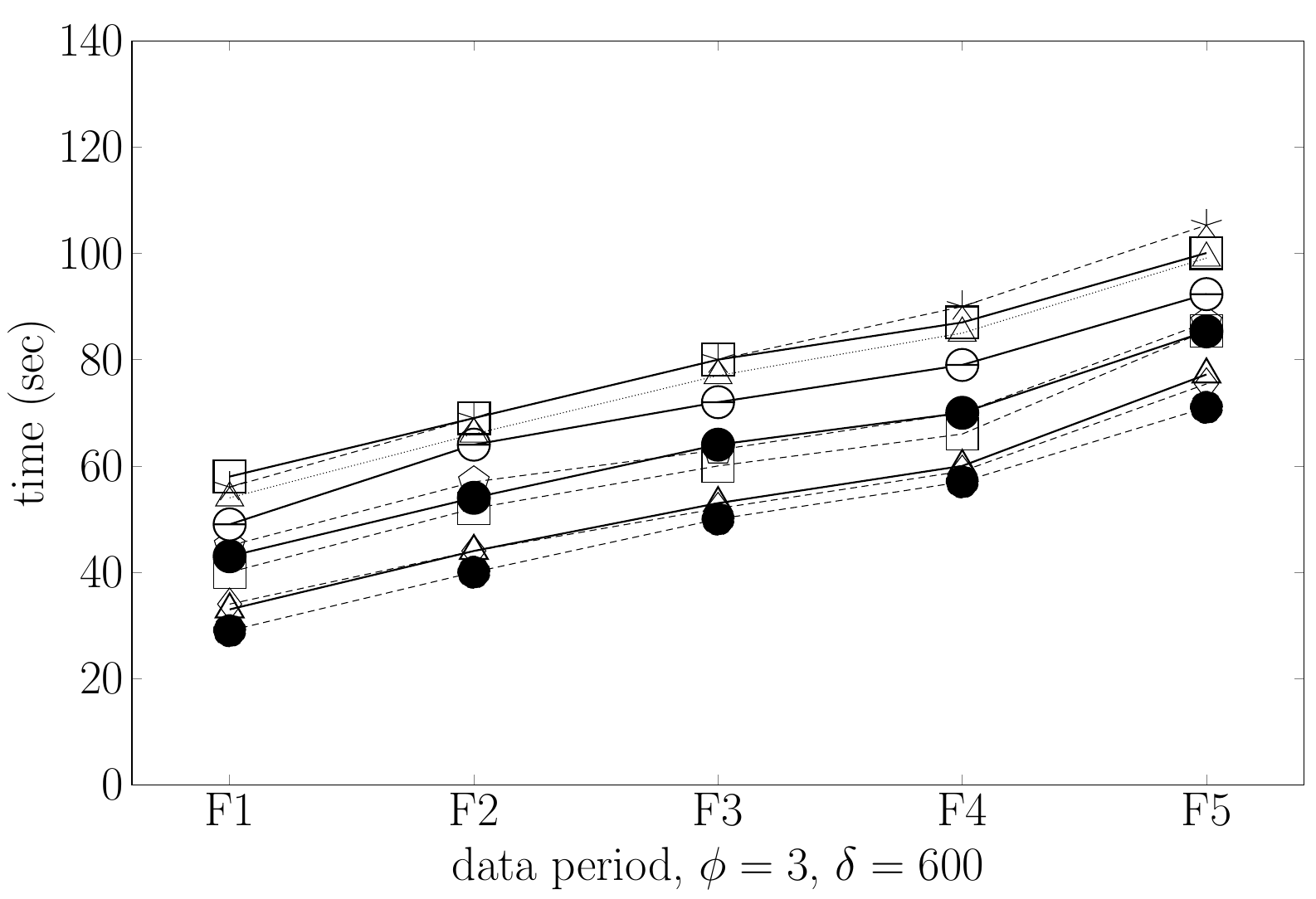}
    }\!\!\!\!
  \subfigure[Passenger: time per dataset]{
    \label{fig:exp:scala:traffic:time}
    \includegraphics[width=0.32\textwidth]{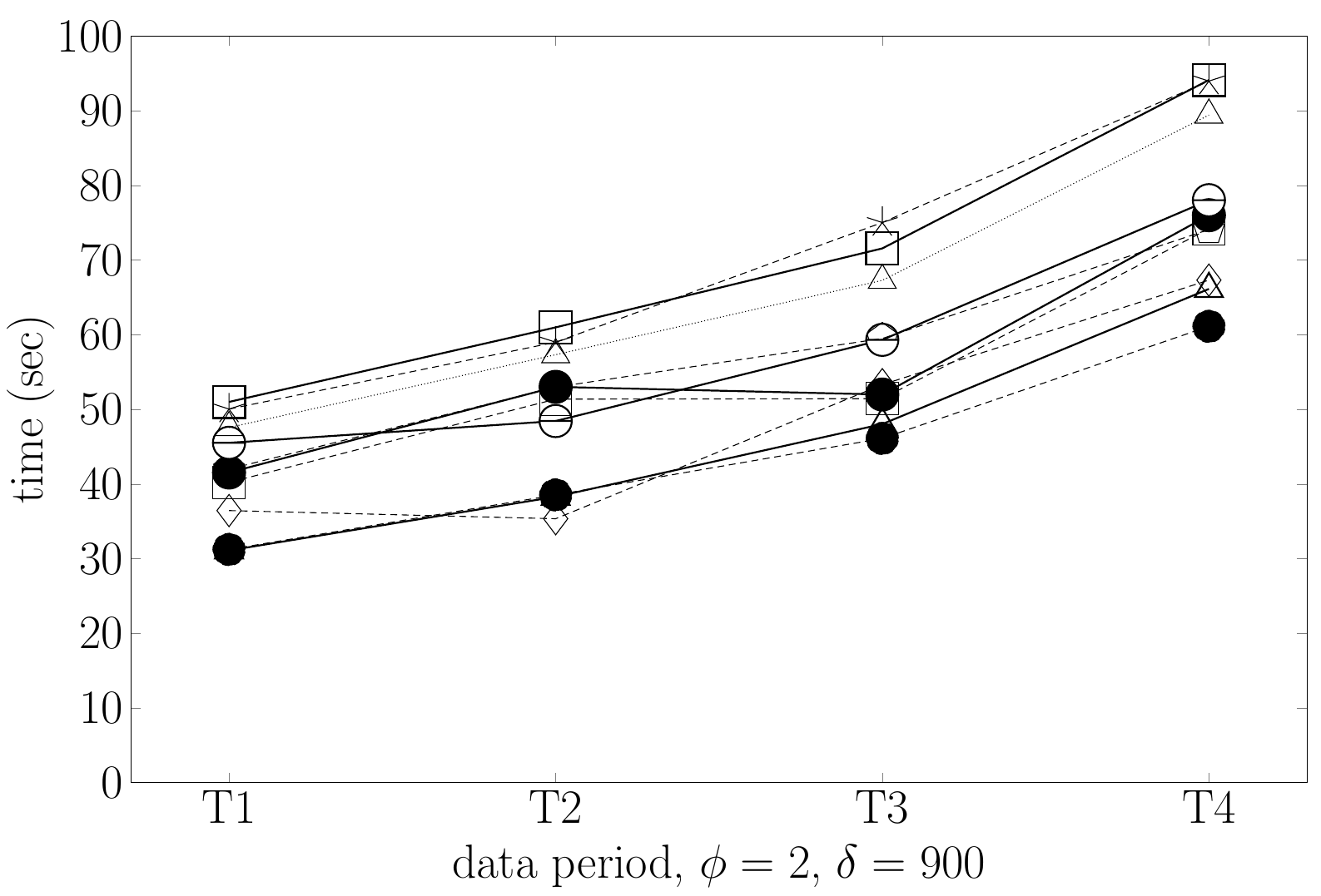}
    }
  \vspace{-0.1in}
  \caption{Scalability to input graph size}
  \label{fig:exp:scala}
\end{figure*}

\subsubsection{Scalability to the dataset size}\label{sec:scalability}
In the next experiment, we test the performance of our algorithm on
samples of the original datasets having different sizes.
For each of the three datasets, we take samples defined by prefixes of the total
period covered by the timestamps of the edges included in the sample.
Specifically, for the Bitcoin network we define 5 samples: B1, B2, B3, B4, B5. B1 includes
all transactions happened in the first month of the 9-month period of the complete dataset. B2, B3, B4, and B5 cover the first 2, 4, 6, and 9 months respectively.
Similarly F1, F2, F3, F4, and F5 cover the first 1, 2, 3, 4, and 6 months of the entire dataset respectively.
Lastly, T1, T2, T3, and T4 cover the first 8, 16, 24, and 31 days of January 2018 respectively.
Figure \ref{fig:exp:scala} shows the growth in the number of instances and in the
runtime of the algorithm for the different motifs. Observe that the algorithm scales
well as its cost grows at a slower pace compared to the number of
instances and the size of the input data.

\begin{figure*}[t!]
\subfigure[Bitcoin Network]{
    \label{fig:exp:boxplot_bitcoin}
    \includegraphics[width=0.32\textwidth]{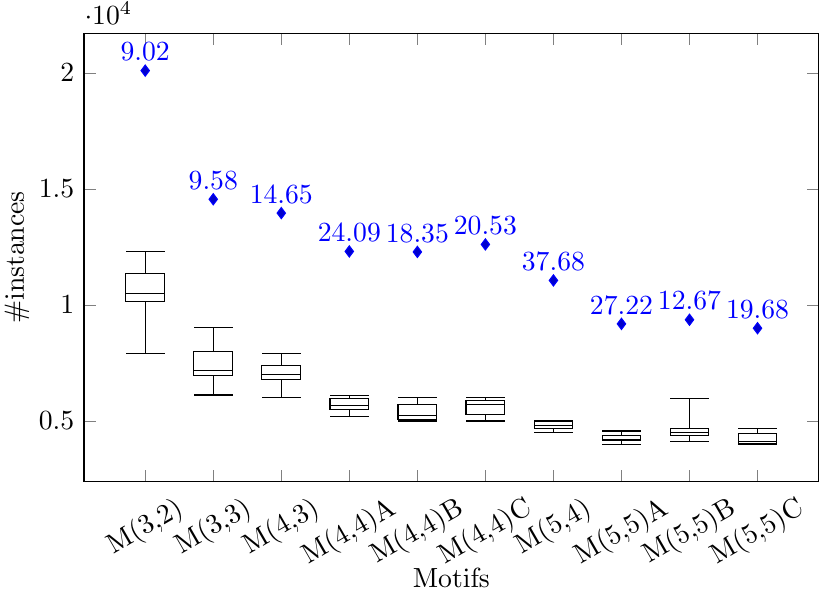}
    }\!\!\!\!
  \subfigure[Facebook Network]{
    \label{fig:exp:boxplot_fb}
    \includegraphics[width=0.32\textwidth]{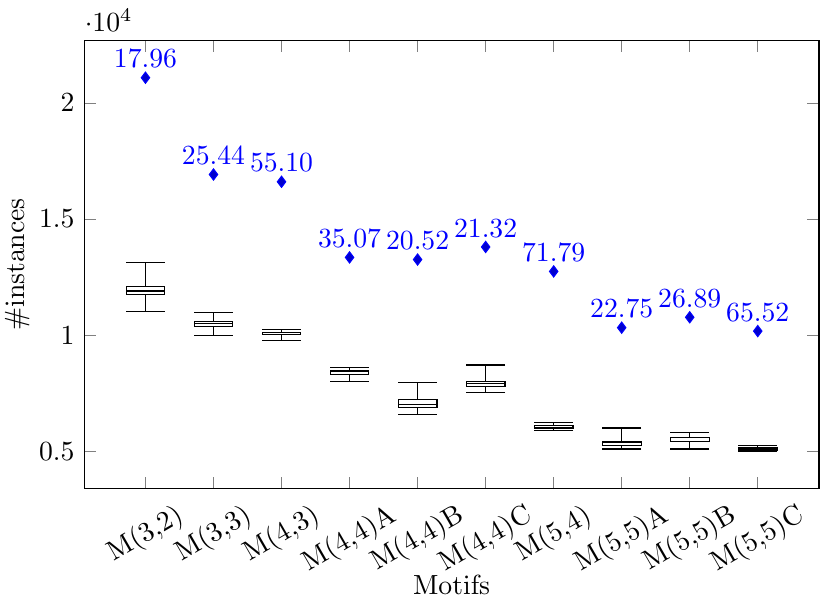}
    }\!\!\!\!
  \subfigure[Passenger Network]{
    \label{fig:exp:boxplot_traffic}
    \includegraphics[width=0.32\textwidth]{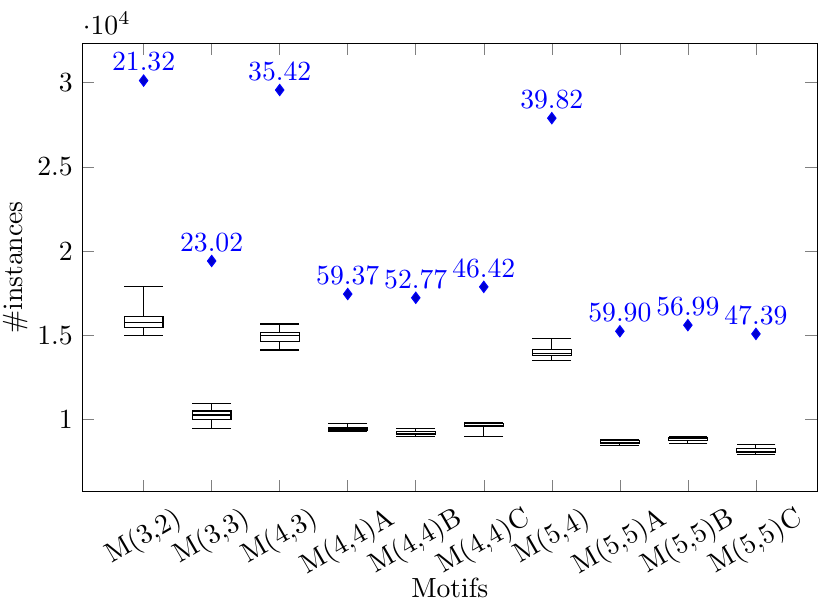}
    }
  \vspace{-0.1in}
  \caption{Number of instances in random networks (box plots), in real networks (diamonds), and z-scores}
  \label{fig:exp:boxplot}
\end{figure*}

\subsection{Significance of Motifs}

In the last experiment, we assess the significance of the different flow motifs
in our networks.
Following the standard practice \cite{DBLP:conf/icde/RanuS09},
we generated randomized versions of our datasets, we computed the number
of instances of each motif in each of these datasets, and we compared it
against the same number for the real dataset.
A large divergence between real and randomized numbers indicates a significant motif.

Specifically, from each dataset (e.g.. Bitcoin network) we generated random
datasets by keeping the
structure of the corresponding graph fixed, and permuting the flows on the edges.
Recall that in the original input multigraph $G = (V,E)$
each edge $e$ is associated with a timestamp $t(e)$ and a flow value $f(e)$.
A pair of nodes $(u,v)$ is connected by a set of edges $E(u,v)$.
Given the entire set of flow values $\{f(e): e\in E\}$, we compute
a random permutation $\pi$ of the flow values and reassign them to the
graph edges in this order.
This generates a randomized dataset $G_r(V,E)$ with the same set of nodes
and the same set of edges;
each edge $e$ has the same timestamp $t(e)$, and flow value $\pi(f(e))$.
Hence, $G_r$ is derived from $G$ by randomly ``shuffling'' the flow values
on the edges.

The random graph $G_r$ has the exact same structure as $G$ and
the edges in the graph appear at the same timestamps.
Therefore, all structural matches of the motifs in $G$
will also appear in $G_r$.
In addition, putting aside the flow constraint $\phi$,
the motif instances in the two graphs will be the same, when considering only $\delta$.
What changes is the flow value of each motif instance, which will
result in a different number
of flow motif instances in $G_r$ compared to $G$, for non-zero values of $\phi$.
Our goal is to study whether the motif instances that satisfy the $\phi$ constraint
in the real data are statistically significantly more than those in the randomized data.

We generated 20 different random graphs for each real network
according to the procedure we described above.
We found the instances of each motif in all these random datasets.
In addition, we computed the
mean and standard deviation of the number of motif instances
in all 20 random graphs per real dataset.
To assess the significance of a motif in the real data,
we compared the number of instances in the real data
with those in the random data.
Figure \ref{fig:exp:boxplot} shows, for each dataset and motif,
the distribution of the numbers of instances for all random graphs
in a box plot, and the corresponding number
in the real graph (marked by a diamond). Each real value is also associated with the \emph{$z$-score} (shown above the corresponding diamond), which is computed as follows.
For some motif $M$, let $r_M$ denote the number of instances of the motif
in the real data, let $\mu_M$ denote the mean number of motif instances
in the randomized data, and let $\sigma_M$ denote the standard deviation.
The $z$-score $z_M$ of the motif is computed as
\[
z_M = \frac{r_M - \mu_M}{\sigma_M}
\]
The higher the $z$-score, the further the value $r_M$ from $\mu_M$.

The first observation is that the number of instances in all random graphs is much lower compared to that in the corresponding real network and these values do not deviate much from their mean.
The empirical $p$-value (the fraction of random datasets with number of instances greater than that of the real data) is zero,
indicating statistical significance of the motif occurrences in all cases.
This is consistent with the intuition that the flow is not arbitrarily generated or
consumed at the vertices of the network, but it is transferred 
from one node to another.
To discriminate between the different motifs we look at the $z$-scores.
We observe that for the Bitcoin network,
two out of the three top $z$-scores are
for motifs that contain cycles, 
indicating that large flow movements that close a cycle are
statistically over-represented in the bitcoin network.
A similar observation holds for the Passenger flow network, where three out of the top-three motifs contain a cycle.
A different pattern emerges in the Facebook dataset,
where two out of the three highest $z$-scores are for chains of nodes.
We conjecture that this due to propagation trees of information
in the Facebook network, which result in chains with significantly
high flow movement.
It is interesting that the significance of the discovered motifs varies
in the different types of interaction networks, indicating differences in the way flow is distributed in such networks.

%% file: conclusion.tex
\section{Conclusion}\label{sec:conclusion}

In this paper, we introduced the novel concept of \textit{network flow motifs}.
To the best of our knowledge we are the first to define and study motifs in interaction networks, which consider both the temporal and flow information of the interactions. 
We proposed an efficient algorithm for enumerating flow motif instances in large graphs and variants of that find the top-$k$ instances of maximal flow. 
We evaluated our algorithm on three real datasets and demonstrated its scalability. 
In addition, we compared it to a baseline motif instance finding method based on joining instances of motif components and showed its superiority. Finally, we studied the statistical significance of a wide range of representative motifs on the real graphs and 
showed that they indeed appear more frequently than in random networks with the same characteristics. This indicates that the flow is transferred from one node to another (as opposed to being arbitrarily consumed or generated) and that there are subgraphs in the network where significant flow is transferred at certain periods of time.

In the future, we plan to investigate in more detail the distribution of motif instances in
the real networks. For example, we can group the motif instances per structural match, 
in order to identify the structural matches (i.e., sets of vertices in the graph $G$) with 
the largest activity and how this activity is spread along the timeline.
Another direction is to improve the efficiency of the algorithm, by processing 
multiple structural instances together in phase P2. Since two or more structural matches
may share the same prefix, we can compute the flow instances of their common prefix simultaneously before expanding these instances to complete ones for the different motifs.
In addition, we will work towards a version of the algorithm which focuses on counting 
instances of (possibly multiple) motifs without constructing them (along the direction of
previous work \cite{DBLP:conf/wsdm/ParanjapeBL17}). Finally, we will generalize the 
definition of flow motifs to capture other graph structures besides paths 
(e.g., directed acyclic graphs with forks and joins) and study their search in large networks.

%% file: main.bbl

\begin{thebibliography}{22}


\ifx \showCODEN    \undefined \def \showCODEN     #1{\unskip}     \fi
\ifx \showDOI      \undefined \def \showDOI       #1{#1}\fi
\ifx \showISBNx    \undefined \def \showISBNx     #1{\unskip}     \fi
\ifx \showISBNxiii \undefined \def \showISBNxiii  #1{\unskip}     \fi
\ifx \showISSN     \undefined \def \showISSN      #1{\unskip}     \fi
\ifx \showLCCN     \undefined \def \showLCCN      #1{\unskip}     \fi
\ifx \shownote     \undefined \def \shownote      #1{#1}          \fi
\ifx \showarticletitle \undefined \def \showarticletitle #1{#1}   \fi
\ifx \showURL      \undefined \def \showURL       {\relax}        \fi
\providecommand\bibfield[2]{#2}
\providecommand\bibinfo[2]{#2}
\providecommand\natexlab[1]{#1}
\providecommand\showeprint[2][]{arXiv:#2}

\bibitem[\protect\citeauthoryear{Cazabet, Baccour, and Latapy}{Cazabet
  et~al\mbox{.}}{2017}]%
        {DBLP:conf/complexnetworks/RemyRM17}
\bibfield{author}{\bibinfo{person}{R{\'{e}}my Cazabet}, \bibinfo{person}{Rym
  Baccour}, {and} \bibinfo{person}{Matthieu Latapy}.}
  \bibinfo{year}{2017}\natexlab{}.
\newblock \showarticletitle{Tracking Bitcoin Users Activity Using Community
  Detection on a Network of Weak Signals}. In
  \bibinfo{booktitle}{\emph{{COMPLEX} {NETWORKS}}}. \bibinfo{pages}{166--177}.
\newblock


\bibitem[\protect\citeauthoryear{Gomez{-}Rodriguez, Leskovec, and
  Krause}{Gomez{-}Rodriguez et~al\mbox{.}}{2012}]%
        {Gomez-RodriguezLK12}
\bibfield{author}{\bibinfo{person}{Manuel Gomez{-}Rodriguez},
  \bibinfo{person}{Jure Leskovec}, {and} \bibinfo{person}{Andreas Krause}.}
  \bibinfo{year}{2012}\natexlab{}.
\newblock \showarticletitle{Inferring Networks of Diffusion and Influence}.
\newblock \bibinfo{journal}{\emph{{TKDD}}} \bibinfo{volume}{5},
  \bibinfo{number}{4} (\bibinfo{year}{2012}), \bibinfo{pages}{21:1--21:37}.
\newblock


\bibitem[\protect\citeauthoryear{Gurukar, Ranu, and Ravindran}{Gurukar
  et~al\mbox{.}}{2015}]%
        {DBLP:conf/sigmod/GurukarRR15}
\bibfield{author}{\bibinfo{person}{Saket Gurukar}, \bibinfo{person}{Sayan
  Ranu}, {and} \bibinfo{person}{Balaraman Ravindran}.}
  \bibinfo{year}{2015}\natexlab{}.
\newblock \showarticletitle{{COMMIT:} {A} Scalable Approach to Mining
  Communication Motifs from Dynamic Networks}. In
  \bibinfo{booktitle}{\emph{{SIGMOD}}}. \bibinfo{pages}{475--489}.
\newblock


\bibitem[\protect\citeauthoryear{Holme}{Holme}{2015}]%
        {DBLP:journals/corr/Holme15a}
\bibfield{author}{\bibinfo{person}{Petter Holme}.}
  \bibinfo{year}{2015}\natexlab{}.
\newblock \showarticletitle{Modern temporal network theory: {A} colloquium}.
\newblock \bibinfo{journal}{\emph{CoRR}}  \bibinfo{volume}{abs/1508.01303}
  (\bibinfo{year}{2015}).
\newblock
\showeprint[arxiv]{1508.01303}
\urldef\tempurl%
\url{http://arxiv.org/abs/1508.01303}
\showURL{%
\tempurl}


\bibitem[\protect\citeauthoryear{Kempe, Kleinberg, and Kumar}{Kempe
  et~al\mbox{.}}{2002}]%
        {DBLP:journals/jcss/KempeKK02}
\bibfield{author}{\bibinfo{person}{David Kempe}, \bibinfo{person}{Jon~M.
  Kleinberg}, {and} \bibinfo{person}{Amit Kumar}.}
  \bibinfo{year}{2002}\natexlab{}.
\newblock \showarticletitle{Connectivity and Inference Problems for Temporal
  Networks}.
\newblock \bibinfo{journal}{\emph{J. Comput. Syst. Sci.}} \bibinfo{volume}{64},
  \bibinfo{number}{4} (\bibinfo{year}{2002}), \bibinfo{pages}{820--842}.
\newblock


\bibitem[\protect\citeauthoryear{Kondor, P{\'{o}}sfai, Csabai, and
  Vattay}{Kondor et~al\mbox{.}}{2013}]%
        {DBLP:journals/corr/KondorPCV13}
\bibfield{author}{\bibinfo{person}{D{\'{a}}niel Kondor},
  \bibinfo{person}{M{\'{a}}rton P{\'{o}}sfai}, \bibinfo{person}{Istv{\'{a}}n
  Csabai}, {and} \bibinfo{person}{G{\'{a}}bor Vattay}.}
  \bibinfo{year}{2013}\natexlab{}.
\newblock \showarticletitle{Do the rich get richer? An empirical analysis of
  the BitCoin transaction network}.
\newblock \bibinfo{journal}{\emph{PLoS ONE}} \bibinfo{volume}{9},
  \bibinfo{number}{2} (\bibinfo{year}{2013}), \bibinfo{pages}{e86197}.
\newblock


\bibitem[\protect\citeauthoryear{Kovanen, Karsai, Kaski, Kert{\'{e}}sz, and
  Saram{\"{a}}ki}{Kovanen et~al\mbox{.}}{2011}]%
        {DBLP:journals/corr/abs-1107-5646}
\bibfield{author}{\bibinfo{person}{Lauri Kovanen},
  \bibinfo{person}{M{\'{a}}rton Karsai}, \bibinfo{person}{Kimmo Kaski},
  \bibinfo{person}{J{\'{a}}nos Kert{\'{e}}sz}, {and} \bibinfo{person}{Jari
  Saram{\"{a}}ki}.} \bibinfo{year}{2011}\natexlab{}.
\newblock \showarticletitle{Temporal motifs in time-dependent networks}.
\newblock \bibinfo{journal}{\emph{CoRR}}  \bibinfo{volume}{abs/1107.5646}
  (\bibinfo{year}{2011}).
\newblock
\showeprint[arxiv]{1107.5646}
\urldef\tempurl%
\url{http://arxiv.org/abs/1107.5646}
\showURL{%
\tempurl}


\bibitem[\protect\citeauthoryear{Leskovec, McGlohon, Faloutsos, Glance, and
  Hurst}{Leskovec et~al\mbox{.}}{2007}]%
        {DBLP:conf/sdm/LeskovecMFGH07}
\bibfield{author}{\bibinfo{person}{Jure Leskovec}, \bibinfo{person}{Mary
  McGlohon}, \bibinfo{person}{Christos Faloutsos}, \bibinfo{person}{Natalie~S.
  Glance}, {and} \bibinfo{person}{Matthew Hurst}.}
  \bibinfo{year}{2007}\natexlab{}.
\newblock \showarticletitle{Patterns of Cascading Behavior in Large Blog
  Graphs}. In \bibinfo{booktitle}{\emph{SDM}}. \bibinfo{pages}{551--556}.
\newblock


\bibitem[\protect\citeauthoryear{Li, Lou, Shi, and Han}{Li
  et~al\mbox{.}}{2018}]%
        {li2018temporal}
\bibfield{author}{\bibinfo{person}{Yuchen Li}, \bibinfo{person}{Zhengzhi Lou},
  \bibinfo{person}{Yu Shi}, {and} \bibinfo{person}{Jiawei Han}.}
  \bibinfo{year}{2018}\natexlab{}.
\newblock \showarticletitle{Temporal Motifs in Heterogeneous Information
  Networks}. In \bibinfo{booktitle}{\emph{MLG Workshop @ KDD}}.
\newblock


\bibitem[\protect\citeauthoryear{McAuley and Leskovec}{McAuley and
  Leskovec}{2012}]%
        {DBLP:conf/nips/McAuleyL12}
\bibfield{author}{\bibinfo{person}{Julian~J. McAuley} {and}
  \bibinfo{person}{Jure Leskovec}.} \bibinfo{year}{2012}\natexlab{}.
\newblock \showarticletitle{Learning to Discover Social Circles in Ego
  Networks}. In \bibinfo{booktitle}{\emph{NIPS}}. \bibinfo{pages}{548--556}.
\newblock


\bibitem[\protect\citeauthoryear{Meiklejohn, Pomarole, Jordan, Levchenko,
  McCoy, Voelker, and Savage}{Meiklejohn et~al\mbox{.}}{2013}]%
        {DBLP:conf/imc/MeiklejohnPJLMVS13}
\bibfield{author}{\bibinfo{person}{Sarah Meiklejohn}, \bibinfo{person}{Marjori
  Pomarole}, \bibinfo{person}{Grant Jordan}, \bibinfo{person}{Kirill
  Levchenko}, \bibinfo{person}{Damon McCoy}, \bibinfo{person}{Geoffrey~M.
  Voelker}, {and} \bibinfo{person}{Stefan Savage}.}
  \bibinfo{year}{2013}\natexlab{}.
\newblock \showarticletitle{A fistful of bitcoins: characterizing payments
  among men with no names}. In \bibinfo{booktitle}{\emph{IMC}}.
  \bibinfo{pages}{127--140}.
\newblock


\bibitem[\protect\citeauthoryear{Milo, Shen-Orr, Itzkovitz, Kashtan,
  Chklovskii, and Alon1}{Milo et~al\mbox{.}}{2004}]%
        {Milo02networkmotifs:}
\bibfield{author}{\bibinfo{person}{R. Milo}, \bibinfo{person}{S. Shen-Orr},
  \bibinfo{person}{S. Itzkovitz}, \bibinfo{person}{N. Kashtan},
  \bibinfo{person}{D. Chklovskii}, {and} \bibinfo{person}{U. Alon1}.}
  \bibinfo{year}{2004}\natexlab{}.
\newblock \showarticletitle{Network Motifs: Simple Building Blocks of Complex
  Networks}.
\newblock \bibinfo{journal}{\emph{Science}} \bibinfo{volume}{298},
  \bibinfo{number}{5594} (\bibinfo{year}{2004}), \bibinfo{pages}{824--827}.
\newblock


\bibitem[\protect\citeauthoryear{Nakamoto}{Nakamoto}{2007}]%
        {Nakamoto_bitcoin:a}
\bibfield{author}{\bibinfo{person}{Satoshi Nakamoto}.}
  \bibinfo{year}{2007}\natexlab{}.
\newblock \bibinfo{title}{Bitcoin: A peer-to-peer electronic cash system
  http://bitcoin.org/bitcoin.pdf}.
\newblock
\newblock


\bibitem[\protect\citeauthoryear{Paranjape, Benson, and Leskovec}{Paranjape
  et~al\mbox{.}}{2017}]%
        {DBLP:conf/wsdm/ParanjapeBL17}
\bibfield{author}{\bibinfo{person}{Ashwin Paranjape},
  \bibinfo{person}{Austin~R. Benson}, {and} \bibinfo{person}{Jure Leskovec}.}
  \bibinfo{year}{2017}\natexlab{}.
\newblock \showarticletitle{Motifs in Temporal Networks}. In
  \bibinfo{booktitle}{\emph{{WSDM}}}. \bibinfo{pages}{601--610}.
\newblock


\bibitem[\protect\citeauthoryear{Ranu and Singh}{Ranu and Singh}{2009}]%
        {DBLP:conf/icde/RanuS09}
\bibfield{author}{\bibinfo{person}{Sayan Ranu} {and} \bibinfo{person}{Ambuj~K.
  Singh}.} \bibinfo{year}{2009}\natexlab{}.
\newblock \showarticletitle{GraphSig: {A} Scalable Approach to Mining
  Significant Subgraphs in Large Graph Databases}. In
  \bibinfo{booktitle}{\emph{ICDE}}. \bibinfo{pages}{844--855}.
\newblock


\bibitem[\protect\citeauthoryear{Roy, Zeng, Bagga, Porter, and Snoeren}{Roy
  et~al\mbox{.}}{2015}]%
        {DBLP:journals/ccr/RoyZBPS15}
\bibfield{author}{\bibinfo{person}{Arjun Roy}, \bibinfo{person}{Hongyi Zeng},
  \bibinfo{person}{Jasmeet Bagga}, \bibinfo{person}{George Porter}, {and}
  \bibinfo{person}{Alex~C. Snoeren}.} \bibinfo{year}{2015}\natexlab{}.
\newblock \showarticletitle{Inside the Social Network's (Datacenter) Network}.
\newblock \bibinfo{journal}{\emph{Computer Communication Review}}
  \bibinfo{volume}{45}, \bibinfo{number}{5} (\bibinfo{year}{2015}),
  \bibinfo{pages}{123--137}.
\newblock


\bibitem[\protect\citeauthoryear{Semertzidis and Pitoura}{Semertzidis and
  Pitoura}{2016}]%
        {DBLP:conf/icde/SemertzidisP16}
\bibfield{author}{\bibinfo{person}{Konstantinos Semertzidis} {and}
  \bibinfo{person}{Evaggelia Pitoura}.} \bibinfo{year}{2016}\natexlab{}.
\newblock \showarticletitle{Durable graph pattern queries on historical
  graphs}. In \bibinfo{booktitle}{\emph{ICDE}}. \bibinfo{pages}{541--552}.
\newblock


\bibitem[\protect\citeauthoryear{Wernicke and Rasche}{Wernicke and
  Rasche}{2006}]%
        {DBLP:journals/bioinformatics/WernickeR06}
\bibfield{author}{\bibinfo{person}{Sebastian Wernicke} {and}
  \bibinfo{person}{Florian Rasche}.} \bibinfo{year}{2006}\natexlab{}.
\newblock \showarticletitle{{FANMOD:} a tool for fast network motif detection}.
\newblock \bibinfo{journal}{\emph{Bioinformatics}} \bibinfo{volume}{22},
  \bibinfo{number}{9} (\bibinfo{year}{2006}), \bibinfo{pages}{1152--1153}.
\newblock


\bibitem[\protect\citeauthoryear{Xiang, Neville, and Rogati}{Xiang
  et~al\mbox{.}}{2010}]%
        {DBLP:conf/www/XiangNR10}
\bibfield{author}{\bibinfo{person}{Rongjing Xiang}, \bibinfo{person}{Jennifer
  Neville}, {and} \bibinfo{person}{Monica Rogati}.}
  \bibinfo{year}{2010}\natexlab{}.
\newblock \showarticletitle{Modeling relationship strength in online social
  networks}. In \bibinfo{booktitle}{\emph{WWW}}. \bibinfo{pages}{981--990}.
\newblock


\bibitem[\protect\citeauthoryear{Yavero\u{g}lu, Malod-Dognin, Davis, Levnajic,
  Janjic, Karapandza, and Pr\v{z}ulj}{Yavero\u{g}lu et~al\mbox{.}}{2014}]%
        {Yaverolu2014RevealingTH}
\bibfield{author}{\bibinfo{person}{\"{O}mer Yavero\u{g}lu},
  \bibinfo{person}{No\"{e}l Malod-Dognin}, \bibinfo{person}{Darren Davis},
  \bibinfo{person}{Zoran Levnajic}, \bibinfo{person}{Vuk Janjic},
  \bibinfo{person}{Aleksandar Karapandza, Rasa~Stojmirovic}, {and}
  \bibinfo{person}{Nata\v{s}a Pr\v{z}ulj}.} \bibinfo{year}{2014}\natexlab{}.
\newblock \showarticletitle{{Revealing the Hidden Language of Complex
  Networks}}.
\newblock \bibinfo{journal}{\emph{Scientific Reports}}  \bibinfo{volume}{4}
  (\bibinfo{year}{2014}), \bibinfo{pages}{4547}.
\newblock


\bibitem[\protect\citeauthoryear{Zhao, Tian, He, Oliver, Jin, and Lee}{Zhao
  et~al\mbox{.}}{2010}]%
        {DBLP:conf/cikm/ZhaoTHOJL10}
\bibfield{author}{\bibinfo{person}{Qiankun Zhao}, \bibinfo{person}{Yuan Tian},
  \bibinfo{person}{Qi He}, \bibinfo{person}{Nuria Oliver},
  \bibinfo{person}{Ruoming Jin}, {and} \bibinfo{person}{Wang{-}Chien Lee}.}
  \bibinfo{year}{2010}\natexlab{}.
\newblock \showarticletitle{Communication motifs: a tool to characterize social
  communications}. In \bibinfo{booktitle}{\emph{CIKM}}.
  \bibinfo{pages}{1645--1648}.
\newblock


\bibitem[\protect\citeauthoryear{Z{\"{u}}fle, Renz, Emrich, and
  Franzke}{Z{\"{u}}fle et~al\mbox{.}}{2018}]%
        {DBLP:conf/edbt/ZufleREF18}
\bibfield{author}{\bibinfo{person}{Andreas Z{\"{u}}fle},
  \bibinfo{person}{Matthias Renz}, \bibinfo{person}{Tobias Emrich}, {and}
  \bibinfo{person}{Maximilian Franzke}.} \bibinfo{year}{2018}\natexlab{}.
\newblock \showarticletitle{Pattern Search in Temporal Social Networks}. In
  \bibinfo{booktitle}{\emph{{EDBT}}}. \bibinfo{pages}{289--300}.
\newblock


\end{thebibliography}
